\documentclass[LaM, oneside, noexaminfo]{sapthesis}

\usepackage{mathtools}
\usepackage{amssymb}
\usepackage{tensor}
\usepackage{subfig}
\usepackage{float}
\usepackage{cite}
\usepackage{comment}

\title{Instability of Schwarzschild Black Holes in Einstein-scalar-Gauss-Bonnet Gravity: Perturbative Approach and Time-Domain Analysis}
\author{Fabrizio Corelli}
\IDnumber{1706626}
\course{Fisica}
\courseorganizer{Facolt\`a di Scienze Matematiche, Fisiche e Naturali}
\AcademicYear{2019/2020}
\copyyear{2020}
\advisor{Prof. Paolo Pani}
\coadvisor{Dr. Taishi Ikeda}
\authoremail{fabrizio.corelli@uniroma1.it}
\versiondate{22nd December 2021}

\newcommand{\G}{\mathcal{G}}
\newcommand{\OO}{\mathcal{O}}

\newcommand{\eg}{\textit{e.g. }}
\newcommand{\ie}{\textit{i.e. }}
\newcommand{\ea}{\textit{et al.} }
\newcommand{\abs}[1]{\lvert #1 \rvert}

\begin{document}

\frontmatter

\maketitle

\abstract{
We study the instability of Schwarzschild black holes and the appearance of scalarized solutions in Einstein-scalar-Gauss-Bonnet gravity performing a time-domain analysis in a perturbative scheme. First we consider a quadratic coupling function and we perform an expansion for a small perturbation of the scalar field around the Schwarzschild solution up to the second order; we do not observe any stable scalarized configuration, in agreement with previous studies. We then consider the cases of quartic and exponential coupling, using an expansion for small values of the Newton's constant, in order to  include the nonlinear terms introduced by the coupling in the field equations; in this case we observe the appearance of stable scalarized solutions different from those found in literature. The discrepancy can be an artifact of the perturbative approach. \\

	% Thesis defended on 20th October 2020.\\
}

\tableofcontents

\mainmatter

\chapter{Introduction}
\label{chapter:introduction}

General Relativity was introduced by Albert Einstein in 1915. It represents one of the most important theories of the last century, and has been widely tested in the range of lenghts $1 \, \mu m \lesssim l \lesssim 10^{11} \, m$. In particular the recent observations of binary black hole mergers and binary neutron star mergers have provided excellent confirmations of the predictions of the Einstein's theory of gravity. \\ 
However there are some critical aspects from a theoretical and cosmological point of view that suggest to investigate alternative theories of gravity. For instance, it can be shown that General Relativity is not renormalizable and therefore it cannot be easily treated from quantum field theory point of view. Moreover, the accelerated expansion of the universe leads to the introduction of dark energy, which can be explained with a nonzero cosmological constant. However, the measurements of the cosmological constant from experimental observations are in contrast with the estimations of the zero-point energy obtained by quantum field theory computations.\\ 

An interesting way to extend General Relativity is to introduce a dynamical scalar field in the action; in this case the equations for the metric contain also the scalar field and they are different from the Einstein's equations. When the scalar field is nonminimally coupled to gravity the theory goes under the name of scalar-tensor gravity. \\ 
A critical issue of scalar-tensor theories is that the field equations can be of order higher than two. This can lead to the appearance of a ghost-like instability called Ostrogradsky instability. However this phenomenon can be avoided by considering the so called Horndeski theories, which are the most general scalar-tensor theories whose field equations are of order two. \\ 
Another interesting theory is quadratic gravity, in which the action contains quadratic terms in the curvature, and it can be considered as a low-energy approximation of a more general action in a quantum field theory model that includes gravity. \\ 

In this work we are going to study Einstein-scalar-Gauss-Bonnet (EsGB) gravity, in which the coupling between the scalar field and the metric is given by a coupling function multiplied by the Gauss-Bonnet invariant $\G = R^2 - 4 R_{\mu\nu}R^{\mu\nu} + R_{\mu\nu\alpha\beta}R^{\mu\nu\alpha\beta}$. This theory belongs simultaneously to the classes of Horndeski and quadratic gravity. \\
The corrections of EsGB gravity to General Relativity are in the high curvature/energy regime and it is interesting to analyze the characteristics of compact objects, in particular black holes.\\ 

Thanks to the no-hair theorem in General Relativity black holes are described by the Kerr-Newman metric, which is defined using only three quantities: mass, angular momentum and electric charge. This theorem is very stringent and expresses a conceptual simplicity in the description of black holes, that can be used to test the validity of an alternative theory of gravity: the eventual observation of deviations from the predictions of the no-hair theorem can give a confirmation to a theory that introduces auxiliary quantities for the description of the black hole metric. Moreover, even if these deviations are absent, the prediction of a theory of modified gravity can be useful to understand from an observational point of view how the no-hair theorem operates and why it is not violated. \\

In some EsGB gravities when the scalar field assumes a trivial configuration, the black hole solution reduces to the one predicted by General Relativity. Nevertheless there are choices of the coupling function for which the scalar field can assume nontrivial configurations that require additional constants (usually referred as ``hairs'') to be described. \\
Recently a new phenomenon called \textit{spontaneous black hole scalarization} has been found to occur in EsGB gravity (refs. \cite{EGBNoHair, DonevaYazadjiev}). This phenomenon is the analog of the spontaneous scalarization introduced in the 90's by Damour and Esposito-Far\`ese for neutron stars (refs. \cite{DEF1, DEF2}), and it consists in the development of an instability in the trivial solution of the scalar field and in the appearance of stable nontrivial configurations. The additional parameter required to describe the black hole solutions is called scalar charge, while the nontrivial configurations of the scalar field are called scalarized solutions. \\
From an observational point of view, the effects of the presence of the scalar charge may be seen in gravitational waves. In fact a binary system of scalarized black holes possesses a scalar dipole moment that, rotating, emits scalar dipole radiation, affecting the gravitational wave signal (see \eg ref. \cite{GWcorrections}). \\
Spontaneous scalarization has been studied for quadratic, quartic and exponential coupling functions in the case of massless scalar field. It has been found that for certain choices of the parameters in the coupling function the Schwarzschild solution is unstable, and there are scalarized solutions that can be classified by the number of nodes and the scalar charge. In the case of quadratic coupling function there are not stable scalarized solutions, while in the other two cases there are stable nontrivial configurations of the scalar field, and the scalarized solutions with $n \ge 1$ nodes are unstable. According to ref. \cite{Silva+Stability} this behavior is related to the presence of nonlinear terms in the equation for the scalar field. In fact the quadratic term of the coupling function contributes with a linear term in the equation, and the higher order terms in the coupling function provide nonlinear terms that can quench the instability. In the case of quadratic coupling these nonlinear terms are absent and the scalarized solutions are unstable. \\
In recent years spontaneous scalarization has also been studied for charged black holes (see refs. \cite{DonevaCharged, Charged2}), for massive scalar field (see refs. \cite{Macedo+, DonevaMassive}) and for spinning black holes (see refs. \cite{Dima+, Hod}). We will not analyze these studies, but we will focus only on the spherically symmetric case with massless scalar field. \\

In this work we will study the instability of a Schwarzschild black hole and the appearance of stable scalarized solutions in some EsGB gravities performing a time-domain analysis in a perturbative scheme. The time-domain approach allows us to delineate the evolution of the solutions, to study how the instability appears, and how the stable configurations are reached. \\
In order to compute the time evolution of the system we developed a 1+1 code in C for the numerical integration of time dependent partially differential equations with a constraint.\\
In the first part we are going to consider a quadratic coupling function $F[\phi] = \lambda \phi^2$, where $\phi$ is the scalar field and $\lambda$ is a coupling constant. We will assume $\phi$ as a small perturbation around the Schwarzschild solution and then we will expand up to the second order in the field equations. In this approximation we can consider the effects of the backreaction of the scalar field on the metric, and the equation for $\phi$ can be integrated using the Schwarzschild metric as background. This remarkable simplification in the field equations is the point of strength of the perturbative approach. \\
Our intent is to determine the values of the coupling parameter $\lambda$ for which the Schwarzschild solution is stable and to characterize the instability. We will also discuss the possibility of studying the evolution of the position of the apparent horizon in this perturbative scheme.\\

Then we are going to consider the instability of a Schwarzschild black hole with a quartic coupling function $F[\phi] = \lambda \phi^2 + \gamma \phi^4$. In this case the equation for the scalar field contains nonlinear terms in $\phi$, and stable nodeless scalarized solutions have been found, therefore it is interesting to study how the instability of the Schwarzschild solution is quenched and how the scalar field reaches a nontrivial stable configuration. \\
However the perturbative scheme we are going to use for the quadratic coupling function cannot be used with the quartic coupling. In fact, expanding in a perturbation of the scalar field, the terms that come from $\phi^4$ do not appear in the equations at the second order. Therefore we will perform an expansion for a small Newton's constant $G$, in such a way that we can treat the linear and the nonlinear terms in the equation for the scalar field at the same order in perturbation theory. \\
We will determine with a time-domain analysis the ranges of values for the parameters $\lambda$ and $\gamma$ where the Schwarzschild solution is unstable, we will discuss the appearance of stable scalarized solutions and compute their scalar charge. \\
We will also consider the case of exponential coupling $F[\phi] = \frac{\lambda}{3} \Bigl( 1 - e^{-3\phi^2} \Bigr)$ using the expansion in the Newton's constant. As in the case of quartic coupling, the equation for the scalar field contains nonlinearities and stable scalarized solutions appear.\\

The structure of the thesis is the following. \\
In chapter \ref{chapter:review} we will discuss the reasons to introduce modifications in General Relativity and a scheme to classify the alternative theories on the basis of the violation of the Lovelock's theorem, following the structure of ref. \cite{Berti+}. We will introduce scalar-tensor theories, Horndeski theories and quadratic gravity, arriving then to Einstein-scalar-Gauss-Bonnet gravity. We will explore the no-hair theorem in General Relativity, and some extensions in scalar-tensor theories, analyzing in detail the case of EsGB gravity. Finally we will discuss the phenomenon of spontaneous scalarization, and we will review some results obtained in recent years. \\
In chapter \ref{chapter:QuadraticCoupling} we will analyze the case of quadratic coupling function. We will perform the expansion for a perturbation around the trivial configuration of the scalar field, and we will integrate numerically the equations analyzing the stability of the solutions. We will then compute the evolution of the position of the apparent horizon and discuss the possibility of studying it in the perturbative scheme. \\
In chapter \ref{chapter:QuarticCoupling} and \ref{chapter:ExponentialCoupling} we will consider respectively the quartic and the exponential couplings with the expansion in the Newton's constant, analyzing the stability of the scalarized solutions and computing their scalar charge. \\
The thesis ends with the conclusions in chapter \ref{chapter:conclusions}.

\chapter{Modifications of General Relativity and Hairy Black Holes} 
\label{chapter:review}

In this chapter we will discuss some possible approaches for extending General Relativity. In particular we will describe scalar-tensor theories and quadratic gravity, and we will focus on EsGB gravity. We will discuss about possible extensions of the no-hair theorems in EsGB gravity, and the occurrence of the so called spontaneous scalarization, reporting some results obtained in recent years. \\
In the first part of the chapter we will follow the presentation in ref. \cite{Berti+}.
\section{Necessity for a Modification of General Relativity}

General Relativity (GR) has received several confirmations during the last century, from the deflection of light discovered by Arthur Eddington in 1919 \cite{Eddington} to the recent detections of gravitational waves from binary systems by the LIGO-Virgo collaboration starting from 2015 \cite{LIGO1, LIGO2, LIGO3, LIGO4, LIGO5, LIGO6, LIGO7, LIGO8, LIGO9, LIGO10} and the image of a black hole with its shadow obtained in 2019 by the EHT collaboration \cite{EHTC1,EHTC2,EHTC3,EHTC4,EHTC5,EHTC6}.\\
However, there are reasons to investigate possible modifications of GR, and they come mainly from theoretical and cosmological arguments. \\
From a theoretical perspective the main issue is the classical nature of General Relativity. Using the renormalizability criterion it can be shown that GR is a non-renormalizable theory; in fact the dimension of the coupling constant in natural units ($\hbar = c = 1$) are given by the dimensions of the gravitational constant, which are $[G] = [E^{-2}]$ (see ref. \cite{Weinberg1}). Therefore GR cannot be included in a quantum field theory description of the universe. \\
At the same time, the discovery of the accelerated expansion of the universe in 1998 \cite{CosmologicalConstantMeasure} highlighted new criticalities from a cosmological point of view. In GR this phenomenon can be described with a positive cosmological constant\footnote{See also ref. \cite{DarkEnergy} for a review on the Dark Energy and the accelerated expansion of the universe.} \cite{CosmologicalConstantCarroll}, whose evaluation poses new critical issues in the context of the General Relativity and the Standard Model of particle physics, such as the \textit{cosmological constant problem} (or \textit{vacuum catastrophe}), which is the discrepancy between the measured value of the cosmological constant and the quantum field theory prediction of the zero-point energy \cite{WeinbergCosmologicalConstant}.\\
A modified version of General Relativity must include correction at low or high energies, while it must reduce to GR in the intermediate energy regime, in order to be consistent with the current experimental results. In particular, the length scale where GR has been tested is $1 \, \mu m \lesssim \mathit{l} \lesssim 10^{11} \, m$ (see ref. \cite{Berti+}) and the corrections must appear out of this range. Corrections at high energy and small length scales are called \textit{ultraviolet (UV) corrections}, whose effects can appear in astrophysical compact objects, such as Black Holes (BH) and Neutron Stars (NS), while corrections at low energy and long length scales are called \textit{infrared (IR) corrections}, and they can be investigated by examining cosmological phenomena and also from the gravitational waves (see \eg \cite{NS-NSmerger}). 

\section{Lovelock's Theorem and Possible Modifications of GR}

Let us now delineate a possible scheme to introduce modifications in GR. This scheme is described in detail in ref. \cite{Berti+}, and we are going to summarize it without dwelling on the illustration of all the theories referable to it.
According to Lovelock's theorem \cite{Lovelock}, in four dimensions any rank-2 symmetric tensor $A^{\mu\nu} = A^{\nu\mu}$ with null divergence $\tensor{A}{^{\mu\nu}_{;\nu}}$, constructed only from the metric tensor $g^{\mu\nu}$ and its first and second order derivatives, and which preserves diffeomorphism invariance, can be written as a linear combination of the Einstein tensor $G^{\mu\nu} = R^{\mu\nu} - \frac{1}{2} R g^{\mu\nu}$ and $g^{\mu\nu}$:
\begin{equation}
	A^{\mu\nu} = \alpha G^{\mu\nu} + \beta g^{\mu\nu}.
	\label{eq:Lovelock}
\end{equation}
This leads to the Einstein's equations with the addiction of a cosmological term
\begin{equation}
	G_{\mu\nu} + \Lambda g_{\mu\nu} = \frac{8 \pi G}{c^4} T_{\mu\nu},
	\label{eq:EinsteinCosmological}
\end{equation}
where $T_{\mu\nu}$ is the matter stress-energy tensor and $\Lambda$ is the cosmological constant.\\
In order to modify GR we have to violate one or more hypotheses of the Lovelock's theorem. We are now going to list some possible violations of the postulates, following ref. \cite{Berti+}, and some theories proposed.
\begin{enumerate}
	\item \textbf{Violation of the hypothesis that $A^{\mu\nu}$ must be constructed only with $g^{\mu\nu}$ and its first and second derivatives.}

		This hypothesis can be violated by adding a field, which can be either dynamical and nondynamical. In the case of dynamical field we can add in the left hand side of the Einstein's equations some couplings between the metric tensor and the field \cite{Berti+}. \textit{Einstein-scalar-Gauss-Bonnet} (EsGB) gravity belongs to this class of theories. Instead in the case of nondynamical field the aim is to write the Einstein's equations with $G_{\mu\nu}$ on the left hand side, and, on the right hand side, a nonlinear combination of a stress-energy tensor that satisfies the condition $\tensor{T}{^{\mu\nu}_{;\nu}} = 0$, in such a way that the weak equivalence principle holds. A theory of this type is the \textit{Palatini} $f(\mathcal{R})$ gravity \cite{Berti+}.

	\item \textbf{Violation of the hypothesis of a 4 dimensional spacetime.}

		This hypothesis is violated by considering the number of dimensions as a parameter of the theory. This could be interesting in order to study how the theory depends on the number of dimensions, and also in the context of string theory. An example of higher dimensional theory is the \textit{Kaluza-Klein} gravity \cite{Berti+}. 

	\item \textbf{Violation of the diffeomorphism invariance hypothesis.}

		Two common ways of violating the diffeomorphism invariance are to break Lorentz invariance or to consider a model of massive gravity. In the first case the idea is that Lorentz symmetry is broken at high energy and hence this type of theory can provide UV corrections to GR. A Lorentz violating theory is, for instance, \textit{Einstein-\AE ther} gravity \cite{Berti+}. Instead in the second case the idea is to consider a massive graviton, \ie a massive spin-2 field. An example of massive gravity theory is \textit{de Rham-Gabadadze-Tolley} (dRGT) gravity \cite{Berti+}.

	\item \textbf{Violation of the hypothesis $\tensor{A}{^{\mu\nu}_{;\nu}} = 0$.}

		In this case the general idea is that if we violate the \textit{Weak Equivalence Principle} (WEP), then condition $\tensor{T}{^{\mu\nu}_{;\nu}} = 0$ does not hold anymore and the hypothesis $\tensor{A}{^{\mu\nu}_{;\nu}} = 0$ can be released \cite{Berti+}.

\end{enumerate}

\section{EsGB Gravity Introduction and General Aspects}

Now we are going to discuss about the theory of modified gravity we want to study in this work. To explain the general aspects of the theory we are following ref. \cite{Berti+}.
From now we will use geometrized units ($c = 8 \pi G = 1$).

\subsection{Scalar-Tensor Theories and Horndeski Gravity}
Let us start writing the Einstein-Hilbert action, whose variation yields the Einstein's equations in vacuum:
\begin{equation}
	S_{EH} = \frac{1}{2} \int_\Omega d^4 x \, \sqrt{-g} \, R.
	\label{eq:E-H}
\end{equation}
Eq. \eqref{eq:E-H} is the starting point for introducing modifications to GR.

Our approach to avoid the Lovelock's theorem is to introduce a dynamical scalar field $\phi$, with scalar potential $V(\phi)$, and add interaction terms between the metric and the field.
The theory we want to consider belongs to the class called \textit{Scalar-Tensor gravity}, whose theories are characterized by an action that contains a nonminimal coupling between the scalar field $\phi$ and the metric as follows \cite{Berti+, HorndeskiOriginal, QuirosScalarTensor}
\begin{equation}
	S_{ST} = \frac{1}{2} \int_\Omega \, \sqrt{-g} \biggl\{ \phi R - \frac{\omega[\phi]}{\phi} (\nabla \phi)^2 - V[\phi] \biggr\} + S_M[\psi, g_{\mu\nu}],
	\label{eq:ScalarTensorDefBD}
\end{equation}
where $(\nabla \phi)^2 = \nabla_\mu \phi \nabla^\mu \phi = \partial_\mu \phi \partial^\mu \phi$, and $S_M[\psi, g_{\mu\nu}]$ is the action of matter. \\
Alternatively, with a redefinition of the $\phi$ and $\omega[\phi]$, the action can be written as \cite{QuirosScalarTensor}
\begin{equation}
	S_{ST} = \frac{1}{2} \int_\Omega \, \sqrt{-g} \biggl\{ F[\phi] R - \omega[\phi] (\nabla \phi)^2 - V[\phi] \biggr\} + S_M[\psi, g_{\mu\nu}].
	\label{eq:ScalarTensorDef}
\end{equation}
These actions are defined in the so called \textit{Jordan Frame}. Another representation of the theory can be given in the \textit{Einstein Frame} and it can be obtained from \eqref{eq:ScalarTensorDefBD} redefining the scalar field and performing a conformal transformation:
\begin{gather}
	\varphi = \varphi[\phi], \notag \\
	g^*_{\mu\nu} = A^{-2}[\varphi]g_{\mu\nu}.
	\label{eq:conformal}
\end{gather}
After these transformations the action can be rewritten as \cite{Berti+}
\begin{equation}
	S_{ST} = \frac{1}{2} \int_\Omega \, \sqrt{-g^*} \biggl\{ R^* - (\nabla^* \varphi)^2 - V[\varphi] \biggr\} + S_M[\psi, A^2[\varphi] g^*_{\mu\nu}],
	\label{eq:ScalarTensorEinstein}
\end{equation}
where $(\nabla^* \varphi)^2 = g^*_{\mu\nu}\partial^\mu \varphi \partial^\nu \varphi$.\\
In the Einstein frame the scalar field is minimally coupled to gravity while in $S_M$ there is a nonminimal coupling between the matter and the scalar field \cite{Berti+, Representations}.\\

Scalar-tensor theories are subject to the so called \textit{Ostrogradsky instability}, which is a ghost-like instability related to the presence of a nondegenerate lagrangian whose Euler-Lagrange equations are of order higher than two. 
We will not go through the details of Ostrogradsky's construction, that can be found in ref. \cite{WoodardOstrogradski}, but we will show how this instability appears using the example in the context of classical mechanics that can be found in ref. \cite{KobayashiHorndeski}. \\
Let $a$ be a non-null real constant, $\phi$ a lagrangian variable and $V(\phi)$ a potential. Let us consider a system described by the lagrangian
\begin{equation}
	L = \frac{a}{2} \ddot{\phi}^2 - V(\phi).
	\label{eq:lagrangianclassicOstrograsky}
\end{equation}
When the lagrangian contains a time derivative of order two the Lagrange equation is
\begin{equation}
	-\frac{d^2}{d t^2} \frac{\partial L}{\partial \ddot \phi} + \frac{d}{dt} \frac{\partial L}{\partial \dot \phi} - \frac{\partial L}{\partial \phi} = 0.
	\label{eq:LagrangeSecondDerivative}
\end{equation}
For the lagrangian \eqref{eq:lagrangianclassicOstrograsky} we obtain
\begin{equation}
	a \ddddot{\phi} - \frac{\partial V(\phi)}{\partial \phi} = 0.
	\label{eq:lagrangeequationclassic}
\end{equation}
To solve this Cauchy problem we should impose four initial conditions: $\phi(t = 0)$, $\dot \phi(t = 0)$, $\ddot \phi(t = 0)$, $\dddot \phi(t = 0)$. Defining the variable $\psi = \ddot \phi$ the lagrangian can be written as
\begin{equation}
	L = a \psi \ddot \phi - \frac{a}{2} \psi^2 - V(\phi) = a \frac{d}{dt}\bigl( \psi \dot \phi) - a \dot \psi \dot \phi - \frac{a}{2} \psi^2 - V(\phi) 
\end{equation}
The lagrangian is defined up to the total time derivative of a function, and therefore the first term in the second step can be neglected. In the new lagrangian the two dynamical degrees of freedom have been made explicit in two lagrangian variables, for each of which two initial conditions are required. \\
Defining the new variables
\begin{equation}
	q = \frac{\phi + \psi}{\sqrt{2}}, \qquad \qquad Q = \frac{\phi - \psi}{\sqrt{2}},
\end{equation}
we can rewrite the lagrangian as
\begin{equation}
	L = -\frac{a}{2} \dot{q}^2 + \frac{a}{2} \dot{Q}^2 - U(q, Q),
	\label{eq:NewLagrangianQ}
\end{equation}
where $U(q, Q) = V\Bigl(\frac{q + Q}{\sqrt{2}}\Bigr) + \frac{a}{2} \Bigl( \frac{q - Q}{\sqrt{2}} \Bigr)^2$. As we can see in this coordinates one of the two degrees of freedom (depending on the sign of $a$) has a negative kinetic term, and therefore it generates the Ostrogradsky instability. Such a degree of freedom is called \textit{ghost}.\\

In order to avoid the Ostrogradsky instability we should restrict ourselves to \textit{Horndeski gravity}, which is defined as the most general scalar-tensor theory with second order field equations \cite{Berti+,KobayashiHorndeski}. The action of Horndeski gravity is
\begin{multline}
	S_H = \int_\Omega d^4 \, \sqrt{-g} \biggl\{ G_2[\phi, X] - G_3[\phi, X] \Box \phi + \\
	+ G_4[\phi, X] R + G_{4, X}[\phi, X] \Bigl[ (\Box \phi)^2 - (\nabla_\mu \nabla_\nu \phi) (\nabla^\mu \nabla^\nu \phi) \Bigr] + \\
	+ G_5[\phi, X] G_{\mu\nu} \nabla^\mu \nabla^\nu \phi - \frac{G_{5, X}[\phi, X]}{6} \Bigl[ (\Box \phi)^3 - 3(\Box \phi)(\nabla_\mu \nabla_\nu \phi)(\nabla^\mu \nabla^\nu \phi) + \\
	+ 2 (\nabla_\mu \nabla_\nu \phi)(\nabla^\mu \nabla_\sigma \phi)(\nabla^\nu \nabla^\sigma \phi) \Bigr] \biggr\},
	\label{eq:Horndeski}
\end{multline}
where $X:= -\frac{1}{2} (\nabla \phi)^2$, $G_i[\phi, X]$ with $i \in \{2, 3, 4, 5\}$ are functions of $\phi$ and $X$, and $G_{i, X} = \frac{\partial G_i}{\partial X}$.

\subsection{Renormalizability and Quadratic Gravity}

Let us now focus on a different aspect: renormalizability.\\
As we said earlier in this chapter, it has been proven that the Einstein-Hilbert action is not renormalizable. A possible approach to treat this problem is to include in the action terms quadratic in curvature. In fact it has been proven by Stelle in ref. \cite{StelleRenormalization} that such an action would be renormalizable. A reason for the introduction of quadratic terms in the action comes from low-energy effective string theories and is to consider the Einstein-Hilbert action as the first term of an expansion that contains other curvature invariants \cite{Berti+}. \\

In the context of scalar-tensor theories, the interaction term of a quadratic gravity with a dynamical scalar field $\phi$ can be written multiplying all the algebraic independent quadratic curvature invariants by some functions of the field $f_1[\phi], \dots, f_4[\phi]$, obtaining the action\footnote{We are using a different normalization from the one used by Berti \ea in ref. \cite{Berti+}. In particular, if $\phi_B$ is the scalar field used in \cite{Berti+} then $\phi = \sqrt{2}\phi_B$. The normalization we are using is the one used in \cite{RipleyPretorius}.} \cite{Berti+}
\begin{multline}
	S_Q = \frac{1}{2} \int_\Omega d^4 x \, \sqrt{-g} \biggl\{ R - (\nabla \phi)^2 - V[\phi] + f_1[\phi] R^2 + f_2[\phi] R_{\mu\nu}R^{\mu\nu} + \\
	f_3[\phi] R_{\mu\nu\rho\sigma}R^{\mu\nu\rho\sigma} + f_4[\phi] \frac{1}{2} R_{\mu\nu\rho\sigma}\epsilon^{\mu\nu\lambda\kappa}\tensor{R}{^{\rho\sigma}_{\lambda\kappa}} \biggr\} + S_M\Bigl[\psi, \gamma[\phi]g_{\mu\nu}\Bigr],
	\label{eq:Quadratic}
\end{multline}
where $\epsilon^{\mu\nu\lambda\kappa}$ is the totally antisymmetric Levi-Civita symbol. \\
In \eqref{eq:Quadratic} the function $\gamma[\phi]$ is a nonminimal coupling between the matter and $\phi$.\\
The field equation derived from this action can in general be of order higher than two and thus can appear ghosts related to Ostrogradsky instability. However for some choices of the functions $f_1[\phi], \dots, f_4[\phi]$ the theory belongs to the class of Horndeski gravity, and the instability does not appear \cite{Berti+}. \\

In the other cases a possible approach could be to treat the theory as an \textit{Effective Field Theory} (EFT), which we will describe following ref. \cite{Berti+}. This approach can be applied when a theory involves two well separated fundamental energy scales $M_1 << M_2$. In this case the action can be written as a function of the field $h$ in the high energy regime $E \sim M_2$ and the field $l$ in the low energy regime $E \sim M_1$. If we consider measurements at the lower scale, using functions that depend only on $l$ and not on $h$, then the observables can be written as
\begin{multline}
	\langle \mathcal{O}_1 \cdots \mathcal{O}_n \rangle = \int \mathcal{D}h \, \mathcal{D}l \, [\mathcal{O}_1(l) \cdots \mathcal{O}_n(l)] \, e^{i S(h, l)} = \\
	= \int \mathcal{D}l \, [\mathcal{O}_1(l) \cdots \mathcal{O}_n(l)] \, e^{i S_{\text{eff}}(l)},
	\label{eq:EffectiveObservables}
\end{multline}
where $S_{\text{eff}}(l)$ is called \textit{effective action} and it can be defined by the relation
\begin{equation}
	e^{i S_{\text{eff}}(l)} = \int \mathcal{D}h \, e^{i S(h, l)}.
	\label{eq:EffectiveAction}
\end{equation}
Then $S_{\text{eff}}$ is expanded in powers of $\frac{1}{M_2}$. The effective action can now be interpreted also as an expansion in derivatives of $l$.\\
Using this formalism the action \eqref{eq:Quadratic} can be read as an expansion of the action of an effective field theory that derives from a more general quantum gravity theory. In this way the field equations can be treated perturbatively eliminating terms of order higher than two and thus avoiding Ostrogradsky instability. 

\subsection{Einstein-scalar-Gauss-Bonnet Gravity}

The theory we want to study in this work is \textit{Einstein-scalar-Gauss-Bonnet} (EsGB) gravity, whose action in vacuum is given by \cite{Berti+}
\begin{equation}
	S_{EsGB} = \frac{1}{2} \int_\Omega d^4 x \, \sqrt{-g} \, \biggl\{ R - (\nabla \phi)^2 - V[\phi] + F[\phi] \G \biggr\},
	\label{eq:EsGB}
\end{equation}
where $\G = \frac{1}{4} \delta^{\alpha\beta\mu\nu}_{\rho\sigma\lambda\omega} \tensor{R}{^{\rho\sigma}_{\alpha\beta}} \tensor{R}{^{\lambda\omega}_{\mu\nu}} = R^2 - 4 R_{\mu\nu}R^{\mu\nu} + R_{\mu\nu\alpha\beta}R^{\mu\nu\alpha\beta}$ is the \textit{Gauss-Bonnet invariant} and $\delta^{\alpha\beta\mu\nu}_{\rho\sigma\lambda\omega} = \epsilon^{\alpha\beta\mu\nu} \epsilon_{\rho\sigma\lambda\omega}$ is the generalized Kronecker delta. \\
This action has the following features \cite{Berti+}:
\begin{itemize}
	\item it is a scalar-tensor theory whose Euler-Lagrange equations are of second order, therefore it is a Horndeski theory and it is free from Ostrogradsky instability and the relative ghosts;
	\item it can be obtained from \eqref{eq:Quadratic} imposing $f_1[\phi] = -\frac{1}{4}f_2[\phi] = f_3[\phi] = F[\phi]$ and $f_4[\phi] = 0$, which means that it is a quadratic gravity and it can be considered as a low-energy approximation of a more general theory.
\end{itemize}
These characteristics make Einstein-scalar-Gauss-Bonnet gravity a very interesting GR modification to investigate. Moreover, black holes in this theories can evade \textit{no-hair theorem} and exhibit a phenomenon called \textit{spontaneous scalarization}. We will discuss these topics in the following sections.\\

We conclude this section with a small classification of EsGB theories on the basis of the choice of $F[\phi]$. The most common coupling functions are: 
\begin{itemize}
	\item $F[\phi] = \lambda e^{\gamma \phi}$

		Theories with this exponential coupling function go under the name of \textit{Einstein-dilaton-Gauss-Bonnet} (EdGB). This action is motivated by low-energy effective string theory, and in this context the scalar field is called \textit{dilaton}.

	\item $F[\phi] = \lambda \phi$

		In this case the action is invariant under the transformation $\phi \to \phi + \alpha$, where $\alpha$ is a constant, and for this reason these theories are called \textit{shift-symmetric Gauss-Bonnet}.

	\item $F[\phi] = \lambda \phi^2$

		In this case the action is invariant under $\mathbb{Z}_2$-symmetry, \ie under the transformation $\phi \to -\phi$. In chapter \ref{chapter:QuadraticCoupling} we are going to study this theory.
\end{itemize}

\section{Black Holes and No-Hair Theorems}

EsGB gravity provides corrections to GR in the high curvature/energy regime and for this reason an interesting research subject to study the modifications introduced is represented by black holes (BH) \cite{Berti+}. We are now going to discuss about BHs and no-hair theorems starting from GR and then moving to EsGB.

\subsection{General Relativity}

In GR a static, stationary and spherically symmetric black hole is described by the Schwarzschild metric\footnote{The mass variable in eq. \eqref{eq:SchwarzschildGR} in geometrized units with $8\pi G = c = 1$ is actually $M = \frac{M_0}{8\pi}$, where $M_0$ is the (geometrical) mass of the BH. This choice allows to write the Einstein's equations as $G_{\mu\nu} = T_{\mu\nu}$. In the rest of the thesis we will refer to $M$ as the mass of the BH.}

\begin{equation}
	ds^2 = - \biggl( 1 - \frac{2M}{r} \biggr) dt^2 + \frac{1}{1 - \frac{2M}{r}} dr^2 + r^2 (d\theta^2 + \sin^2{\theta} d\varphi^2).
	\label{eq:SchwarzschildGR}
\end{equation}

The Birkhoff's theorem guarantees that under the hypotheses of spherical symmetry and asymptotic flatness the metric in eq. \eqref{eq:SchwarzschildGR} is the unique solution of the Einstein's equations in vacuum \cite{Ferrari-Gualtieri-Pani}.\\
Instead, for a spinning BH with angular momentum $J$, the line element is given by the Kerr metric and in Boyer-Lindquist coordinates has the form
\begin{multline}
	ds^2 = -dt^2 + \Sigma \biggl(\frac{dr^2}{\Delta} + d\theta^2 \biggr) + (r^2 + a^2)\sin^2{\theta} \, d\varphi^2 + \\
	+ \frac{2Mr}{\Sigma} \Bigl(a \sin^2{\theta} \, d\varphi - dt \Bigr)^2,
	\label{eq:KerrGR}
\end{multline}
where $a = \frac{J}{M}$, $\Delta = r^2 - 2Mr + a^2$ and $\Sigma = r^2 + a^2 \cos^2{\theta}$. In this case the metric is not static and the spherical symmetry of eq. \eqref{eq:SchwarzschildGR} has been replaced by axial symmetry. However for $a = 0$ eq. \eqref{eq:KerrGR} reduces to the Schwarzschild metric \cite{Ferrari-Gualtieri-Pani}.\\
Another possible extension is to consider electric charge. This can be done in the Einstein-Maxwell theory, whose action in electric units ($4\pi\varepsilon_0 = 1$) is \cite{Ferrari-Gualtieri-Pani} 
\begin{equation}
	S_{EM} = \int_\Omega d^4 x \, \sqrt{-g} \biggl\{ \frac{1}{2} R - \frac{1}{16 \pi} F_{\mu\nu}F^{\mu\nu} \biggr\},
	\label{eq:EinsteinMaxwell}
\end{equation}
where $F_{\mu\nu}$ is the electromagnetic tensor. In this context a charged spinning black hole is described by the so called Kerr-Newman metric, which in Boyer-Lindquist coordinates is given by \cite{Ferrari-Gualtieri-Pani}
\begin{equation}
	ds^2 = -\frac{\Delta}{\Sigma}(dt - a \sin^2{\theta} \, d\varphi)^2 + \frac{\sin^2{\theta}}{\Sigma}[a dt - (r^2 + a^2) \, d\varphi]^2 + \frac{\Sigma}{\Delta} dr^2 + \Sigma \, d\theta^2,
	\label{eq:KerrNewmanGR}
\end{equation}
where $\Delta = r^2 - 2Mr + a^2 + Q^2$, $\Sigma = r^2 + a^2 \cos^2{\theta}$ and $Q$ is the BH electric charge. However this metric is interesting from a conceptual perspective rather than from a practical point of view. In fact the $Q/M$ of a BH must be smaller than the mass-to-charge ratio of the electron, which is extremely small, and therefore astrophysical BH are assumed to be neutral \cite{Ferrari-Gualtieri-Pani}.\\
For the sake of completeness we cite the Reissner-Nordstr\"om metric, which describes a charged BH with zero angular momentum \cite{Ferrari-Gualtieri-Pani}
\begin{equation}
	ds^2 = - \biggl( 1 - \frac{2M}{r} + \frac{Q^2}{r^2} \biggr) dt^2 + \frac{1}{1 - \frac{2M}{r} + \frac{Q^2}{r^2}} dr^2 + r^2 (d\theta^2 + \sin^2{\theta} d\varphi^2).
	\label{eq:RessnerNordstromGR}
\end{equation}\\

In GR a uniqueness theorem holds, the no-hair theorem, which we report in the form written in \cite{CardosoGualtieri}: \\

\textit{An isolated, stationary and regular BH in Einstein-Maxwell theory is described by the Kerr-Newman family.}\\

This means that in a Einstein-Maxwell theory from the hypotheses of asymptotic flatness and stationarity derives the axial symmetry, and the metric of the BH can be determined using only three parameters: mass, angular momentum and electric charge \cite{CardosoGualtieri, Ferrari-Gualtieri-Pani}.

\subsection{Scalar-Tensor Theories}

An interesting point in modified gravity is whether the no-hair theorem can be extended to these theories or not. Some results have been found in the context of scalar-tensor theories \cite{SotiriouNoHair, GrahamNoHair}. Here we report two theorems as written in \cite{CardosoGualtieri}.\\
The first is the following: \\

\textit{An isolated, stationary and regular BH in the Einstein-Klein-Gordon or Einstein-Proca theory with a \emph{time-independent boson} is described by the Kerr family.}\\

The Einstein-Klein-Gordon gravity is the theory with a scalar field minimally coupled to gravity, therefore the action is
\begin{equation}
	S_{EKG} = \frac{1}{2} \int_\Omega d^4 x \, \sqrt{-g} \biggl\{ R - (\nabla \phi)^2 - V[\phi]  \biggr\},
	\label{eq:EinsteiKleinGordon}
\end{equation}
which is the action of a scalar-tensor theory in absence of a matter field.\\
In the case of Einstein-Klein-Gordon the hypothesis of a time-independent boson can be relaxed if the scalar field is real, and this is the content of the second theorem we report from \cite{CardosoGualtieri}:\\

\textit{An isolated, stationary and regular BH in the Einstein-Klein-Gordon theory with one \emph{real} scalar is described by the Kerr family.}\\

In Horndeski gravity a no-hair theorem has been proved by Hui and Nicolis in \cite{ShiftSymmetricNoHair} with the additional hypothesis that the theory must be shift-symmetric with respect to the scalar field, \ie the action must be invariant under the transformation $\phi \to \phi + \alpha$ with $\alpha$ constant. However it has been found \cite{ShiftControesempio} that in shift-symmetric Gauss-Bonnet gravity the scalar field can assume some nontrivial configuration, and this could mean that to describe the black hole metric a scalar hair must be introduced, \ie a charge related to the scalar field. Nevertheless in this case the scalar charge depends on the BH mass and for this reason the hair is said to be of the second kind \cite{Berti+}.\\
The case of EsGB gravity is particularly interesting and hairy solutions have been found also in EdGB \cite{Kanti}.\\

\subsection{Einstein-scalar-Gauss-Bonnet Gravity}

In EsGB gravity a no-hair theorem has been proved by Silva \ea in \cite{EGBNoHair}. In what follows we are going to repeat the proof presented in the paper.\\
We consider the action of EsGB gravity with null scalar potential written as
\begin{equation}
	S_{EsGB} = \frac{1}{2} \int_\Omega d^4 x \, \sqrt{-g} \, \biggl\{ R - (\nabla \phi)^2 + 2 F[\phi] \G \biggr\}.
	\label{eq:EsGBTheorem}
\end{equation}
We write the field equations that can be obtained from the action \eqref{eq:EsGBTheorem}
\begin{gather}
	\nabla_\mu \nabla^\mu \phi + \frac{\delta F[\phi]}{\delta \phi} \G = 0, \label{eq:EsGB:field} \\
	G_{\mu\nu} = T_{\mu\nu}, \label{eq:EsGB:metric}
\end{gather}
where the stress-energy tensor $T_{\mu\nu}$ is given by
\begin{equation}
	\tensor{T}{_{\mu\nu}} = - \frac{1}{2}\bigl(\nabla \phi \bigr)^2 g_{\mu\nu} + \bigl( \nabla_\mu \phi \bigr) \bigl( \nabla_\nu \phi \bigr) - 2 \Bigl( \nabla_\gamma \nabla^\alpha F[\phi] \Bigr) \delta^{\gamma \delta \kappa \lambda}_{\alpha \beta \rho \sigma} \tensor{R}{^{\rho \sigma}_{\kappa \lambda}} \delta^\beta_\mu g_{\nu \delta}.
	\label{eq:EsGB:StressEnergyTensor}
\end{equation}\\
We are going to derive the field equations and the expression of the stress-energy tensor $T_{\mu\nu}$ in the appendix.\\

The hypotheses of the theorem are:
\begin{enumerate}
	\item the BH is stationary and asymptotically flat, therefore there exists a Killing vector $\xi^\mu$ which is timelike at infinity and at the event horizon acts as a generator of the horizon itself;
	\item the scalar field satisfies the stationarity condition
		\begin{equation}
			\xi^\mu \nabla_\mu \phi = 0;
			\label{eq:StationarietyHypotesis}
		\end{equation}
	\item there exists a constant $\phi_0$ such that
		\begin{equation}
			\frac{\delta F[\phi]}{\delta \phi} \biggr|_{\phi = \phi_0} = 0;
			\label{eq:FieldHypotesis}
		\end{equation}
	\item the following condition holds:
		\begin{equation}
			\frac{\delta^2 F[\phi]}{\delta \phi^2} \G < 0.
			\label{eq:GHypotesis}
		\end{equation}
\end{enumerate}
The aim is to show that the field equations are only satisfied by a constant configuration of the scalar field. In this case the stress-energy tensor becomes null and eq. \eqref{eq:EsGB:metric} reduces to $G_{\mu\nu} = 0$, therefore the BH solution coincides with the GR one. \\

Let us start the proof by defining a volume $\mathcal{V}$ in the spacetime whose boundaries are the event horizon, the spatial infinity and two partial Cauchy surfaces. If we multiply the Euler-Lagrange equation for the scalar field \eqref{eq:EsGB:field} by $\frac{\delta F[\phi]}{\delta \phi}$ and we integrate over $\mathcal{V}$ we obtain

\begin{multline}
	0 = \int_{\mathcal{V}} d^4 x \, \sqrt{-g} \biggl\{ \frac{\delta F[\phi]}{\delta \phi} \nabla_\mu \nabla^\mu \phi + \biggl( \frac{\delta F[\phi]}{\delta \phi} \biggr)^2 \G \biggr\} = \\
	= \int_{\mathcal{V}} d^4 x \, \sqrt{-g} \biggl\{ \nabla_\mu \biggl( \frac{\delta F[\phi]}{\delta \phi} \nabla^\mu \phi \biggr) - \nabla_\mu \biggl( \frac{\delta F[\phi]}{\delta \phi} \biggr) \nabla^\mu \phi + \biggl( \frac{\delta F[\phi]}{\delta \phi} \biggr)^2 \G \biggr\} = \\
	= \int_{\mathcal{V}} d^4 x \, \sqrt{-g} \biggl\{ - \frac{\delta^2 F[\phi]}{\delta \phi^2} (\nabla_\mu \phi) (\nabla^\mu \phi) + \biggl( \frac{\delta F[\phi]}{\delta \phi} \biggr)^2 \G \biggr\} + \\
	+ \int_{\partial \mathcal{V}} d^3 y \, \sqrt{\abs{h}} \frac{\delta F[\phi]}{\delta \phi} n_\mu \nabla^\mu \phi,
	\label{eq:NoHairProof}
\end{multline}
where in the last step we used the divergence theorem in curved space.\\
In the integral over $\partial \mathcal{V}$ the contribution from the spatial infinity vanishes for the asymptotic flatness, the contributions from the Cauchy surfaces cancel each other because they can be generated by an isometry, and at the event horizon $\xi_\mu = n_\mu$ for the first hypothesis, therefore $n_\mu \nabla^\mu \phi = \xi_\mu \nabla^\mu \phi = 0$ for the second hypothesis, thus the integral over the horizon is zero. Hence $\int_{\partial \mathcal{V}} d^3 y \, \sqrt{\abs{h}} \frac{\delta F[\phi]}{\delta \phi} n_\mu \nabla^\mu \phi = 0$ and we are left with
\begin{equation}
	\int_{\mathcal{V}} d^4 x \, \sqrt{-g} \biggl\{ - \frac{\delta^2 F[\phi]}{\delta \phi^2} (\nabla_\mu \phi) (\nabla^\mu \phi) + \biggl( \frac{\delta F[\phi]}{\delta \phi} \biggr)^2 \G \biggr\} = 0. \\
	\label{eq:NoHairProof2}
\end{equation}
Now for the condition \eqref{eq:GHypotesis} the quantities $\frac{\delta^2 F[\phi]}{\delta \phi^2}$ and $\G$ have opposite sign, and since $(\nabla_\mu \phi) (\nabla^\mu \phi) > 0$ then $- \frac{\delta^2 F[\phi]}{\delta \phi^2} (\nabla_\mu \phi) (\nabla^\mu \phi)$ and $\Bigl( \frac{\delta F[\phi]}{\delta \phi} \Bigr)^2 \G$ have the same definite sign. This means that the two terms of the integrand in eq. \eqref{eq:NoHairProof2} must vanish separately in every point of $\mathcal{V}$ and therefore $\phi = \phi_0$. In other words the Euler-Lagrange equations for a BH are satisfied only with a trivial configuration of the scalar field and the proof is completed.\\

Before moving on the possible violation of this theorem and the so-called \textit{spontaneous scalarization}, we want to make some comments on the third and the fourth hypotheses of the theorem.\\
The third hypothesis requires the existence of a trivial configuration of the scalar field that is solution of the Euler-Lagrange equations. Thus it is a condition for the existence of a solution without hairs. Coupling functions such $F[\phi] = \lambda \phi^2$ or $F[\phi] = \lambda \phi^2 + \gamma \phi^4$ satisfy this requirement, on the other hand EdGB gravity, with $F[\phi] = \lambda e^{\gamma \phi}$, and shift-symmetric Gauss-Bonnet gravity, with $F[\phi] = \lambda \phi$, do not admit the existence of a constant $\phi_0$ that satisfies \eqref{eq:FieldHypotesis} \cite{EGBNoHair}.\\
The fourth hypothesis was used in the proof as a technical requirement to characterize the sign of the terms involved. Nevertheless it has a physical meaning that can be understood by linearizing the equation of the scalar field around $\phi_0$, obtaining
\begin{equation}
	\biggl[ \nabla_\mu \nabla^\mu + \frac{\delta^2 F[\phi]}{\delta \phi^2}\biggr|_{\phi = \phi_0} \G \biggr] \delta \phi = 0,
	\label{eq:linearized}
\end{equation}
where $\delta \phi$ is the perturbation of the scalar field, defined by the relation $\phi = \phi_0 + \delta \phi$.\\
With the signature that we are using, eq. \eqref{eq:linearized} can be seen as a Klein-Gordon equation with effective mass $m_{\text{eff}}^2 = - \frac{\delta^2 F[\phi]}{\delta \phi^2}\Bigr|_{\phi = \phi_0} \G$. Therefore the hypothesis requires the positivity of the effective mass \cite{EGBNoHair}.

\section{Spontaneous Scalarization of a BH}

In EsGB gravity hairy BH solutions can appear when some of the hypotheses of the no-hair theorem are violated. A particularly interesting case is when only the fourth requirement does not hold. In this case the BH admits a GR solution, but $m_{\text{eff}}^2$ in the linearized equation \eqref{eq:linearized} is negative and this can trigger a tachyonic\footnote{The \textit{tachyon} is an hypothetical particle which travels faster than light and has an imaginary mass. The name has been proposed by Feinberg in 1967 \cite{tachyon}.} instability. The scalar field can assume a nontrivial configuration and the BH can acquire a hair \cite{EGBNoHair}. This phenomenon, called \textit{spontaneous scalarization}, is the analog for black holes of a phenomenon introduced by Damour and Esposito-Far\`ese for neutron stars \cite{DEF1, DEF2}. However, a remarkable difference is that for black holes the spontaneous scalarization can occur in vacuum and it is generated by the curvature, while for neutron stars it is related to the presence of a coupling between the scalar field and the matter \cite{EGBNoHair, DonevaYazadjiev, Silva+Stability}.\\
In recent years this spontaneous black hole scalarization has been studied with different choices of the coupling function. In particular the theory with $F[\phi] = \frac{\lambda}{3} \Bigl( 1 - e^{- 3 \phi^2} \Bigr)$ has been analyzed in refs. \cite{DonevaYazadjiev, BlazquezSalcedo+}, the coupling $F[\phi] = \lambda \phi^2$ in refs. \cite{EGBNoHair, BlazquezSalcedo+, Silva+Stability, MasatoTaishi, Macedo+}, and the case with $F[\phi] = \lambda \phi^2 + \gamma \phi^4$ in refs. \cite{MasatoTaishi, Macedo+}.\\

Before discussing about some of the results that have been found in the mentioned papers, we want to mention that, although EsGB theories are in the class of Horndeski gravity and they are not subject Ostrogradsky instability, the effective field theory approach could be invoked to treat the theories from a quantum field theory point of view. In this context the Gauss-Bonnet term can be considered as a coupling term between a scalar field and a massless spin-2 field and the spontaneous scalarization can be seen as a phenomenon that occurs in EsGB theories whose action is invariant under $\mathbb{Z}_2$ transformations of the scalar field \cite{Macedo+}. As mentioned in \cite{EGBNoHair}, the $\mathbb{Z}_2$-symmetry is required in order to avoid the appearance of hairy solutions related to the presence of the coupling term $\lambda \phi \G$, as those found in ref. \cite{ShiftControesempio}.

\subsection{Instability of Schwarzschild BH}
\label{subsection:SchwarzschildInstability}

The tachyonic instability of Schwarzschild BH has been studied in refs. \cite{EGBNoHair, DonevaYazadjiev, MasatoTaishi} with a frequency-domain analysis. We are now going to describe the procedure used by Doneva and Yazadjiev in \cite{DonevaYazadjiev}.\\
Let us start by writing the coupling function as $F[\phi] = \lambda f[\phi]$, where $\lambda$ is a coupling constant and $\frac{\delta^2 f[\phi]}{\delta \phi^2}\Bigr|_{\phi = \phi_0} > 0$. In order to simplify the equations we rescale $\lambda$ and $f[\phi]$ in such a way that $\frac{\delta^2 f[\phi]}{\delta \phi^2}\Bigr|_{\phi = \phi_0} = 1$.\\
Eq. \eqref{eq:linearized} can now be rewritten as\footnote{In ref. \cite{DonevaYazadjiev} is used a different normalization and a different coupling constant. In this work we have modified the equations in order to be consistent with the action \eqref{eq:EsGBTheorem}.}
\begin{equation}
	\biggl[ \nabla_\mu \nabla^\mu + \lambda \G \biggr] \delta \phi = 0,
	\label{eq:linearizedDoneva}
\end{equation}
where $\G = \frac{48 M^2}{r^6}$ is the Gauss-Bonnet invariant for the Schwarzschild metric.\\
Since the background spacetime is static and spherically symmetric, the perturbation of the scalar field can be written as
\begin{equation}
	\delta \phi = \frac{u(r)}{r} e^{-i \omega t} Y_l^m (\theta, \varphi),
	\label{eq:deltaphiSchwarzschild}
\end{equation}
where $Y_l^m (\theta, \varphi)$ are the spherical harmonics. Substituting this expression for $\delta \phi$ in eq. \eqref{eq:linearizedDoneva} it reduces to a Schr\"odinger-like form
\begin{equation}
	\biggl[ -\frac{d^2}{d r_*^2} + V_{\text{eff}}(r) \biggr] u(r) = \omega^2 u(r),
	\label{eq:SchwarzschildStabilitySchroedinger}
\end{equation}
where $dr_* = \Bigl( 1 - \frac{2M}{r} \Bigl)^{-1} dr$ is the tortoise coordinate and $V_{\text{eff}}(r)$ is the effective potential
\begin{equation}
	V_{\text{eff}}(r) = \biggl( 1 - \frac{2M}{r} \biggr) \biggl[ \frac{2M}{r^3} + \frac{l(l+1)}{r^2} - \lambda \frac{48 M^2}{r^6} \biggr].
	\label{eq:SchwarzschildEffective}
\end{equation}
Eq. \eqref{eq:SchwarzschildStabilitySchroedinger} can now be treated as an eingenvalue problem for $\omega^2$. In particular we are interested in modes with imaginary $\omega$ since they can indicate the presence of unstable modes. In fact if $\omega = i \omega_I$ the exponent in \eqref{eq:deltaphiSchwarzschild} becomes real and for $\omega_I > 0$ the perturbation $\delta \phi$ grows exponentially and the mode is unstable.\\
Pure imaginary modes are absent as long as $V_{\text{eff}}(r) \ge 0$ outside the horizon \cite{MasatoTaishi}. For $F[\phi] = \lambda \phi^2$ this condition holds for $\frac{\lambda}{M^2} < \frac{1}{6} \sim 0.167$ and the Schwarzschild BH is linearly stable under radial perturbations ($l = 0$) \cite{BlazquezSalcedo+, MasatoTaishi}. The same condition holds for $F[\phi] = \frac{\lambda}{3} \Bigl[ 1 - e^{-3\phi^2} \Bigr]$, and in both cases the instability starts to appear for $\frac{\lambda}{M^2} > 0.363$ \cite{BlazquezSalcedo+}.

\subsection{Stability of Scalarized Solutions}

Now we are going to discuss the procedure and the results obtained in refs. \cite{BlazquezSalcedo+, Silva+Stability, MasatoTaishi} for the analysis of the stability of the scalarized solutions.\\
In this case the metric was written as \cite{BlazquezSalcedo+}
\begin{equation}
	ds^2 = \exp{\Bigl[ 2\Phi(r) + \epsilon F_t(r, t)\Bigr]} \, dt^2 + \exp{\Bigl[2 \Lambda(r) + \epsilon F_r(r, t) \Bigr]} + r^2 \, d\Omega^2,
	\label{eq:ScalarizedMetric}
\end{equation}
where $d\Omega^2 = d\theta^2 + \sin^2{\theta} \, d\varphi^2$, and $\epsilon$ is a bookkeeping parameter. The scalar field was expanded as \cite{BlazquezSalcedo+}
\begin{equation}
	\phi = \phi_0(r) + \epsilon \phi_1(r, t),
	\label{eq:ScalarizedFieldExpansion}
\end{equation}
where $\phi_0$ is a nontrivial solution of eq. \eqref{eq:EsGB:field}.\\
The scalarized BH solution must be regular at the horizon and must behave as the Schwarzschild solution at the infinity. Therefore at the horizon $r_h$ the conditions are \cite{Silva+Stability, MasatoTaishi}:
\begin{gather}
	\exp{\Bigl[ 2 \Phi(r) \Bigr]} = \alpha (r - r_h) + \mathcal{O}\bigl[(r - r_h)^2 \bigr], \\
	\exp{\Bigl[ -2 \Lambda(r) \Bigr]} = \beta (r - r_h) + \mathcal{O}\bigl[(r - r_h)^2 \bigr], \\ 
	\phi_0(r) = \phi_0(r_h) + \frac{d \phi_0}{dr}\Bigl|_{r = r_h} (r - r_h) + \mathcal{O}\bigl[(r - r_h)^2 \bigr],
	\label{eq:ScalarizedRegularityHorizon}
\end{gather}
where $\alpha$ and $\beta$ are constants.
At spatial infinity the conditions to impose are \cite{Silva+Stability, MasatoTaishi}:
\begin{gather}
	\exp{\Bigl[ 2 \Phi(r) \Bigr]} = 1 - \frac{2M}{r} + \mathcal{O}\biggl(\frac{1}{r^2} \biggr), \\
	\exp{\Bigl[ -2 \Lambda(r) \Bigr]} = 1 - \frac{2M}{r} + \mathcal{O}\biggl(\frac{1}{r^2} \biggr), \\
	\phi_0(r) = \phi_0(r_\infty) + \frac{Q}{r} + \mathcal{O}\biggl(\frac{1}{r^2} \biggr), 
	\label{eq:ScalarizedRegularityInfinity}
\end{gather}
where $M$ is the mass and $Q$ is the scalar charge. Using these conditions $M$ and $Q$ can be computed as \cite{MasatoTaishi}
\begin{gather}
	M = \lim_{r \to +\infty} \frac{r}{2}\biggl( 1 - \exp{\Bigl[ -2 \Lambda(r) \Bigr]} \biggr), \\
	Q = - \lim_{r \to +\infty} r^2 \frac{d \phi_0(r)}{dr}.
	\label{eq:MQ}
\end{gather}
From the field equations it can be obtained a second order differential equation which has the form \cite{BlazquezSalcedo+}
\begin{equation}
	g^2(r) \frac{\partial^2 \phi_1}{\partial t^2} - \frac{\partial^2 \phi_1}{\partial r^2} + C_1(r) \frac{\partial \phi_1}{\partial r} + U(r) \phi_1 = 0.
	\label{eq:BSPerturbations}
\end{equation}
Then the perturbation of the scalar field is written as\footnote{In ref. \cite{BlazquezSalcedo+} eq. \eqref{eq:BSTimeDecomposition} is written as $\phi_1 (r, t) = \phi_1(r) e^{i \omega t}$. We used the minus sign at the exponent in order to maintain the convention of eq. \eqref{eq:deltaphiSchwarzschild}.}
\begin{equation}
	\phi_1 (r, t) = \phi_1(r) e^{-i \omega t}, 
	\label{eq:BSTimeDecomposition}
\end{equation}
and $\phi_1(r) = C_0(r) Z(r)$, where $C_0(r)$ satisfies the equation
\begin{equation}
	\frac{1}{C_0} \frac{d C_0}{dr} = C_1 - \frac{1}{g} \frac{dg}{dr}.
	\label{eq:BSCondition}
\end{equation}
The final equation in tortoise coordinates $dr_* = g \, dr$ can be written in a Schr\"odinger-like form as \cite{BlazquezSalcedo+}
\begin{equation}
	\biggl[ -\frac{d^2}{d r_*^2} + V_{\text{eff}}(r_*) \biggr] Z = \omega^2 Z.
	\label{eq:ScalarizedStabilitySchroedinger}
\end{equation}
The expressions of $g$, $U$, $C_0$ and $V_{\text{eff}}$ can be found in ref. \cite{BlazquezSalcedo+}, here we only showed the main passages contained in the paper, in order to summarize the procedure used to analyze the stability of the scalarized solutions. \\

Using the coupling function $F[\phi] = \frac{\lambda}{3} \Bigl( 1 - e^{- 3 \phi^2} \Bigr)$ several scalarized solutions have been found in ref. \cite{BlazquezSalcedo+}, and they can be classified using the number of nodes $n$ and the scalar charge $Q$. The scalarized solutions with $n = 0$ nodes exist for $\frac{\lambda}{M^2} > 0.363$, where the Schwarzschild solution is unstable, while the branches with $n \ge 1$ nodes exist in some intervals of $\frac{\lambda}{M^2}$, and they extend from the Schwarzschild branch ($Q = 0$) up to finite values of the scalar charge symmetrically in positive and negative directions, due to the $\mathbb{Z}_2$-invariance of the action.\\
Studying the frequencies it has been found (ref. \cite{BlazquezSalcedo+}) that there are different sets of unstable Schwarzschild modes from which the unstable scalarized modes bifurcate. In particular each set starts from a bifurcation point with a scalarized solution with $\tilde n$ nodes and continues up to infinite values of $\frac{\lambda}{M^2}$, containing the bifurcations with the modes of the solutions with $n > \tilde n$ nodes.\\
Moreover, analyzing the effective potential, Bl\'azquez-Salcedo \ea (ref. \cite{BlazquezSalcedo+}) found that the scalarized solutions with $n \ge 1$ nodes are unstable. Instead for the nodeless branch the authors observed that $V(r_*)$ is positive for $0.426 \lesssim \frac{\lambda}{M^2} \lesssim 1.54$, and the scalarized solutions are stable. They also found strong indications that the nodeless configurations can be stable for $\frac{\lambda}{M^2} \lesssim 4.27$.\\ 
In the same paper is also analyzed the case with $F[\phi] = \lambda \phi^2$ and the behavior of the scalarized solutions with $n \ge 1$ nodes is the same as with exponential coupling. The significant difference appears for the nodeless solutions: in this case the branch exists only in a finite interval of $\frac{\lambda}{M^2}$ and it is unstable.\\
A theory with a quartic coupling function $F[\phi] = \lambda \phi^2 + \gamma \phi^4$ was considered in \cite{Silva+Stability, MasatoTaishi}. 
It has been shown in ref. \cite{MasatoTaishi} that for the scalarized configurations with $n \ge 1$ the effective potential is negative in some regions and the solutions are unstable against radial perturbations. For the nodeless branch the bifurcation point from the Schwarzschild solution is independent of $\gamma$ and it is at $\frac{\lambda}{M^2} = 0.363$, as for the other couplings. For $\alpha := \frac{\gamma}{\lambda} > 0$ the nodeless solutions are unstable, while for large negative values of $\alpha$ the effective potential becomes non-negative and the solutions are stable. These results are in agreement with ref. \cite{Silva+Stability}, in which the authors conclude that the nodeless configurations are stable for $\alpha < -0.8$. \\
According to ref. \cite{Silva+Stability} the linear term in the equation for the scalar field affects the onset of the tachyonic instability, while the nonlinear terms can provide a quenching mechanism and some stable scalarized solutions can appear. The linear term in the equation is provided by $\lambda \phi^2$ in the coupling function, while $\gamma \phi^4$ contributes with a nonlinear term that can quench the instability. For this reason in the case of quartic coupling function there are stable scalarized solutions. On the other hand, when the coupling function is quadratic, there are not nonlinear terms in the equation for the scalar field and the scalarized configurations are unstable.

\chapter{Schwarzschild BHs in EsGB Gravity with Quadratic Coupling}
\label{chapter:QuadraticCoupling}
In this chapter we will study the Schwarzschild BHs in Einstein-scalar-Gauss-Bonnet gravity with quadratic coupling using a perturbative approach. We will start by writing the equations that can be obtained from a perturbative expansion of the field equations in the scalar field around the Schwarzschild solution, then we will use a numerical integration to study the stability of the BH. Finally we will discuss about the possibility of analyzing the behavior of the eventual apparent horizon with perturbative methods.

\section{Perturbative Approach in EsGB for a Schwarzchild BH}

The conceptual scheme that we will use for deriving the equations of a perturbed Schwarzschild BH in EsGB is the following:
\begin{enumerate}
	\item we will start from the field equations of EsGB gravity in a 1+1 formalism;
	\item we will find the constant configuration of the scalar field that reduces the field equations to the Schwarzschild case;
	\item we will expand perturbatively the field equations around the trivial configuration of the scalar field up to the second order, and reduce to a set of two equations and two constraints;
	\item finally we will make a coordinate transformation to write the equations in tortoise coordinates.
\end{enumerate}

\subsection{1+1 Decomposition and Schwarzschild Solution}

The equation for the scalar field is\footnote{See the appendix for the derivation.}
\begin{equation}
	E^{(\phi)} := \nabla_\mu \nabla^\mu \phi + \frac{\delta F[\phi]}{\delta \phi} \mathcal{G} = 0,
	\label{eq:fieldeq}
\end{equation}
while the equation for the metric is
\begin{multline}
	E^{(g)}_{\mu\nu} := R_{\mu\nu} - \frac{1}{2} g_{\mu\nu} R + \frac{1}{2} \bigl( \nabla \phi \bigr)^2 g_{\mu\nu} - \bigl( \nabla_\mu \phi \bigr) \bigl( \nabla_\nu \phi \bigr) + \\
	+ 2 \Bigl( \nabla_\gamma \nabla^\alpha F[\phi] \Bigr) \delta^{\gamma\delta\kappa\lambda}_{\alpha\beta\rho\sigma} \tensor{R}{^{\rho\sigma}_{\kappa\lambda}} \delta^\beta_\mu g_{\nu\delta} = 0.
	\label{eq:metriceq}
\end{multline}
Eq. \eqref{eq:metriceq} can be rewritten in the standard form of the Einstein's equations
\begin{equation}
	G_{\mu\nu} = T_{\mu\nu},
\end{equation}
where $G_{\mu\nu} = R_{\mu\nu} - \frac{1}{2} g_{\mu\nu} R$ is the Einstein tensor, and the stress-energy tensor $T_{\mu\nu}$ is given by
\begin{equation}
	\tensor{T}{_{\mu\nu}} = - \frac{1}{2}\bigl(\nabla \phi \bigr)^2 g_{\mu\nu} + \bigl( \nabla_\mu \phi \bigr) \bigl( \nabla_\nu \phi \bigr) - 2 \Bigl( \nabla_\gamma \nabla^\alpha F[\phi] \Bigr) \delta^{\gamma \delta \kappa \lambda}_{\alpha \beta \rho \sigma} \tensor{R}{^{\rho \sigma}_{\kappa \lambda}} \delta^\beta_\mu g_{\nu \delta}.
	\label{eq:StressEnergyTensor}
\end{equation}\\

Since we are interested in considering a system with spherical symmetry, we use a 1+1 decomposition writing the scalar field as a function of the radial and the time coordinates $r$ and $t$
\begin{equation}
	\phi = \phi(r, t),
\end{equation}
and the metric tensor as
\begin{equation}
g_{\mu\nu} = 
\begin{bmatrix}
   -A(r, t) & 0 & 0 & 0 \\
   0 & B(r, t) & 0 & 0 \\
   0 & 0 & r^2 & 0 \\
   0 & 0 & 0 & r^2 (\sin \theta)^2
\end{bmatrix}.
\end{equation} \\

The coupling function is $F[\phi] = \lambda \phi^2$, and therefore the condition in eq. \eqref{eq:FieldHypotesis} for a trivial configuration of the scalar field is given by
\begin{equation}
	0 = \frac{\delta F[\phi]}{\delta \phi} \biggr|_{\phi = \phi_0} = 2 \lambda \phi_0,
\end{equation}
and therefore $\phi_0 = 0$.

\subsection{Perturbative Expansion of the Scalar Field Around $\phi_0$}

Let us now write $\phi$ as $\phi(r, t) = \epsilon \varphi_1(r, t)$, where $\epsilon$ is a bookkeeping parameter, substitute it into the field equations, and expand up to the second order in $\epsilon$.\\
We will proceed order by order.

\subsubsection*{Order $0$ in Perturbation Theory}

At the order $0$ in perturbation theory $\phi(r, t) = \phi_0 + \OO(\epsilon) = \OO(\epsilon)$ and therefore the equation for the scalar field is trivially satisfied.\\
The equation for the metric is given by:
\begin{equation}
	G_{\mu\nu} = R_{\mu\nu} - \frac{1}{2} g_{\mu\nu} R = \OO(\epsilon).
	\label{eq:Order0MetricPhi}
\end{equation}
Therefore the solution of eq. \eqref{eq:Order0MetricPhi} is given by the Schwarzschild metric.

\subsubsection*{First Order in Perturbation Theory}

Since the stress-energy tensor depends quadratically on the field, then $T_{\mu\nu} = \OO(\epsilon^2)$ and hence the equation for the metric is
\begin{equation}
	G_{\mu\nu} = R_{\mu\nu} - \frac{1}{2} g_{\mu\nu} R = \OO(\epsilon^2).
	\label{eq:Order1MetricPhi}
\end{equation}
Therefore the metric is given by the Schwarzschild solution, without corrections at the first order:
\begin{equation}
	g_{\mu\nu} = 
	\begin{bmatrix}
	   -(1-\frac{2 M_0}{r}) & 0 & 0 & 0 \\
	   0 & \frac{1}{1-\frac{2 M_0}{r}} & 0 & 0 \\
	   0 & 0 & r^2 & 0 \\
	   0 & 0 & 0 & r^2 (\sin \theta)^2
	\end{bmatrix} + \OO(\epsilon^2),
	\label{eq:FirstOrderMetricPhi}
\end{equation}
where $M_0$ is the mass of the pure Schwarzschild solution.\\
Substituting the metric \eqref{eq:FirstOrderMetricPhi} in the Gauss-Bonnet invariant we obtain
\begin{equation}
	\G = \frac{48 M_0^2}{r^6} + \OO(\epsilon^2).
	\label{eq:FirstOrderGBPhi}
\end{equation}
The equation for the perturbation of the scalar field is given by
\begin{equation}
	\epsilon \nabla_\mu \nabla^\mu \varphi_1 + 2 \epsilon \lambda \mathcal{G} \varphi_1 = 0.
	\label{eq:Order1FieldPhi}
\end{equation}
At the first order we have the perturbation of the scalar field around the trivial solution $\phi_0$ evolving on a Schwarzschild background. This is the case considered in refs. \cite{EGBNoHair, DonevaYazadjiev, MasatoTaishi} for the study of the tachyonic instability of a Schwarzschild BH. \\
Substituting the metric \eqref{eq:FirstOrderMetricPhi} in eq. \eqref{eq:Order1FieldPhi} we obtain
\begin{multline}
  E^{(\phi)} = \epsilon  \Biggl(\frac{96 \lambda  M_0^2 \varphi_1(r,t)}{r^6} + \frac{2 (r-M_0) \varphi_1^{(1,0)}(r,t)+r (r-2 M_0) \varphi_1^{(2,0)}(r,t)}{r^2}+ \\
  + \frac{r \varphi_1^{(0,2)}(r,t)}{2 M_0-r}\Biggr)+\OO\left(\epsilon ^2\right) = 0,
  \label{eq:Order1FieldPhiDerived}
\end{multline}
where with the notation $\varphi_1^{(i, j)}(r, t)$ we indicate the $i$-th derivative with respect to $r$ and the $j$-th derivative with respect to $t$
\begin{equation}
	\varphi_1^{(i, j)}(r, t) = \partial_r^i \partial_t^j \varphi_1(r, t) = \frac{\partial^i}{\partial r^i} \frac{\partial^j}{\partial t^j} \varphi_1(r, t).
\end{equation}

The equation for the scalar field at the first order has been obtained using Mathematica \cite{Mathematica}, however eq. \eqref{eq:Order1FieldPhiDerived} can also be derived with few passages, since at this order the terms in eq. \eqref{eq:Order1FieldPhi} can be computed using the Schwarzschild metric.\\
The operator $\nabla_\mu \nabla^\mu$ can be expressed as \cite{LaplaceDeRahm}
\begin{equation}
	\nabla_\mu \nabla^\mu = \frac{1}{\sqrt{\abs{g}}} \partial_\mu \biggl( \sqrt{\abs{g}} g^{\mu\nu} \partial_\nu \biggr).
	\label{eq:LaplaceDeRahm}
\end{equation}
For the Schwarzschild metric $\sqrt{\abs{g}} = r^2 \sin{\theta}$ and eq. \eqref{eq:LaplaceDeRahm} can be rewritten as
\begin{equation}
	\nabla_\mu \nabla^\mu = \sum_{\mu = 0}^3 \, \frac{1}{r^2 \sin{\theta}} \partial_\mu \biggl( r^2 \sin{\theta} g^{\mu\mu} \partial_\mu \biggr).
\end{equation}
Since $\nabla_\mu \nabla^\mu$ is applied to $\varphi_1$, which depends only on $r$ and $t$, we obtain that
\begin{multline}
	\nabla_\mu \nabla^\mu \varphi_1 =  - \partial_t \Biggl(\frac{1}{1 - \frac{2M_0}{r}} \partial_t \varphi_1 \Biggr) + \frac{1}{r^2} \partial_r \Biggl(r^2 \Bigl( 1 - \frac{2M_0}{r} \Bigr) \partial_r \varphi_1 \Biggl) = \\
	= - \frac{1}{1 - \frac{2M_0}{r}} \partial_t^2 \varphi_1 + \frac{2(r - M_0)}{r^2} \partial_r \varphi_1 + \Bigl(1 - \frac{2M_0}{r} \Bigr) \partial_r^2 \varphi_1 = \\
	= \frac{r \, \partial_t^2 \varphi_1}{2M_0 - r} + \frac{2(r - M_0) \, \partial_r \varphi_1 + r(r - 2M_0) \, \partial_r^2 \varphi_1}{r^2}.
	\label{eq:LaplacianOrder1Phi}
\end{multline}
Computing the Gauss-Bonnet term with the Schwarzschild metric (see eq. \eqref{eq:FirstOrderGBPhi}) we can write $2 \lambda \G \varphi_1$ as
\begin{equation}
	2 \lambda \G \varphi_1 = \frac{96 M_0^2}{r^6} \varphi_1.
	\label{eq:FirstOrderGBTerm}
\end{equation}
Substituting \eqref{eq:LaplacianOrder1Phi} and \eqref{eq:FirstOrderGBTerm} in the equation for the scalar field at the first order we obtain eq. \eqref{eq:Order1FieldPhiDerived}.

\subsubsection*{Second Order in Perturbation Theory}

In order to compute the second order corrections of the metric, we write the $tt$ and the $rr$ components of $g_{\mu\nu}$ in the following way:
\begin{equation}
	\begin{cases}
		g_{tt} = -A(r, t) = -\Bigl(1 - \frac{2M_0}{r} + \epsilon^2 A_2(r, t)\Bigr) \\
		g_{rr} = B(r, t) = \frac{1}{1 - \frac{2M(r, t)}{r}}
	\end{cases},
\end{equation}
where $M(r, t) = M_0 + \epsilon^2 M_2(r, t)$ has the dimensions of a mass.\\
Expanding the field equation \eqref{eq:fieldeq} up to the second order in $\epsilon$ we obtain
\begin{multline}
  E^{(\phi)} = \epsilon  \Biggl(\frac{96 \lambda  M_0^2 \varphi_1(r,t)}{r^6}+\frac{2 (r-M_0) \varphi_1^{(1,0)}(r,t)+r (r-2 M_0) \varphi_1^{(2,0)}(r,t)}{r^2}+ \\
  + \frac{r \varphi_1^{(0,2)}(r,t)}{2 M_0-r}\Biggr)+\OO\left(\epsilon ^3\right) = 0.
  \label{eq:second_order_field}
\end{multline}
This is the same equation obtained at the first order. This is due to the fact that the metric and the Gauss-Bonnet invariant have not first order corrections, and therefore when substituted in eq. \eqref{eq:Order1FieldPhi} give a $\OO(\epsilon)$ term, which we considered before, plus a $\OO(\epsilon^3)$ term, which at this order is neglected.\\
Instead in the equations for the metric the stress-energy tensor has non-null components. In fact $T_{\mu\nu}$ is quadratic in the field and therefore it contributes with $\OO(\epsilon^2)$ terms in the equations. This means that at this order we are considering the backreaction of the scalar field to the metric.\\
Performing the computations for the expansion of $E^{(g)}_{\mu\nu}$ up to the second order we obtain

\begin{multline}
  E^{(g)}_{tt} = \frac{1}{2} \epsilon ^2 \Biggl(\frac{(r-2 M_0)}{r^6} \biggl(r (2 M_0-r) \left(64 \lambda  M_0+r^3\right) \varphi_1^{(1,0)}(r,t)^2+ \\
  + 64 \lambda  M_0 \varphi_1(r,t) \left((r-3 M_0) \varphi_1^{(1,0)}(r,t)+r (2 M_0-r) \varphi_1^{(2,0)}(r,t)\right)+ \\
  + 4 r^3 M_2^{(1,0)}(r,t)\biggr)-\varphi_1^{(0,1)}(r,t)^2\Biggr)+\OO\left(\epsilon ^3\right) = 0,
\end{multline}
\begin{multline}
  E^{(g)}_{tr} =  \frac{\epsilon ^2}{r^4 (2 M_0-r)} \Biggl(-32 \lambda  M_0^2 \varphi_1(r,t) \varphi_1^{(0,1)}(r,t)+ \\
  + r (r-2 M_0) \biggl(\left(32 \lambda  M_0+r^3\right) \varphi_1^{(0,1)}(r,t) \varphi_1^{(1,0)}(r,t)+ \\
  + 32 \lambda  M_0 \varphi_1(r,t) \varphi_1^{(1,1)}(r,t)\biggr) - 2 r^3 M_2^{(0,1)}(r,t)\Biggr)+\OO\left(\epsilon ^3\right) = 0,
\end{multline}
\begin{equation}
  E^{(g)}_{t\theta} = 0 = 0,
\end{equation}
\begin{equation}
  E^{(g)}_{t\varphi} = 0 = 0,
\end{equation}
\begin{multline}
  E^{(g)}_{rr} = \epsilon ^2 \Biggl(\frac{A_2^{(1,0)}(r,t)}{r-2 M_0}-\frac{1}{2 r (r-2 M_0)^2}\biggl(4 M_0 A_2(r,t)+ \\
  + \left(64 \lambda  M_0+r^3\right) \varphi_1^{(0,1)}(r,t)^2 + 64 \lambda  M_0 \varphi_1(r,t) \varphi_1^{(0,2)}(r,t)+4 M_2(r,t)\biggr)+ \\
  - \frac{32 \lambda  M_0 (r-3 M_0) \varphi_1(r,t) \varphi_1^{(1,0)}(r,t)}{r^4 (r-2 M_0)}-\frac{1}{2} \varphi_1^{(1,0)}(r,t)^2\Biggr)+\OO\left(\epsilon ^3\right) = 0,
\end{multline}
\begin{equation}
  E^{(g)}_{r\theta} = 0 = 0,
\end{equation}
\begin{equation}
  E^{(g)}_{r\varphi} = 0 = 0,
\end{equation}
\begin{multline}
  E^{(g)}_{\theta\theta} = \frac{\epsilon ^2}{2 r^3 (r-2 M_0)^2}\Biggl(r^3 \biggl((2 M_0-r) \Bigl(r^2 (2 M_0-r) A_2^{(2,0)}(r,t)+ \\
  + r (5 M_0-2 r) A_2^{(1,0)}(r,t)+(4 r-6 M_0) M_2^{(1,0)}(r,t)\Bigr)+ \\
  + 2 M_0^2 A_2(r,t)+2 M_0 M_2(r,t)-2 r^3 M_2^{(0,2)}(r,t)\biggr)+ \\
  - r^3 (2 M_0-r) \Bigl((2 M_0-r) \left(r A_2^{(1,0)}(r,t)-2 M_2^{(1,0)}(r,t)\right)+ \\
  + 2 M_0 A_2(r,t) + 2 M_2(r,t)\Bigr)+ \\
  - 32 \lambda  M_0 (2 M_0-r) \Bigl(\varphi_1(r,t) \bigl((2 M_0-r) \bigl((6 M_0-2 r) \varphi_1^{(1,0)}(r,t)+ \\
  + r (r-2 M_0) \varphi_1^{(2,0)}(r,t)\bigr)+r^3 \varphi_1^{(0,2)}(r,t)\bigr) + \\
  - r (r-2 M_0)^2 \varphi_1^{(1,0)}(r,t)^2+r^3 \varphi_1^{(0,1)}(r,t)^2\Bigr)+ \\
  + r^4 (2 M_0-r) \left(r^2 \varphi_1^{(0,1)}(r,t)^2-(r-2 M_0)^2 \varphi_1^{(1,0)}(r,t)^2\right)\Biggr)+\OO\left(\epsilon ^3\right) = 0,
\end{multline}
\begin{equation}
  E^{(g)}_{\theta\varphi} = 0 = 0,
\end{equation}
\begin{multline}
  E^{(g)}_{\varphi\varphi} = \frac{\epsilon ^2 \sin ^2(\theta )}{2 r^3 (r-2 M_0)^2} \Biggl(r^3 \biggl((2 M_0-r) \Bigl(r^2 (2 M_0-r) A_2^{(2,0)}(r,t)+ \\
  + r (5 M_0-2 r) A_2^{(1,0)}(r,t)+(4 r-6 M_0) M_2^{(1,0)}(r,t)\Bigr)+ \\
  + 2 M_0^2 A_2(r,t)+2 M_0 M_2(r,t) - 2 r^3 M_2^{(0,2)}(r,t)\biggl) + \\
  - r^3 (2 M_0-r) \Bigl((2 M_0-r) \left(r A_2^{(1,0)}(r,t)-2 M_2^{(1,0)}(r,t)\right) + 2 M_0 A_2(r,t)+ \\
  + 2 M_2(r,t)\Bigr) - 32 \lambda  M_0 (2 M_0-r) \Bigl(\varphi_1(r,t) \Bigl((2 M_0-r) \bigl((6 M_0-2 r) \varphi_1^{(1,0)}(r,t)+ \\
  + r (r-2 M_0) \varphi_1^{(2,0)}(r,t)\bigr)+r^3 \varphi_1^{(0,2)}(r,t)\Bigr) + \\
  - r (r-2 M_0)^2 \varphi_1^{(1,0)}(r,t)^2+r^3 \varphi_1^{(0,1)}(r,t)^2\Bigr)+ \\
  + r^4 (2 M_0-r) \left(r^2 \varphi_1^{(0,1)}(r,t)^2-(r-2 M_0)^2 \varphi_1^{(1,0)}(r,t)^2\right)\Biggr)+\OO\left(\epsilon ^3\right) = 0.
\end{multline}
Let us now manipulate these equations in order to obtain a system of two equations and two constraints.\\
The equation for the scalar field can be rewritten as
\begin{multline}
  \varphi_1^{(0,2)}(r,t) = -\frac{(2 M_0-r) }{r^7}\Biggl(96 \lambda  M_0^2 \varphi_1(r,t) + \\
  + r^4 \left(r (r-2 M_0) \varphi_1^{(2,0)}(r,t)-2 (M_0-r) \varphi_1^{(1,0)}(r,t)\right)\Biggr).
\label{eq:Order2FieldPhi}
\end{multline}
Then we replace this expression for $\varphi_1^{(0,2)}(r,t)$ in the equations for the metric and isolate $M_2^{(0,1)}(r,t)$ in $E^{(g)}_{tr} = 0$, obtaining
\begin{multline}
  M_2^{(0,1)}(r,t) = \frac{1}{2 r^3}\Biggl(r (r-2 M_0) \left(32 \lambda  M_0+r^3\right) \varphi_1^{(0,1)}(r,t) \varphi_1^{(1,0)}(r,t)+ \\
  - 32 \lambda  M_0 \varphi_1(r,t) \left(M_0 \varphi_1^{(0,1)}(r,t)+r (2 M_0-r) \varphi_1^{(1,1)}(r,t)\right)\Biggr).
\label{eq:Order2M2Phi}
\end{multline}
We can now find two constraints.\\
The equation $E^{(g)}_{tt} = 0$ can be rewritten as
\begin{multline}
  M_2^{(1,0)}(r,t) = \frac{1}{4 r^3 (r-2 M_0)}\Biggl(r^6 \varphi_1^{(0,1)}(r,t)^2 +\\
  + (2 M_0-r) \biggl(r (2 M_0-r) \left(64 \lambda  M_0+r^3\right) \varphi_1^{(1,0)}(r,t)^2+ \\
  + 64 \lambda  M_0 \varphi_1(r,t) \left((r-3 M_0) \varphi_1^{(1,0)}(r,t)+r (2 M_0-r) \varphi_1^{(2,0)}(r,t)\right)\biggr)\Biggr),
\label{constraint:Order2M2Phi}
\end{multline}
while, isolating $A_2^{(1,0)}(r,t)$ in the equation $E^{(g)}_{rr} = 0$, we obtain
\begin{multline}
  A_2^{(1,0)}(r,t) = \frac{1}{2} (r-2 M_0) \Biggl(\frac{1}{r (r-2 M_0)^2}\biggl(4 M_0 A_2(r,t)+ \\
  - \frac{64}{r^7} \lambda  M_0 (2 M_0-r) \varphi_1(r,t) \Bigl(96 \lambda  M_0^2 \varphi_1(r,t)+r^4 \bigl(2 (r-M_0) \varphi_1^{(1,0)}(r,t)+ \\
  + r (r-2 M_0) \varphi_1^{(2,0)}(r,t)\bigr)\Bigr)+\left(64 \lambda  M_0+r^3\right) \varphi_1^{(0,1)}(r,t)^2+4 M_2(r,t)\biggr)+ \\
  + \frac{64 \lambda  M_0 (r-3 M_0) \varphi_1(r,t) \varphi_1^{(1,0)}(r,t)}{r^4 (r-2 M_0)}+\varphi_1^{(1,0)}(r,t)^2\Biggr).
\label{constraint:Order2A2Phi}
\end{multline}

To summarize, we obtained two equations for the evolution of $\varphi_1$ and $M_2$, and two constraints for $M_2$ and $A_2$.

\subsection{Transformation into Tortoise Coordinates}

Let us now perform a coordinate transformation to write the equations and the constraints in tortoise coordinates, which are defined by the relations \cite{Ferrari-Gualtieri-Pani} 
\begin{gather}
  z[r, t] = r + 2M_0\ln\Bigl(\frac{r}{2M_0} - 1 \Bigr), \\
  r[z, t] = 2M_0\biggl(1 + W\Bigl(e^{\frac{z}{2M_0}-1}\Bigr) \biggr),
  \label{eq:Tortoise}
\end{gather}
where $W$ is the Lambert W function.\\
Before doing the computations it is worth mentioning that, in order for $r = 2M_0$ to be a coordinate singularity, the curvature invariants must be regular when $r \to 2M_0$ \cite{Ferrari-Gualtieri-Pani}. If we compute the Kretschmann scalar $K = R^{\mu\nu\rho\sigma} R_{\mu\nu\rho\sigma}$ we obtain
\begin{multline}
	K = \frac{48 M_0^2}{r^6}+\frac{8 M_0 \epsilon^2}{r^{12}} \Biggl(9216 \lambda^2 M_0^3 \varphi_1(r,t)^2 + r^4 \biggl(12 r^2 M_2(r,t) + \\
	+ \left(r-2 M_0\right) \varphi_1^{(1,0)}(r,t) \left(r \left(r^3-96 \lambda  M_0\right) \varphi_1^{(1,0)}(r,t)+384 \lambda  M_0 \varphi_1(r,t)\right)+ \\
	- \frac{\left(r^6-96 \lambda  M_0 r^3\right) \varphi_1^{(0,1)}(r,t)^2}{r-2 M_0}\biggr)\Biggr)+\OO\left(\epsilon ^3\right).
	\label{eq:Kretschmann}
\end{multline}
As expected $K$ is formed by a zero order term, which is the Kretschmann scalar of the Schwarzschild metric, plus a second order contribution, which has a singularity at $r = 2M_0$ that comes from the term
\begin{equation}
	- \frac{8 M_0 \epsilon^2}{r^{8}} \frac{\left(r^6-96 \lambda  M_0 r^3\right) \varphi_1^{(0,1)}(r,t)^2}{r-2 M_0}.
\end{equation}
However if we impose condition
\begin{equation}
	\lim_{r \to 2M_0} \partial_t \varphi_1(r, t) = \lim_{r \to 2M_0} \biggl( 1 - \frac{2M_0}{r} \biggr) \partial_r \varphi_1(r, t),
	\label{eq:IngoingSchwarzschild}
\end{equation}
then the Kretschmann scalar becomes regular at $r = 2M_0$. In tortoise coordinates the condition \eqref{eq:IngoingSchwarzschild} is written as
\begin{equation}
	\lim_{z \to -\infty} \partial_t \varphi_1(z, t) = \lim_{z \to -\infty} \partial_z \varphi_1(z, t),
	\label{eq:IngoingTortoise}
\end{equation}
which is an ingoing boundary condition for $\varphi_1$.\\
We can now rewrite the equations in tortoise coordinates obtaining
\begin{multline}
  \varphi_1^{(0,2)}(z,t) = -\frac{96 \lambda  M_0^2 \varphi_1(z,t) (2 M_0-r(z,t))}{r(z,t)^7}+ \\
  +\frac{2 \varphi_1^{(1,0)}(z,t) (r(z,t)-2 M_0)}{r(z,t)^2}+\varphi_1^{(2,0)}(z,t),
  \label{eq:Order2FieldPhiTortoise}
\end{multline}
\begin{multline}
  M_2^{(0,1)}(z,t) = \frac{1}{2 r(z,t)^3}\biggl(r(z,t)^2 \varphi_1^{(0,1)}(z,t) \varphi_1^{(1,0)}(z,t) \left(r(z,t)^3+32 \lambda  M_0\right)+ \\
  - 32 \lambda  M_0 \varphi_1(z,t) \left(M_0 \varphi_1^{(0,1)}(z,t)-r(z,t)^2 \varphi_1^{(1,1)}(z,t)\right)\biggr).
  \label{eq:Order2MassPhiTortoise}
\end{multline}
Instead for the constraints we obtain
\begin{multline}
  M_2^{(1,0)}(z,t) = \frac{1}{4 r(z,t)^3}\biggl(64 \lambda  M_0^2 \varphi_1(z,t) \varphi_1^{(1,0)}(z,t)+ \\
  + 64 \lambda  M_0 r(z,t)^2 \left(\varphi_1^{(1,0)}(z,t)^2+\varphi_1(z,t) \varphi_1^{(2,0)}(z,t)\right) + \\
  - 64 \lambda  M_0 r(z,t) \varphi_1(z,t) \varphi_1^{(1,0)}(z,t)+r(z,t)^5 \left(\varphi_1^{(0,1)}(z,t)^2+\varphi_1^{(1,0)}(z,t)^2\right)\biggr),
  \label{constraint:Order2MassPhiTortoise}
\end{multline}
\begin{multline}
  A_2^{(1,0)}(z,t) = \frac{1}{2 r(z,t)^9}\biggl(4 M_0 A_2(z,t) r(z,t)^7-12288 \lambda ^2 M_0^4 \varphi_1(z,t)^2+ \\
  + 6144 \lambda ^2 M_0^3 r(z,t) \varphi_1(z,t)^2-448 \lambda  M_0^2 r(z,t)^5 \varphi_1(z,t) \varphi_1^{(1,0)}(z,t)+ \\
  + 64 \lambda  M_0 r(z,t)^7 \varphi_1^{(0,1)}(z,t)^2+64 \lambda  M_0 r(z,t)^7 \varphi_1(z,t) \varphi_1^{(2,0)}(z,t)+ \\
  + 192 \lambda  M_0 r(z,t)^6 \varphi_1(z,t) \varphi_1^{(1,0)}(z,t)+4 M_2(z,t) r(z,t)^7+ \\
  + r(z,t)^{10} \varphi_1^{(0,1)}(z,t)^2+r(z,t)^{10} \varphi_1^{(1,0)}(z,t)^2\biggr).
  \label{constraint:Order2A2PhiTortoise}
\end{multline}

\section{Numerical Integration}
We integrated the equations with the method of lines \cite{ScholarpediaMOL} in the following way: first we discretized the tortoise coordinate $z$ fixing the boundaries of the domain and the number $N$ of points, then we evolved $\varphi_1$ and $M_2$ in each discretized point using the 4th order Runge-Kutta method \cite{NumericalRecipes}. \\
The spatial derivatives present in the equations were computed using second order accurate finite differences methods \cite{ScholarpediaMOL, NumericalRecipes, NumericDerivatives}:
\begin{gather}
	\partial_z u(z_n, t) = \frac{u(z_{n+1}, t) - u(z_{n-1}, t)}{2 \Delta z}, \label{eq:FirstDerivative}\\
	\partial_z^2 u(z_n, t) = \frac{u(z_{n+1}, t) - 2u(z_n, t) + u(z_{n-1}, t)}{(\Delta z)^2}, \label{eq:SecondDerivative}
\end{gather}
where $u(z, t)$ is a generic function of $z$ and $t$, $z_n$ indicates the n-th spatial point and $\Delta z$ is the spatial step of the grid.\\

For the integration we have imposed the following boundary conditions at the horizon:
\begin{equation}
	\begin{cases}
		\partial_z \varphi_1(z_0, t) = \partial_t \varphi_1(z_0, t)\\
		M_2(z_0, t) = 0
	\end{cases}.
	\label{eq:boundary_cond}
\end{equation}
$z_0$ represents the numerical inner boundary and the first condition is the ingoing boundary condition we discussed in the previous section.\\

The numerical scheme was the following:

\begin{enumerate}
	\item we chose an initial profile for $\varphi_1$ which satisfies the boundary conditions;
	\item we integrated numerically eq. \eqref{constraint:Order2MassPhiTortoise} with the Simpson rule \cite{NumericalRecipes} using the scalar field at t=0 to obtain the initial profile of $M_2$;
	\item we integrated eq. \eqref{eq:Order2FieldPhiTortoise} and \eqref{eq:Order2MassPhiTortoise} with the method of lines using the 4th order Runge-Kutta method for time evolution.
\end{enumerate}
As a check we verified at each step that the numerical solution satisfies the constraint on the mass in eq. \eqref{eq:Order2MassPhiTortoise}.\\

Numerically the ingoing boundary condition can be imposed by using the numerical time derivative of $\varphi_1$ at $z_0$, which we now call $v_t$, in place of $\partial_z \varphi_1(z_0, t)$. Instead for the second spatial derivative at $z_0$ we can introduce the variable $\varphi_1(z_{-1}, t)$ and use it to write numerically the ingoing boundary condition as
\begin{equation}
	v_t = \partial_z \varphi_1(z_0, t) = \frac{\varphi_1(z_{1}, t) - \varphi_1(z_{-1}, t)}{2 \Delta z}.
	\label{eq:FirstDerivativez0}
\end{equation}
From \eqref{eq:FirstDerivativez0} we can obtain an expression for $\varphi_1(z_{-1}, t)$ that can be used to compute $\partial_z^2\varphi_1(z_0, t)$ with the formula \eqref{eq:SecondDerivative} as follows
\begin{equation}
	\partial_r^2\varphi_1(z_0, t) = 2\frac{\varphi_1(z_1, t) - \varphi_1(z_0, t) - v_t \Delta z}{(\Delta z)^2}.
\end{equation}

\section{Stability Analysis}

\begin{figure}
\centering
	\subfloat[][]
	{\includegraphics[width=0.50\textwidth]{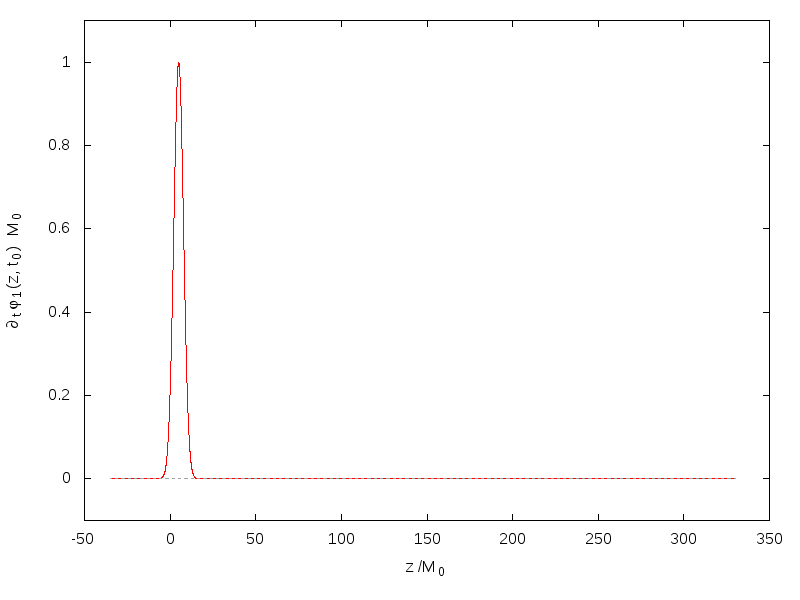}}
	\subfloat[][]
	{\includegraphics[width=0.50\textwidth]{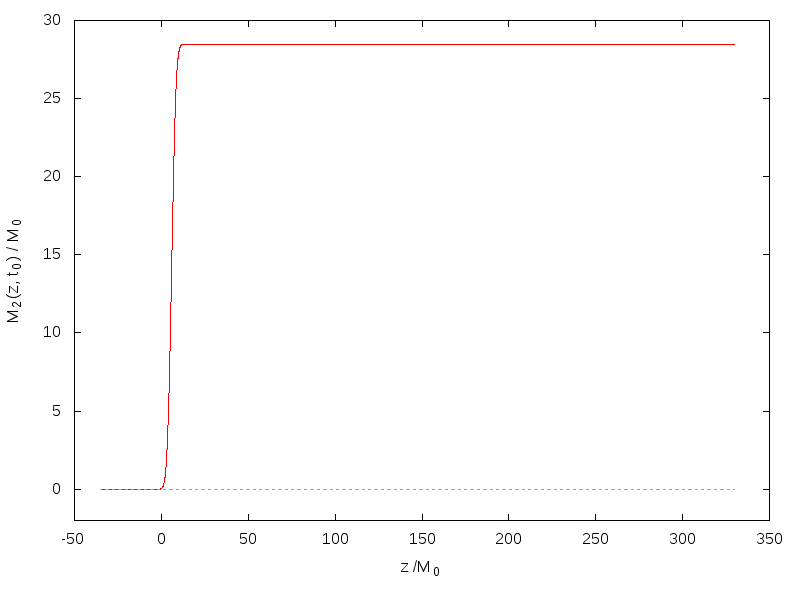}}
	\caption{Initial profiles of the time derivative of the scalar field $\partial_t \varphi_1(z, t_0)$ (on the left) and of the mass correction $M_2(z, t_0)$ (on the right).}
	\label{fig:InitialProfile}
\end{figure}

For the numerical integration we used a grid with 10000 points in a range for $z/M_0$ between $-34.84$ and $330.11$, so that the grid step is $\frac{\Delta z}{M_0} = 0.036$. The time step is $\frac{\Delta t}{M_0} = 0.001$. As initial profile for the scalar field we chose
\begin{gather}
	\varphi_1(z, t) = 0, \notag \\
	\partial_t \varphi_1(z, t) = 
	\begin{cases}
		e^{-\frac{(z-\mu)^2}{\sigma^2}} & \quad \text{if} \quad \mu - 5\sigma < z < \mu + 5\sigma\\
		0 				& \qquad \text{otherwise}
	\end{cases},
\end{gather}
where $\mu = 5 \, M_0$ and $\sigma = 4 \, M_0$.\\
With a null scalar field the constraint \eqref{constraint:Order2MassPhiTortoise} reduces to
\begin{equation}
	M_2^{(1,0)}(z,t) = \frac{r(z,t)^2}{4}\varphi_1^{(0,1)}(z,t)^2,
\end{equation}
and therefore the profile of $M_2$ at $\frac{t}{M_0} = \frac{t_0}{M_0} = 0$  does not depend on the coupling constant $\lambda$. The profiles of $\partial_t \varphi_1(z, t_0)$ and $M_2(z, t_0)$ are shown in fig. \ref{fig:InitialProfile}.
\\

Before studying the stability of the Schwarzschild solution we checked the numerical stability of the code, by evaluating the violation of the constraint on the mass. \\
Let us define $F(z, t)$ as the right hand side of \eqref{constraint:Order2MassPhiTortoise}, where we have substituted the profile of the scalar field obtained with the numerical integration. The constraint violation $V(z, t)$ can be computed as the absolute value of the difference between the numerical derivative of $M_2(z, t)$ and $F(z, t)$:
\begin{equation}
	V(z, t) = \abs{\partial_z M_2(z, t) - F(z, t)}.
\end{equation}
Since the numerical derivatives that we used in the time evolution are second order accurate, we expected that $V(z, t)$ scales quadratically with $\Delta z$. We run the same simulation twice, using 10000 points the first time, and 50000 the second time. In this way in the first case $\frac{\Delta z_1}{M_0} = 0.036$, while in the second $\frac{\Delta z_2}{M_0} = \frac{0.036}{5}$, and we verified that the constraint violations $V_1(z, t)$ and $V_2(z, t)$, obtained respectively using $\Delta z_1$ and $\Delta z_2$, are related by
\begin{equation}
	V_1(z, t) = \biggl( \frac{\Delta z_1}{\Delta z_2} \biggr)^2 V_2(z, t) = 25 V_2(z, t).
	\label{eq:ConstraintScaling}
\end{equation}
In the computation of $V(z, t)$ we used fourth order accurate numerical derivatives, in order to neglect the numerical error introduced in the computation of $V(z, t)$, and to consider only the error that comes from the integration of the time evolution. The formulas which we used are \cite{NumericDerivatives}:
\begin{gather}
	\partial_z u(z_n, t) = \frac{- u(z_{n+2}, t) + 8 u(z_{n+1}, t) - 8 u(z_{n-1}, t) + u(z_{n-2}, t)}{12 \, \Delta z}, \\
	\partial_z^2 u(z_n, t) = \frac{-u(z_{n+2}, t) + 16 u(z_{n+1}, t) - 30 u(z_{n}, t) + 16 u(z_{n-1}, t) - u(z_{n-2}, t)}{12 \, (\Delta z)^2}.
\end{gather}

\begin{figure}
	\centering
	\includegraphics[width=0.6\textwidth]{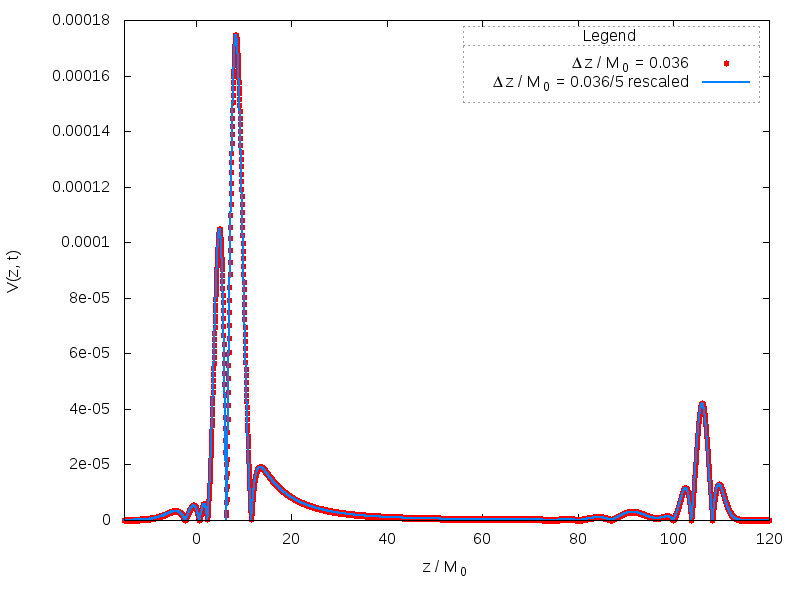}
	\caption{Constraint violation for two different choices of $\Delta z$. In red is represented $V_1(z, t)$, corresponding to $\frac{\Delta z_1}{M_0} = 0.036$, and in blue the constraint violation for the case $\frac{\Delta z_2}{M_0} = \frac{0.036}{5}$, rescaled by $\bigl( \frac{\Delta z_1}{\Delta z_2} \bigr)^2 V_2(z, t) = 25 V_2(z, t)$.}
	\label{fig:constraint_violation}
\end{figure}

Fig. \ref{fig:constraint_violation} shows the scaling of the constraint violations computed at $\frac{t}{M_0} = 100$, in the case of $\frac{\lambda}{M_0^2} = -0.5$, using $\Delta z_1$ (red) and $\Delta z_2$ (blue), where the second is rescaled according to eq. \eqref{eq:ConstraintScaling}. \\
As we can see the constraint violation is sufficiently small and it scales quadratically as expected. \\

\begin{figure}
	\centering
	\subfloat[][$\varphi_1$\label{fig:comparison_field}]
		{\includegraphics[width=0.85\textwidth]{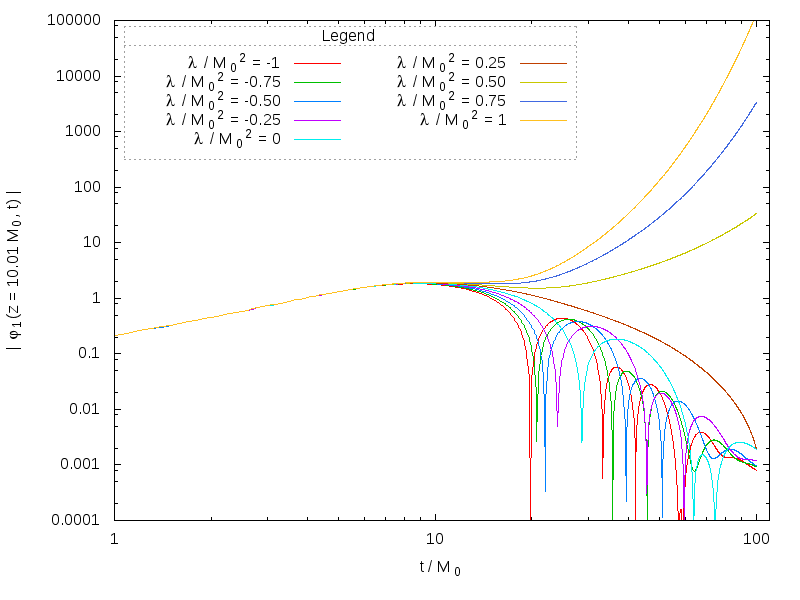}} \\
	\subfloat[][$M_2$\label{fig:comparison_mass}]
		{\includegraphics[width=0.85\textwidth]{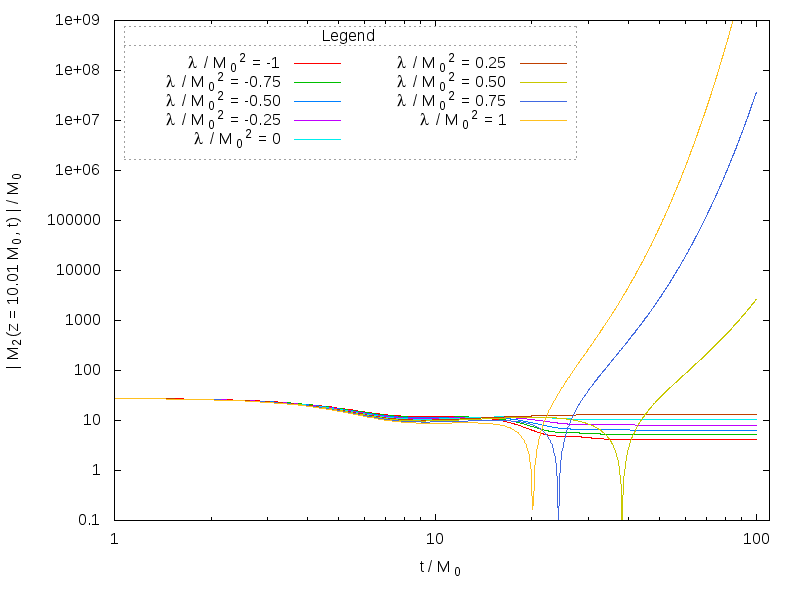}}
	\caption{Time evolution of $\varphi_1$ (upper plot) and $M_2$ (lower plot) at $z/M_0=10.01$ for different values of $\lambda/M_0^2$ from $-1$ to $+1$.}
	\label{fig:comparison}
\end{figure}

\begin{figure}
	\centering
	\includegraphics[width=0.65\textwidth]{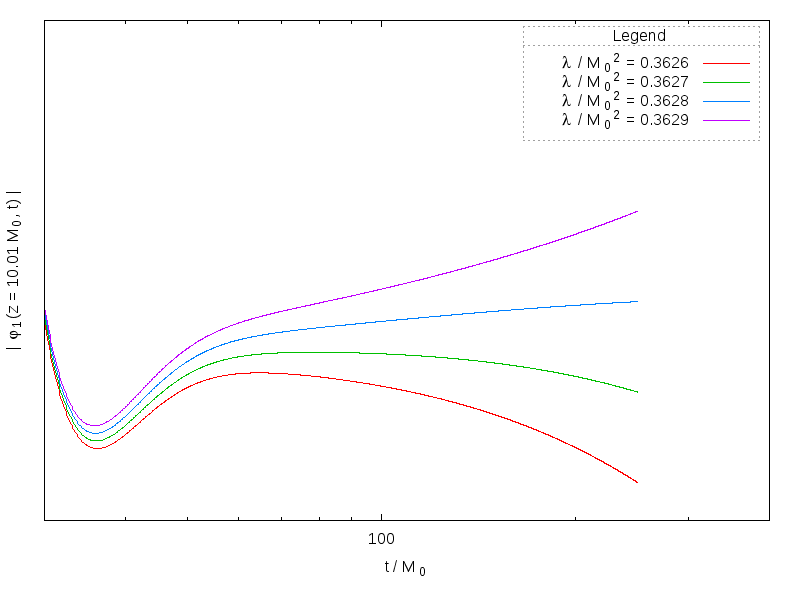}
	\caption{Time evolution of the $\varphi_1$ at $z/M_0 = 10.01$ for large values of $t$ with $\lambda \sim \tilde \lambda$.}
	\label{fig:stability_zoom}
\end{figure}

\begin{figure}
        \centering
	\subfloat[][]
	{\includegraphics[width = 0.5\textwidth]{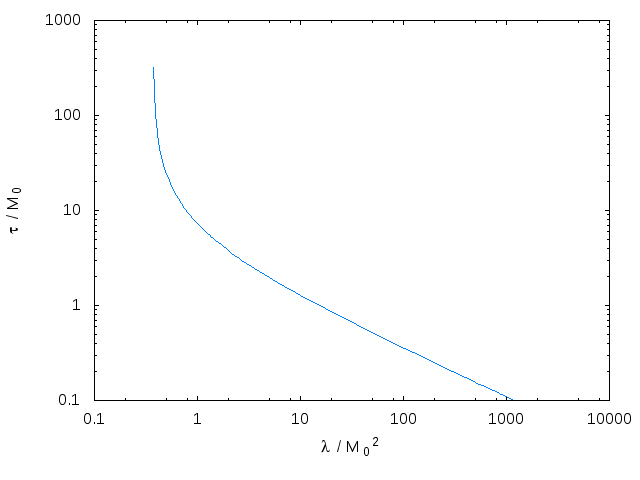}}
	\subfloat[][]
	{\includegraphics[width = 0.5\textwidth]{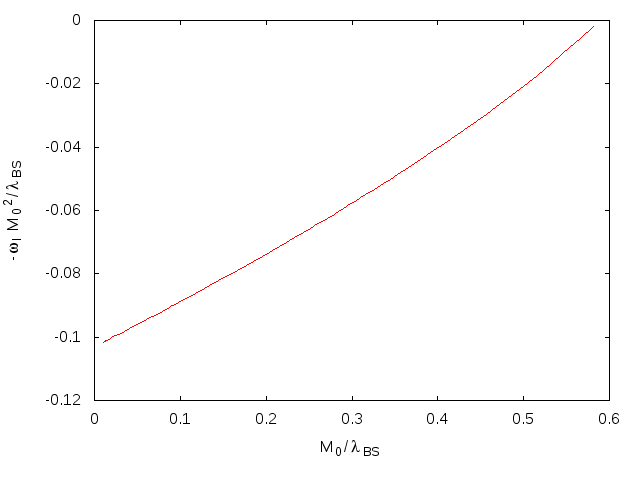}}
	\caption{(left) Behavior of the fit parameter $\tau$ with respect to the coupling constant $\lambda$. (right) Reproduction of the behavior of $-\omega_I \frac{M_0^2}{\lambda_{BS}}$ vs $\frac{M_0}{\lambda_{BS}}$ contained in ref. \cite{BlazquezSalcedo+}, using $\omega_I$ obtained from the fit parameter $\tau$. $\lambda_{BS} = \sqrt{8\lambda}$ is the coupling constant used by Bl\'azquez-Salcedo \ea in ref. \cite{BlazquezSalcedo+}.}
        \label{fig:TauOmega}
\end{figure}

In order to study the stability of the solutions we integrated the equations for several values of $\lambda$ and we analyzed the evolution of $\varphi_1$ and $M_2$ at $z/M_0 = 10.01$ obtaining the results shown in fig. \ref{fig:comparison}.
As we can see the solution becomes unstable for a $\frac{\lambda}{M_0^2} > \frac{\tilde \lambda}{M_0^2} > 0$, and in order to estimate the value of $\tilde \lambda$ we integrated the equations with different values of $\lambda$, searching for the value at which the field starts to diverge for large $t$. In this way we found $0.3627 < \frac{\tilde \lambda}{M_0^2} < 0.3628$, as it can be seen in fig. \ref{fig:stability_zoom}.
This result is consistent with refs. \cite{BlazquezSalcedo+, Silva+Stability}, where it has been found the critical value $\frac{\tilde \lambda}{M_0^2} = 0.363$.\\
As we can see from fig. \ref{fig:comparison}, the instability appears at $\frac{t}{M_0} \sim 10$, when the exponential growth starts. In order to characterize the behavior of the instability we fitted the scalar field at $\frac{z}{M_0}$ with the function
\begin{equation}
        f(x) = \alpha e^{\frac{x}{\tau}}.
\end{equation}
The fit was made in the region with $\frac{t}{M_0} > 30$ and for values of the coupling constant in the range $0.37 \le \frac{\lambda}{M_0^2} \le 1202$. The behavior of the fit parameter $\tau$ with respect to the coupling constant is shown on the left panel of fig. \ref{fig:TauOmega}.\\
We now want to compare our results with the one found by Bl\'azquez-Salcedo \ea (ref. \cite{BlazquezSalcedo+}), who used the procedure that we discussed in chapter \ref{chapter:review}, in which $\varphi_1$ is written as\footnote{We recall that we have chosen a different convention with respect to ref. \cite{BlazquezSalcedo+} for the sign of $\omega$.}
\begin{equation}
	\varphi_1 (r, t) = \varphi_1(r) e^{-i \omega t},
\end{equation}
and the $\omega$ are computed using an eigenvalue equation.
With this decomposition the instability is characterized by pure imaginary frequencies $\omega = i \omega_I$ with positive imaginary part $\omega_I > 0$, since in this case the perturbation of the scalar field grows exponentially. \\
Relating the fit parameter $\tau$ to $\omega_I$ with the relation $\tau = \frac{1}{\omega_I}$, we reproduced the plot of $-\omega_I \frac{M_0^2}{\lambda_{BS}}$ vs $\frac{M_0}{\lambda_{BS}}$ contained in ref. \cite{BlazquezSalcedo+}, where $\lambda_{BS} = \sqrt{8\lambda}$ is the coupling constant used by Bl\'azquez-Salcedo \ea in ref .\cite{BlazquezSalcedo+}. The plot is in the right panel of fig. \ref{fig:TauOmega} and it reproduces the behavior of the first set of Schwarzschild unstable modes shown in ref. \cite{BlazquezSalcedo+}, which is the one that starts from the bifurcation point with the nodeless solutions.\\

\section{Study of the Apparent Horizon}

In this section we want to study the behavior of the apparent horizon.\\
In spherical symmetry the apparent horizons are defined by the condition (see ref. \cite{ApparentHorizonFoliation})
\begin{equation}
	\nabla_\alpha R \nabla^\alpha R = 0,
	\label{eq:ApparentHorizonDefinition}
\end{equation}
where $R$ is the areal radius of the horizon.\\
In the metric that we are using the areal radius is a coordinate and the equation for the horizon is given by \cite{ApparentHorizonFoliation}
\begin{equation}
	g^{rr} = 0.
	\label{eq:ApparentHorizonSchwarzschild}
\end{equation}
Hence at the second order in $\epsilon$ we have to solve
\begin{equation}
	1 - \frac{2M_0}{r} - \epsilon^2 \frac{2 M_2}{r} = 0.
	\label{eq:app_hor}
\end{equation}
Since the term $M(r, t)$ is expanded as $M(r, t) = M_0 + \epsilon^2 M_2(r, t)$, in order for the perturbative expansion to be valid, we imposed the condition 
\begin{equation}
	\epsilon^2 M_2 < M_0.
	\label{eq:ConditionApparent}
\end{equation}
In the stable cases it is sufficient to set a value for $\epsilon$ such that the condition is valid for every timestep, while in the unstable cases this cannot be done because of the growth of $M_2$, and therefore, once set the value of $\epsilon$, the equation \eqref{eq:app_hor} can be solved only for timesteps in which the condition \eqref{eq:ConditionApparent} is valid. In both cases we used $\epsilon = 0.1$.\\
Moreover, as it can be seen from eq. \eqref{eq:Order2MassPhiTortoise}, the boundary condition $M_2(z_0, t)$ can be imposed only when the scalar field is null at the numerical inner boundary, and therefore we computed the apparent horizon only for timesteps such that $\varphi_1(z_0, t) = 0$. In order for this condition to be valid for as sufficiently large time, for the following results we used data obtained from simulations on a grid with 13000 points in a region that extends from $\frac{z}{M_0} = -53.26$ up to $\frac{z}{M_0} = 330.11$.

\subsection{Behavior of the Apparent Horizon}

\begin{figure}
	\centering
	\subfloat[][$\frac{\lambda}{M_0^2} = 0$]
	{\includegraphics[width=0.5\textwidth]{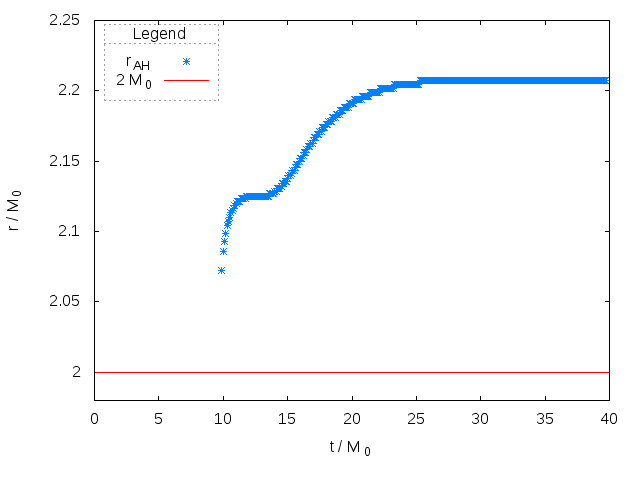}}
	\subfloat[][$\frac{\lambda}{M_0^2} = -0.5$]
	{\includegraphics[width=0.5\textwidth]{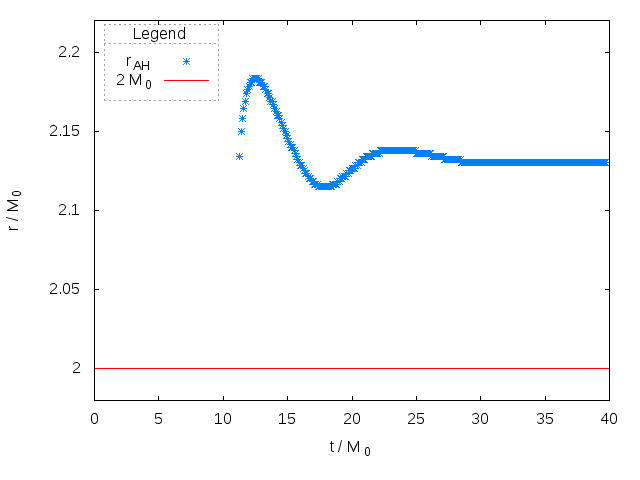}} \\
	\subfloat[][$\frac{\lambda}{M_0^2} = 0.5$]
	{\includegraphics[width=0.5\textwidth]{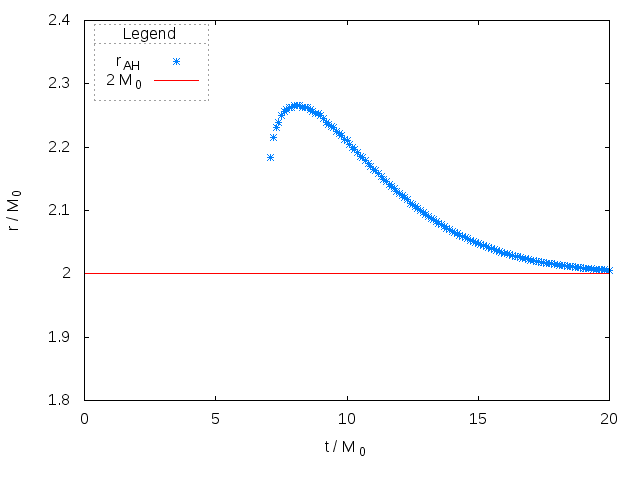}}
	\caption{Behavior of the apparent horizon for $\frac{\lambda}{M_0^2} = 0$ (upper left), $\frac{\lambda}{M_0^2} = -0.5$ (upper right) and $\frac{\lambda}{M_0^2} = 0.5$ (lower).}
	\label{fig:ApparentHorizonBehavior}
\end{figure}

The behaviors of the apparent horizon in the cases $\frac{\lambda}{M_0^2} = 0$, $\frac{\lambda}{M_0^2} = -0.5$ and $\frac{\lambda}{M_0^2} = 0.5$ are shown in fig. \ref{fig:ApparentHorizonBehavior}.\\
In the stable cases we could solve the eq. \eqref{eq:app_hor} only for $\frac{t}{M_0} \lesssim 40$ because for later timesteps the condition $\varphi_1(z_0, t) = 0$ does not hold anymore. For $\frac{\lambda}{M_0^2} = 0$ the position of the apparent horizon increases with time and converges to $\frac{r_{AH}}{M_0} = 2.21$, while for $\frac{\lambda}{M_0^2} = -0.5$ it oscillates before stabilizing at $\frac{r_{AH}}{M_0} = 2.13$.\\
In the unstable case with $\frac{\lambda}{M_0^2} = 0.5$ the apparent horizon could be computed only for $\frac{t}{M_0} \lesssim 25$ in order for the condition \eqref{eq:ConditionApparent} to be valid. In this case $r_{AH}$ initially increases, reaches a maximum at $\frac{r_{AH}}{M_0} = 2.27$ and then decreases converging to $r = 2 M_0$.\\

\subsection{Null Energy Condition in a Perturbative Scheme} 

\begin{figure}
	\centering
	\subfloat[][$\frac{\lambda}{M_0^2} = 0$]
	{\includegraphics[width=0.5\textwidth]{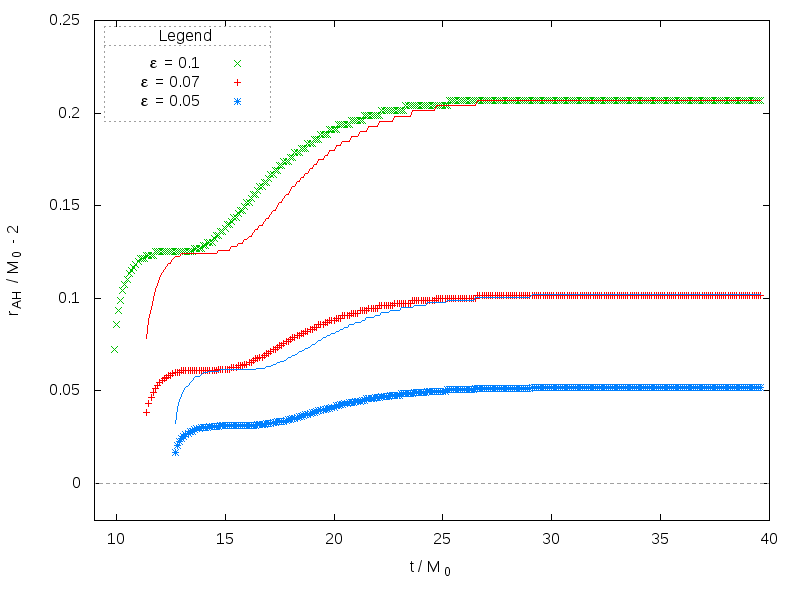}}
	\subfloat[][$\frac{\lambda}{M_0^2} = -0.5$]
	{\includegraphics[width=0.5\textwidth]{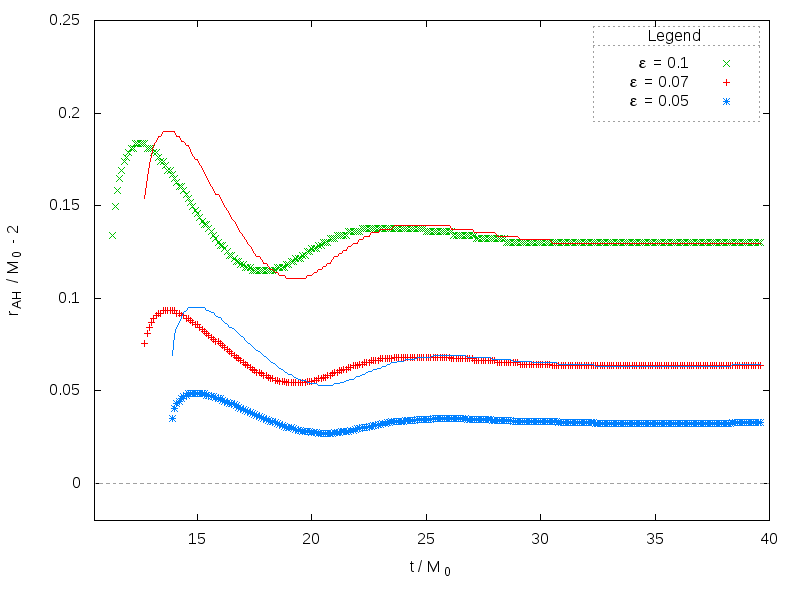}} \\
	\subfloat[][$\frac{\lambda}{M_0^2} = 0.5$]
	{\includegraphics[width=0.5\textwidth]{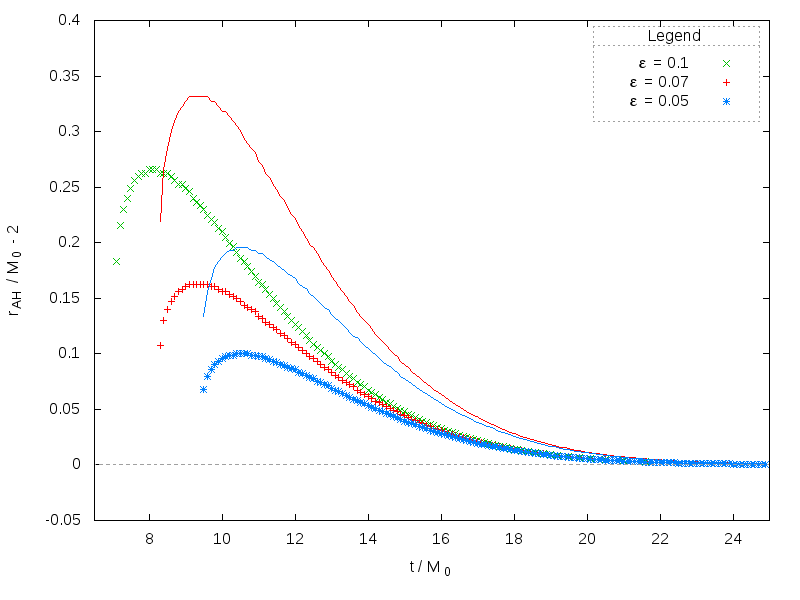}}
	\caption{Scaling of $r_{AH} - 2M_0$ renormalized by $M_0$ in the cases $\frac{\lambda}{M_0^2} = 0$ (upper left), $\frac{\lambda}{M_0^2} = -0.5$ (upper right) and $\frac{\lambda}{M_0^2} = 0.5$ (lower). The points in the plots represent the values of $\frac{r_{AH}}{M_0} - 2$ obtained from the computation of the apparent horizon using three different values of $\epsilon$, while the solid lines are obtained by rescaling quadratically in $\epsilon$ the values with the same color. In particular, the red solid line represents the values of $\frac{r_{AH}}{M_0} - 2$ computed with $\epsilon = 0.07$ and rescaled by $\frac{0.1^2}{0.07^2}$ while the blue solid line is obtained from a computation of the horizon with $\epsilon = 0.05$ with a rescaling of $\frac{0.07^2}{0.05^2}$.}
	\label{fig:AHScaling}
\end{figure}

In order to understand the behaviors of the apparent horizon that appear in fig. \ref{fig:ApparentHorizonBehavior} we want to check whether the Null Energy Condition (NEC) is satisfied or not. When the condition \eqref{eq:NEC} is satisfied the position of the apparent horizon increases, while it could decrease when the NEC is violated \cite{Hayward}.\\
The NEC requires that
\begin{equation}
	T^{\mu\nu}n_\mu n_\nu \ge 0,
	\label{eq:NEC}
\end{equation}
for any null vector $n^{\mu}$ \cite{Ferrari-Gualtieri-Pani}.\\
Since in the coordinates that we are using the definition of the apparent horizon \eqref{eq:ApparentHorizonDefinition} reduces to \eqref{eq:ApparentHorizonSchwarzschild}, then at $r_{AH}$ the vector $n_\mu = (0, 1, 0, 0)$ is null and the NEC reduces to
\begin{equation}
	T^{rr} \ge 0,
	\label{eq:AHNEC}
\end{equation}
where the $T^{rr}$ is given by
\begin{multline}
	T^{rr} = \frac{\epsilon ^2}{2 r^{10}}\Biggl(-6144 \lambda ^2 M_0^3 (2 M_0-r) \varphi_1(r,t)^2 + \\
	+ 64 \lambda  M_0 r^4 (2 M_0-r) \varphi_1(r,t) \biggl((5 M_0-3 r) \varphi_1^{(1,0)}(r,t) + r (2 M_0-r) \varphi_1^{(2,0)}(r,t)\biggr) + \\
	+ r^7 \left(\left(64 \lambda  M_0+r^3\right) \varphi_1^{(0,1)}(r,t)^2+r (r-2 M_0)^2 \varphi_1^{(1,0)}(r,t)^2\right)\Biggr)+\OO\left(\epsilon ^3\right).
	\label{eq:Trr}
\end{multline}
Instead in a generic point of the spacetime we can start from a 4-vector of the form $n_\mu = (n_t, 1, 0, 0)$ and impose that it is null:
\begin{equation}
	0 = n_\mu n^\mu = g^{\mu\nu} n_\mu n_\nu = g^{tt} n_t^2 + g^{rr}.
	\label{eq:nNull}
\end{equation}
Using eq. \eqref{eq:nNull} we can write $n_t^2$ as
\begin{equation}
	n_t^2 = - \frac{g^{rr}}{g^{tt}},
\end{equation}
and the NEC is given by
\begin{multline}
	T^{tt} n_t^2 + T^{rr} =  - \frac{g^{rr}}{g^{tt}} = -g^{rr}g_{tt}T^{tt} + T^{rr} = \\
	= -g^{rr} \tensor{T}{^t_t} + T^{rr} = -\biggl( 1 - \frac{2 M_0}{r} \biggr) \tensor{T}{^t_t} + T^{rr} + \OO(\epsilon^4) \ge 0,
	\label{eq:stNEC}
\end{multline}
where
\begin{multline}
	\tensor{T}{^t_t} = \frac{\epsilon^2}{2 r^5} \biggl[\frac{r^6 \varphi_1^{(0,1)}(r,t)^2}{2 M_0-r}-r \left(r-2 M_0\right) \left(64 \lambda  M_0+r^3\right) \varphi_1^{(1,0)}(r,t)^2 + \\
+ 64 \lambda  M_0 \varphi_1(r,t) \left(\left(r-3 M_0\right) \varphi_1^{(1,0)}(r,t)+r \left(2 M_0-r\right) \varphi_1^{(2,0)}(r,t)\right)\biggr]+\OO\left(\epsilon ^3\right).
	\label{eq:Ttt}
\end{multline}
In eq. \eqref{eq:stNEC} we used the fact that $\tensor{T}{^t}{_t} = \OO(\epsilon^2)$. \\
In principle at the apparent horizon \eqref{eq:AHNEC} and \eqref{eq:stNEC} should coincide, since 
\begin{equation}
	\biggl( 1 - \frac{2 M_0}{r} \biggr) \tensor{T}{^t_t} = \epsilon^2 \frac{2 M_2}{r} \tensor{T}{^t_t} = \OO(\epsilon^4).
\end{equation}
However in numerical computations we found a notable difference between the two expression of the NEC at the apparent horizon.\\
In order to understand this aspect we checked that the perturbative expansion of the apparent horizon behaves correctly. In our scheme the position of the apparent horizon can be written as
\begin{equation}
	r_{AH} = 2M_0 \biggl( 1 + \epsilon^2 \frac{M_2}{M_0} \biggr).
\end{equation}
We verified numerically that the term $r_{AH} - 2M_0$ scales quadratically in $\epsilon$ obtaining the results shown in fig. \ref{fig:AHScaling}. Therefore we conclude that at least in the stable cases $r_{AH}$ behaves correctly in perturbation theory. \\
From the numerical computations of $\tensor{T}{^t_t}$ we found that at the apparent horizon it does not scale with $\epsilon^2$. This behavior is related to the fact that
\begin{equation}
	\tensor{T}{^t_t} = \frac{\epsilon^2}{2} \frac{r \varphi_1^{(0,1)}(r,t)^2}{2 M_0-r} + \dots \quad ,
\end{equation}
and since $r_{AH} - 2M_0 = \OO(\epsilon^2)$, then at the apparent horizon $\tensor{T}{^t_t} = \OO(1)$ and therefore $\Bigl( 1 - \frac{2 M_0}{r} \Bigr) \tensor{T}{^t_t}$ scales with $\epsilon^2$.\\
The analysis we reported in this section has highlighted a criticality in studying the apparent horizon from a perturbative point of view. In fact in the stable cases, while $r_{AH}$ shows the correct behavior in perturbation theory, the NEC cannot be computed with an expansion in $\epsilon$, since near the apparent horizon the terms in $\epsilon^4$ are not negligible with respect to the terms in $\epsilon^2$.\\
This critical issue can be due to the fact that we are using Schwarzschild coordinates, which are not regular. A different choice of coordinates may allow to study the behavior of the apparent horizon in a perturbative framework. \\
However, it is possible to perform a perturbative analysis in Schwarzschild coordinates for timesteps in which the apparent horizon is absent, since in this case $\tensor{T}{^t_t}$ scales correctly. Moreover, decreasing the value of $\epsilon$, the energy of the field decreases, and the process of formation of the apparent horizon is slower (cf. fig. \ref{fig:AHScaling}, the curves corresponding to smaller $\epsilon$ start later). In this way it is possible to perform analysis in larger intervals of time, by setting sufficiently small values of $\epsilon$.

\chapter{Schwarzschild BHs in EsGB Gravity with Quartic Coupling}
\label{chapter:QuarticCoupling}
In this chapter we are going to analyze the stability of a Schwarschild BH in EsGB theory using a quartic coupling function $F[\phi] = \lambda \phi^2 + \gamma \phi^4$. We will start by describing the perturbative scheme we are going to use in this case, then we will derive the field equations in 1+1 variables at the first order around the Schwarzschild solution, and we will study the stability of the BH, with a focus on the case $\frac{\gamma}{\lambda} < 0$.

\section{Quartic EsGB Gravity and Perturbative Scheme}

EsGB gravity with quartic coupling function is particularly interesting since the Euler-Lagrange equation for the scalar field has a nonlinear term that can quench the tachyonic instability leading to the appearance of stable scalarized solutions \cite{Silva+Stability}. However this cannot happen in a perturbative scheme where the scalar field is expanded around the Schwarzschild solution, since at the second order in $\epsilon$ the quartic term does not contribute in the equations. Therefore, since we are interested in treating the quadratic and the quartic term in the scalar field at the same order in perturbation theory, we decided to perform an expansion of the gravitational constant $G$ around 0.

\subsection{Action and Field Equations}

In order to use this perturbative scheme we rewrite the action as
\begin{equation}
	S = \frac{1}{2} \int_\Omega d^4x\sqrt{-g}\biggl\{R - G (\nabla \phi)^2 + 2 G F[\phi] \mathcal{G} \biggr\}.
	\label{eq:action_G}
\end{equation}
In eq. \eqref{eq:action_G} and in the rest of the computations we have absorbed a factor $8 \pi$ in the Newton's constant.\\
The field equations obtained from the action \eqref{eq:action_G} are
\begin{equation}
	E^{(\phi)} := G\nabla_\mu \nabla^\mu \phi + G\frac{\delta F[\phi]}{\delta \phi} \mathcal{G} = 0,
	\label{eq:fieldeqG}
\end{equation}
\begin{multline}
	E^{(g)}_{\mu\nu} := R_{\mu\nu} - \frac{1}{2} g_{\mu\nu} R + G\frac{1}{2} \bigl( \nabla \phi \bigr)^2 g_{\mu\nu} - G\bigl( \nabla_\mu \phi \bigr) \bigl( \nabla_\nu \phi \bigr) + \\
	+ 2 G \Bigl( \nabla_\gamma \nabla^\alpha F[\phi] \Bigr) \delta^{\gamma\delta\kappa\lambda}_{\alpha\beta\rho\sigma} \tensor{R}{^{\rho\sigma}_{\kappa\lambda}} \delta^\beta_\mu g_{\nu\delta} = 0.
	\label{eq:metriceqG}
\end{multline}
The equations for the gravitational field $E^{(g)}_{\mu\nu}$ can be can be rewritten as
\begin{equation}
	G_{\mu\nu} = G T_{\mu\nu},
\end{equation}
where $G_{\mu\nu}$ is the Einstein tensor, and the stress-energy tensor $T_{\mu\nu}$ has the same expression as in eq. \eqref{eq:StressEnergyTensor}
\begin{equation}
	\tensor{T}{_{\mu\nu}} = - \frac{1}{2}\bigl(\nabla \phi \bigr)^2 g_{\mu\nu} + \bigl( \nabla_\mu \phi \bigr) \bigl( \nabla_\nu \phi \bigr) - 2 \Bigl( \nabla_\gamma \nabla^\alpha F[\phi] \Bigr) \delta^{\gamma \delta \kappa \lambda}_{\alpha \beta \rho \sigma} \tensor{R}{^{\rho \sigma}_{\kappa \lambda}} \delta^\beta_\mu g_{\nu \delta}.
	\label{eq:StressEnergyTensorG}
\end{equation}\\
The coupling function is written as $F[\phi] = \lambda \phi^2 + \gamma \phi^4$ and we define $\alpha = \frac{\gamma}{\lambda}$.\\
As we have done in the previous chapter we consider a system with spherical symmetry using a 1+1 formalism, therefore we write the scalar field as
\begin{equation}
	\phi = \phi(r, t),
\end{equation}
and the metric tensor as
\begin{equation}
g_{\mu\nu} = 
\begin{bmatrix}
   -A(r, t) & 0 & 0 & 0 \\
   0 & B(r, t) & 0 & 0 \\
   0 & 0 & r^2 & 0 \\
   0 & 0 & 0 & r^2 (\sin \theta)^2
\end{bmatrix}.
\end{equation}

\subsection{Perturbative Expansion of $G$ around $0$}
Now we are going to write the Newton's constant as $G = \epsilon$ and expand the field equations perturbatively up to the first order. 

\subsubsection*{Order $0$ in Perturbation Theory}
At the order $0$ the gravitational constant is written as $G = 0 + \OO(\epsilon)$. In this limit the action reduces to
\begin{equation}
	S = \frac{1}{2} \int_\Omega d^4x\sqrt{-g}\, R + \OO(\epsilon),
\end{equation}
which is the action of GR in vacuum without a scalar field. Therefore the scalar field equation reduces to
\begin{equation}
	E^{(\phi)} := \OO(\epsilon) = 0.
	\label{eq:Order0FieldG}
\end{equation}
For the metric tensor we have the equation
\begin{equation}
	E^{(g)}_{\mu\nu} := R_{\mu\nu} - \frac{1}{2}g_{\mu\nu}R + \OO(\epsilon) = 0.
	\label{eq:Order0MetricG}
\end{equation}
In the system of coordinates that we are using this equation is solved by the metric \cite{Ferrari-Gualtieri-Pani}
\begin{equation}
	ds^2 = - \biggl( 1 - \frac{K}{r} \biggr) dt^2 + \frac{1}{1 - \frac{K}{r}} dr^2 + r^2 (d\theta^2 + \sin^2{\theta} d\varphi^2),
	\label{eq:GeneralSchwarzschildK}
\end{equation}
where $K$ is an integration constant, that can be determined imposing that in the weak-field limit
\begin{equation}
	g_{tt} \sim - \biggl( 1 - \frac{2 G M}{c^2 r} \biggr).
\end{equation}
Therefore $K = \frac{2 G M}{c^2}$.\\
If we impose only $G = 0$ we obtain $K = 0$ and a flat spacetime, which is consistent with the fact that for a vanishing Newton's constant the gravitational force is absent.\\
However, since we are interested in a perturbative expansion around the Schwarzschild metric, we impose the condition that the term $G M = M_0$ is constant and different from zero when $G \to 0$, and we obtain that the equation \eqref{eq:Order0MetricG} is solved by the metric \eqref{eq:GeneralSchwarzschildK} with $K = 2M_0 \neq 0$.\\
Due to this requirement the perturbative approach that we are using is not fully mathematically consistent. In fact the assumption $\lim_{G \to 0} G M = M_0 \neq 0$ implies that the mass of the black hole diverges as $G \to 0$.

\subsubsection*{First Order in Perturbation Theory}

At the first order in $\epsilon$ the Newton's constant is expanded as $G = \epsilon + \OO(\epsilon^2)$ and the equation for the scalar field is given by
\begin{equation}
	\epsilon \biggl[ \nabla_\mu \nabla^\nu \phi + 2 \G \phi \Bigl( \lambda + 2 \gamma \phi^2  \Bigr) \biggr] + \OO(\epsilon^2) = 0.
	\label{eq:ScalarFieldDale}
\end{equation}
Since all the terms in eq. \eqref{eq:ScalarFieldDale} are multiplied by $\epsilon$, the covariant derivatives and the Gauss-Bonnet term can be computed at the order 0 using the Schwarzschild metric, and the equation for the scalar field can be written as
\begin{multline}
	E^{(\phi)} = \epsilon \biggl(\frac{96 M_0^2 \phi (r,t) \left(2 \gamma  \phi (r,t)^2+\lambda \right)}{r^6} + \frac{r \phi ^{(0,2)}(r,t)}{2 M_0-r} + \\
	+ \frac{2 \left(r-M_0\right) \phi ^{(1,0)}(r,t)+r \left(r-2 M_0\right) \phi ^{(2,0)}(r,t)}{r^2}\biggr)+\OO\left(\epsilon ^2\right) = 0.
\end{multline} \\

Instead for the metric tensor, using the same method as in previous chapter, we write the $tt$ and the $rr$ components of $g_{\mu\nu}$ as
\begin{equation}
	\begin{cases}
		g_{tt} = -A(r, t) = -\Bigl(1 - \frac{2M_0}{r} + \epsilon A_1(r, t)\Bigr) \\
		g_{rr} = B(r, t) = \frac{1}{1 - \frac{2M(r, t)}{r}}
	\end{cases},
\end{equation}
where $M(r, t) = M_0 + \epsilon M_1(r, t)$ has the dimensions of a mass.\\
At the first order in $\epsilon$ the equation for the metric tensor has the form
\begin{equation}
	G_{\mu\nu} = R_{\mu\nu} - \frac{1}{2} g_{\mu\nu} R = \epsilon T_{\mu\nu} + \OO(\epsilon),
	\label{eq:Order1MetricG}
\end{equation}
where the stress-energy tensor can be computed at the order zero. Expliciting the terms in \eqref{eq:Order1MetricG} we obtain the following set of equations
\begin{multline}
	E^{(g)}_{tt} = \epsilon  \biggl[-\frac{\left(2 M_0-r\right)}{2 r^6}\biggl(\phi ^{(1,0)}(r,t) \Bigl(r \left(2 M_0-r\right) \phi ^{(1,0)}(r,t) \bigl(384 \gamma  M_0 \phi (r,t)^2 + \\
	+ 64 \lambda  M_0+r^3\bigr)-64 M_0 \left(3 M_0-r\right) \phi (r,t) \left(2 \gamma  \phi (r,t)^2+\lambda \right)\Bigr)+ \\ 
	+ 4 r^3 M_1^{(1,0)}(r,t)+64 M_0 r \left(2 M_0-r\right) \phi (r,t) \phi ^{(2,0)}(r,t) \left(2 \gamma  \phi (r,t)^2+\lambda \right)\biggr)+ \\ 
	-\frac{1}{2} \phi ^{(0,1)}(r,t)^2\biggr]+\OO\left(\epsilon ^2\right) = 0,
\end{multline}
\begin{multline}
	E^{(g)}_{tr} = \frac{\epsilon }{r^4 \left(2 M_0-r\right)} \biggl(\phi ^{(0,1)}(r,t) \Bigl(r \left(r-2 M_0\right) \phi ^{(1,0)}(r,t) \bigl(192 \gamma  M_0 \phi (r,t)^2+ \\
	+32 \lambda  M_0+r^3\bigr)-32 M_0^2 \phi (r,t) \left(2 \gamma  \phi (r,t)^2+\lambda \right)\bigr)-2 r^3 M_1^{(0,1)}(r,t)+ \\
	-32 M_0 r \left(2 M_0-r\right) \phi (r,t) \phi ^{(1,1)}(r,t) \left(2 \gamma  \phi (r,t)^2+\lambda \right)\biggr)+\OO\left(\epsilon ^2\right) = 0,
\end{multline}
\begin{equation}
	E^{(g)}_{t\theta} = 0,
\end{equation}
\begin{equation}
	E^{(g)}_{t\varphi} = 0,
\end{equation}
\begin{multline}
	E^{(g)}_{rr} = \frac{1}{2} \epsilon  \biggl[-\frac{1}{r \left(r-2 M_0\right)^2} \biggl( 4 \left(M_0 A_1(r,t)+M_1(r,t)\right)+ \\
	- 2 r \left(r-2 M_0\right) A_1^{(1,0)}(r,t)+\phi ^{(0,1)}(r,t)^2 \left(384 \gamma  M_0 \phi (r,t)^2+64 \lambda  M_0+r^3\right)+ \\
	+ 64 M_0 \phi (r,t) \phi ^{(0,2)}(r,t) \left(2 \gamma  \phi (r,t)^2+\lambda \right)\biggr)-\phi ^{(1,0)}(r,t)^2 + \\ 
	- \frac{64 M_0 \left(r-3 M_0\right) \phi (r,t) \phi ^{(1,0)}(r,t) \left(2 \gamma  \phi (r,t)^2+\lambda \right)}{r^4 \left(r-2 M_0\right)}\biggr]+\OO\left(\epsilon ^2\right) = 0,
\end{multline}
\begin{equation}
	E^{(g)}_{r\theta} = 0,
\end{equation}
\begin{equation}
	E^{(g)}_{r\varphi} = 0,
\end{equation}
\begin{multline}
	E^{(g)}_{\theta\theta} = \frac{\epsilon }{2 r^3 \left(r-2 M_0\right)^2} \biggl[-r^3 \left(2 M_0-r\right) \biggl(2 M_0 A_1(r,t)+ \\
	+ \left(2 M_0-r\right) \left(r A_1^{(1,0)}(r,t)-2 M_1^{(1,0)}(r,t)\right)+2 M_1(r,t)\biggr)+ \\
	+ r^3 \biggl(\left(2 M_0-r\right) \Bigl(r^2 \left(2 M_0-r\right) A_1^{(2,0)}(r,t)+r \left(5 M_0-2 r\right) A_1^{(1,0)}(r,t)+ \\
	+ \left(4 r-6 M_0\right) M_1^{(1,0)}(r,t)\Bigr)+2 M_0^2 A_1(r,t)-2 r^3 M_1^{(0,2)}(r,t)+2 M_0 M_1(r,t)\biggr)+ \\
	+ r^4 \left(2 M_0-r\right) \left(r^2 \phi ^{(0,1)}(r,t)^2-\left(r-2 M_0\right)^2 \phi ^{(1,0)}(r,t)^2\right)+ \\ 
	- 16 M_0 \left(2 M_0-r\right) \biggl(2 r^3 \phi ^{(0,1)}(r,t)^2 \left(6 \gamma  \phi (r,t)^2+\lambda \right)+ \\
	- 2 \Bigl(-2 \gamma  \phi (r,t)^3 \bigl(\left(2 M_0-r\right) \left(\left(6 M_0-2 r\right) \phi ^{(1,0)}(r,t)+r \left(r-2 M_0\right) \phi ^{(2,0)}(r,t)\right)+ \\
	+ r^3 \phi ^{(0,2)}(r,t)\bigr)+\lambda  \phi (r,t) \Bigl(-2 \left(-5 M_0 r+6 M_0^2+r^2\right) \phi ^{(1,0)}(r,t)+ \\
	+ r \left(r-2 M_0\right)^2 \phi ^{(2,0)}(r,t)+r^3 \left(-\phi ^{(0,2)}(r,t)\right)\Bigr)+ \\
	+ 6 \gamma  r \left(r-2 M_0\right)^2 \phi ^{(1,0)}(r,t)^2 \phi (r,t)^2+\lambda  r \left(r-2 M_0\right)^2 \phi ^{(1,0)}(r,t)^2\Bigr)\biggr)\biggr]+\OO\left(\epsilon ^2\right) = 0,
\end{multline}
\begin{equation}
	E^{(g)}_{\theta\varphi} = 0,
\end{equation}
\begin{multline}
	E^{(g)}_{\varphi\varphi} = \frac{\epsilon  \sin ^2(\theta )}{2 r^3 \left(r-2 M_0\right)^2} \biggl[-r^3 \left(2 M_0-r\right) \Bigl(2 M_0 A_1(r,t)+ \\
	+ \left(2 M_0-r\right) \left(r A_1^{(1,0)}(r,t)-2 M_1^{(1,0)}(r,t)\right)+2 M_1(r,t)\Bigr)+ \\ 
	+ r^3 \biggl(\left(2 M_0-r\right) \Bigl(r^2 \left(2 M_0-r\right) A_1^{(2,0)}(r,t)+r \left(5 M_0-2 r\right) A_1^{(1,0)}(r,t)+ \\ 
	+ \left(4 r-6 M_0\right) M_1^{(1,0)}(r,t)\Bigr)+2 M_0^2 A_1(r,t)-2 r^3 M_1^{(0,2)}(r,t)+2 M_0 M_1(r,t)\biggr)+ \\ 
	+ r^4 \left(2 M_0-r\right) \left(r^2 \phi ^{(0,1)}(r,t)^2-\left(r-2 M_0\right)^2 \phi ^{(1,0)}(r,t)^2\right)+ \\
	- 16 M_0 \left(2 M_0-r\right) \Biggl(2 r^3 \phi ^{(0,1)}(r,t)^2 \left(6 \gamma  \phi (r,t)^2+\lambda \right)+ \\
	- 2 \biggl(-2 \gamma  \phi (r,t)^3 \Bigl(\left(2 M_0-r\right) \left(\left(6 M_0-2 r\right) \phi ^{(1,0)}(r,t)+r \left(r-2 M_0\right) \phi ^{(2,0)}(r,t)\right)+ \\
	+ r^3 \phi ^{(0,2)}(r,t)\Bigr)+\lambda  \phi (r,t) \Bigl(-2 \left(-5 M_0 r+6 M_0^2+r^2\right) \phi ^{(1,0)}(r,t)+ \\
	+ r \left(r-2 M_0\right)^2 \phi ^{(2,0)}(r,t)+r^3 \left(-\phi ^{(0,2)}(r,t)\right)\Bigr)+ \\
	+ 6 \gamma  r \left(r-2 M_0\right)^2 \phi ^{(1,0)}(r,t)^2 \phi (r,t)^2+\lambda  r \left(r-2 M_0\right)^2 \phi ^{(1,0)}(r,t)^2\biggr)\Biggr)\Biggr]+\OO\left(\epsilon ^2\right) = 0.
\end{multline}

From this system we obtain two equations and two constraints. The equations are
\begin{multline}
	\phi ^{(0,2)}(r,t) =  \frac{\left(r-2 M_0\right)}{r^7} \biggl(r^5 \left(r-2 M_0\right) \phi ^{(2,0)}(r,t)+ \\
	+ 2 r^4 \left(r-M_0\right) \phi ^{(1,0)}(r,t)+96 M_0^2 \phi (r,t) \left(2 \gamma  \phi (r,t)^2+\lambda \right)\biggr),
  \label{eq:Order1FieldG}
\end{multline}
\begin{multline}
	M_1^{(0,1)}(r,t) = \frac{1}{2 r^3}\biggl[\phi ^{(0,1)}(r,t) \biggl(r \left(r-2 M_0\right) \phi ^{(1,0)}(r,t) \Bigl(32 M_0 \left(6 \gamma  \phi (r,t)^2+\lambda \right) + \\
	+ r^3\Bigr)-32 M_0^2 \phi (r,t) \left(2 \gamma  \phi (r,t)^2+\lambda \right)\biggr)+ \\
	+ 32 M_0 r \left(r-2 M_0\right) \phi (r,t) \phi ^{(1,1)}(r,t) \left(2 \gamma  \phi (r,t)^2+\lambda \right)\biggr].
  \label{eq:Order1MassG}
\end{multline}
The constraints are
\begin{multline}
	M_1^{(1,0)}(r,t) = \frac{1}{4 r^3}\biggl[\frac{r^6 \phi ^{(0,1)}(r,t)^2}{r-2 M_0} + \\
	+ \phi ^{(1,0)}(r,t) \biggl(r \left(r-2 M_0\right) \phi ^{(1,0)}(r,t) \left(64 M_0 \left(6 \gamma  \phi (r,t)^2+\lambda \right)+r^3\right) + \\
	- 64 M_0 \left(r-3 M_0\right) \phi (r,t) \left(2 \gamma  \phi (r,t)^2+\lambda \right)\biggr)+ \\ 
	+ 64 M_0 r \left(r-2 M_0\right) \phi (r,t) \phi ^{(2,0)}(r,t) \left(2 \gamma  \phi (r,t)^2+\lambda \right)\biggr],
  \label{constraint:Order1MassG}
\end{multline}
\begin{multline}
	A_1^{(1,0)}(r,t) = \frac{1}{2} \Biggl[\frac{4 \left(M_0 A_1(r,t)+M_1(r,t)\right)}{r \left(r-2 M_0\right)}+ \\ 
	+ \frac{64 M_0 \left(r-3 M_0\right) \phi (r,t) \phi ^{(1,0)}(r,t) \left(2 \gamma  \phi (r,t)^2+\lambda \right)}{r^4} + \\ 
	+ \frac{\phi ^{(0,1)}(r,t)^2 \left(64 M_0 \left(6 \gamma  \phi (r,t)^2+\lambda \right)+r^3\right)}{r \left(r-2 M_0\right)}+ \\ 
	+ \frac{64 M_0 \phi (r,t) \left(2 \gamma  \phi (r,t)^2+\lambda \right)}{r^8} \biggl(r^5 \left(r-2 M_0\right) \phi ^{(2,0)}(r,t)+\\
	+ 2 r^4 \left(r-M_0\right) \phi ^{(1,0)}(r,t)+96 M_0^2 \phi (r,t) \left(2 \gamma  \phi (r,t)^2+\lambda \right)\biggr)+ \\ 
	+ \left(r-2 M_0\right) \phi ^{(1,0)}(r,t)^2 \biggr].
  \label{constraint:Order1A1G}
\end{multline}

We observe that when $\gamma = 0$ the equations are equivalent to the ones obtained in the previous chapter considering a perturbation in the scalar field and expanding up to the second order. In fact when the perturbative approach is applied to quadratic EsGB gravity writing the scalar field as $\phi = \hat \epsilon \varphi_1$, the stress-energy tensor is quadratic in the perturbation, and it can be written as $T_{\mu\nu} = \hat \epsilon^2 \hat T_{\mu\nu}$, where
\begin{equation}
	\tensor{\hat T}{_{\mu\nu}} = - \frac{1}{2}\bigl(\nabla \varphi_1 \bigr)^2 g_{\mu\nu} + \bigl( \nabla_\mu \varphi_1 \bigr) \bigl( \nabla_\nu \varphi_1 \bigr) - 2 \lambda \Bigl( \nabla_\gamma \nabla^\alpha \varphi_1^2 \Bigr) \delta^{\gamma \delta \kappa \lambda}_{\alpha \beta \rho \sigma} \tensor{R}{^{\rho \sigma}_{\kappa \lambda}} \delta^\beta_\mu g_{\nu \delta}.
\end{equation}
The equation for the metric tensor has the form
\begin{equation}
	G_{\mu\nu} = T_{\mu\nu} =  \hat \epsilon^2 \hat T_{\mu\nu},
\end{equation}
and it can equivalently be considered as the equation obtained from a perturbative expansion of the gravitational constant at the first order $G = \epsilon + \OO(\epsilon^2)$, with $\epsilon = \hat \epsilon^2$.

\subsection{Equations in Tortoise Coordinates}

Let us now change system of coordinates and write the equations and the constraints in tortoise coordinates.\\
Before performing the transformation, we compute the Kretschmann scalar $K = R^{\mu\nu\rho\sigma}R_{\mu\nu\rho\sigma}$ obtaining
\begin{multline}
	K = \frac{48 M_0^2}{r^6}+\frac{8 M_0 \epsilon}{r^{12} \left(r-2 M_0\right)} \Biggl(r^7 \phi ^{(0,1)}(r,t)^2 \left(96 M_0 \left(6 \gamma  \phi (r,t)^2+\lambda \right)-r^3\right) + \\
	+ 12 \left(r-2 M_0\right) \left(r^6 M_1(r,t)+768 M_0^3 \phi (r,t)^2 \left(2 \gamma  \phi (r,t)^2+\lambda \right)^2\right) + \\
	+ 384 M_0 r^4 \left(r-2 M_0\right)^2 \phi (r,t) \phi ^{(1,0)}(r,t) \left(2 \gamma  \phi (r,t)^2+\lambda \right)+\\
	+ r^5 \left(r-2 M_0\right)^2 \phi ^{(1,0)}(r,t)^2 \left(r^3-96 M_0 \left(6 \gamma  \phi (r,t)^2+\lambda \right)\right)\Biggr)+\OO\left(\epsilon ^2\right).
	\label{eq:KretschmannG}
\end{multline}
Analogously to the quadratic case that we analyzed with the expansion in the field, the singularity in $K$ for $r = 2 M_0$ is absent if $\phi$ satisfies the condition
\begin{equation}
	\lim_{r \to 2M_0} \partial_t \phi(r, t) = \lim_{r \to 2M_0} \biggl( 1 - \frac{2M_0}{r} \biggr) \partial_r \phi(r, t).
\end{equation} \\

We now proceed with the coordinate transformation obtaining
\begin{multline}
	\phi ^{(0,2)}(z,t) = \frac{2 \left(r(z,t)-2 M_0\right)}{r(z,t)^7} \biggl(48 M_0^2 \phi (z,t) \left(2 \gamma  \phi (z,t)^2+\lambda \right) + \\
	+ r(z,t)^5 \phi ^{(1,0)}(z,t)\biggr)+\phi ^{(2,0)}(z,t),
	\label{eq:Order1FieldGTortoise}
\end{multline}
\begin{multline}
	M_1^{(0,1)}(z,t) = \frac{1}{2 r(z,t)^3}\biggl[32 M_0 r(z,t)^2 \phi (z,t) \phi ^{(1,1)}(z,t) \left(2 \gamma  \phi (z,t)^2+\lambda \right) + \\
	+ \phi ^{(0,1)}(z,t) \biggl(r(z,t)^2 \phi ^{(1,0)}(z,t) \left(32 M_0 \left(6 \gamma  \phi (z,t)^2+\lambda \right)+r(z,t)^3\right) + \\
	- 32 M_0^2 \phi (z,t) \left(2 \gamma  \phi (z,t)^2+\lambda \right)\biggr)\biggr],
	\label{eq:Order1MassGTortoise}
\end{multline}
for the equations, and
\begin{multline}
	M_1^{(1,0)}(z,t) = \frac{1}{4 r(z,t)^3}\biggl[64 M_0 r(z,t)^2 \biggl(\phi ^{(1,0)}(z,t)^2 \left(6 \gamma  \phi (z,t)^2+\lambda \right) + \\
	+ \phi (z,t) \phi ^{(2,0)}(z,t) \left(2 \gamma  \phi (z,t)^2+\lambda \right)\biggr) + \\
	- 64 M_0 r(z,t) \phi (z,t) \phi ^{(1,0)}(z,t) \left(2 \gamma  \phi (z,t)^2+\lambda \right) + \\ 
	+ 64 M_0^2 \phi (z,t) \phi ^{(1,0)}(z,t) \left(2 \gamma  \phi (z,t)^2+\lambda \right) + \\ 
	+ r(z,t)^5 \left(\phi ^{(0,1)}(z,t)^2+\phi ^{(1,0)}(z,t)^2\right)\biggr],
	\label{constraint:Order1MassGTortoise}
\end{multline}
\begin{multline}
	A_1^{(1,0)}(z,t) = \frac{1}{2 r(z,t)^9}\biggl[4 r(z,t)^7 \biggl(M_0 \Bigl(A_1(z,t) + \\
	+ 16 \left(\phi ^{(0,1)}(z,t)^2 \left(6 \gamma  \phi (z,t)^2+\lambda \right)+\phi (z,t) \phi ^{(2,0)}(z,t) \left(2 \gamma  \phi (z,t)^2+\lambda \right)\right)\Bigr) + \\
	+ M_1(z,t)\biggr)+192 M_0 r(z,t)^6 \phi (z,t) \phi ^{(1,0)}(z,t) \left(2 \gamma  \phi (z,t)^2+\lambda \right) + \\
	- 448 M_0^2 r(z,t)^5 \phi (z,t) \phi ^{(1,0)}(z,t) \left(2 \gamma  \phi (z,t)^2+\lambda \right) + \\
	+ 6144 M_0^3 r(z,t) \phi (z,t)^2 \left(2 \gamma  \phi (z,t)^2+\lambda \right)^2 + \\
	- 12288 M_0^4 \phi (z,t)^2 \left(2 \gamma  \phi (z,t)^2+\lambda \right)^2+r(z,t)^{10} \left(\phi ^{(0,1)}(z,t)^2+\phi ^{(1,0)}(z,t)^2\right)\biggr],
	\label{constraint:Order1A1GTortoise}
\end{multline}
for the constraints.

\section{Stability of the Schwarzschild Solution}

The trivial configurations of the scalar field that satisfy the field equations are given by eq. \eqref{eq:FieldHypotesis}, which for quartic EsGB gravity can be written as
\begin{equation}
	0 = \frac{\delta F[\phi]}{\delta \phi}\biggr|_{\phi = \phi_0} = 2 \lambda \phi_0 + 4 \gamma \phi_0^3 = 2 \lambda \phi_0 \biggl( 1 + 2 \frac{\gamma}{\lambda} \phi_0^2 \biggr) = 2 \lambda \phi_0 \bigl( 1 + 2 \alpha \phi_0^2 \bigr).
	\label{eq:QuarticTrivialEq}
\end{equation}
Eq. \eqref{eq:QuarticTrivialEq} has three solutions which are \cite{Silva+Stability, MasatoTaishi}
\begin{equation}
	\phi_0 = 0, \qquad \qquad \qquad \qquad \phi_0^{\pm} = \pm \sqrt{-\frac{1}{2\alpha}},
	\label{eq:QuarticTrivialFields}
\end{equation}
where $\phi_0^{\pm}$ exist only if $\alpha < 0$. We considered only the solution $\phi_0 = 0$.\\
We made a time-domain analysis of the stability of the Schwarzschild solution by integrating the field equations with a numerical method.

\subsection{Numerical Integration}

For the numerical integration of the equations we used the method of lines with the 4th order Runge-Kutta method for the time evolution.\\
We used the following boundary conditions at the horizon:
\begin{equation}
	\begin{cases}
		\partial_z \phi(z_0, t) = \partial_t \phi(z_0, t)\\
		M_1(z_0, t) = 0
	\end{cases},
	\label{eq:BoundaryG}
\end{equation}
where $z_0$ is the numerical inner boundary. \\
We used a spatial grid with $10000$ points that extends from $\frac{z_0}{M_0} = -34.84$ up to $\frac{z_\infty}{M_0} = 330.11$, so that the grid step is $\frac{\Delta z}{M_0} = 0.036$. The time step is $\frac{\Delta t}{M_0} = 0.001$.\\

The procedure we used to analyze the stability of the Schwarzschild solution is the following:
\begin{enumerate}
	\item we chose an initial profile for $\phi$ which is a small perturbation of $\phi_0$ and which satisfies the boundary conditions:
		\begin{gather}
			\phi(z, t) = 0, \notag \\
			\partial_t \phi(z, t) = 
			\begin{cases}
				N e^{-\frac{(z-\mu)^2}{\sigma^2}} & \quad \text{if} \quad \mu - 5\sigma < z < \mu + 5\sigma\\
				0 				& \qquad \text{otherwise}
			\end{cases},
			\label{eq:InitialProfileG}
		\end{gather}
		where $N = 0.01$, $\mu = 5 \, M_0$ and $\sigma = 4 \, M_0$;
	\item we integrated numerically eq. \eqref{constraint:Order1MassGTortoise} with the Simpson rule using the scalar field at t=0 to obtain the initial profile of $M_1$;
	\item we integrated eq. \eqref{eq:Order1FieldGTortoise} and \eqref{eq:Order1MassGTortoise} with the method of lines using the 4th order Runge-Kutta method for time evolution, analyzing whether the scalar field converges to the Schwarzschild solution or diverges.
\end{enumerate}

\subsection{Stability Analysis}

\begin{figure}
	\centering
	\includegraphics[width=0.65\textwidth]{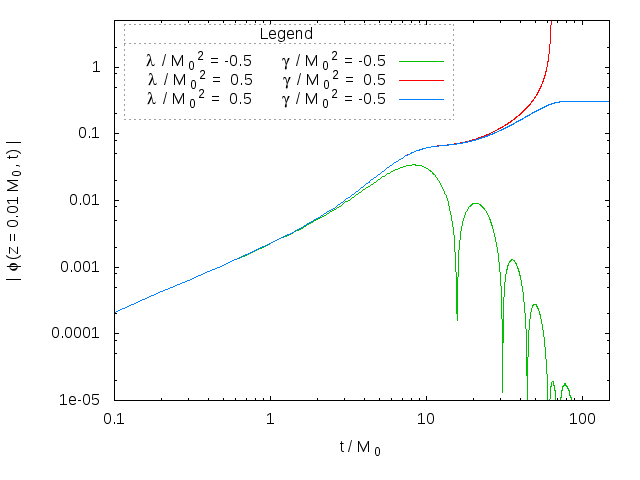}
	\caption{Evolution of the scalar field at $\frac{z}{M_0} = 0.01$ in three different cases: a case in which $\frac{\lambda}{M_0^2} = -0.5$ and $\frac{\gamma}{M_0^2} = -0.5$, and the Schwarzschild solution is stable (green curve); a case in which $\frac{\lambda}{M_0^2} = 0.5$ and $\frac{\gamma}{M_0^2} = 0.5$, in which the Schwarzschild solution in unstable and the scalar field diverges (red curve); a case in which $\frac{\lambda}{M_0^2} = 0.5$ and $\frac{\gamma}{M_0^2} = -0.5$, in which the Schwarschild solution is unstable and $\phi(z = 0.01 M_0, t)$ stabilize to a constant value (blue curve).}
	\label{fig:equilibrium_comparison}
\end{figure}

\begin{figure}
	\centering
	\subfloat[][]
	{\includegraphics[width=0.50\textwidth]{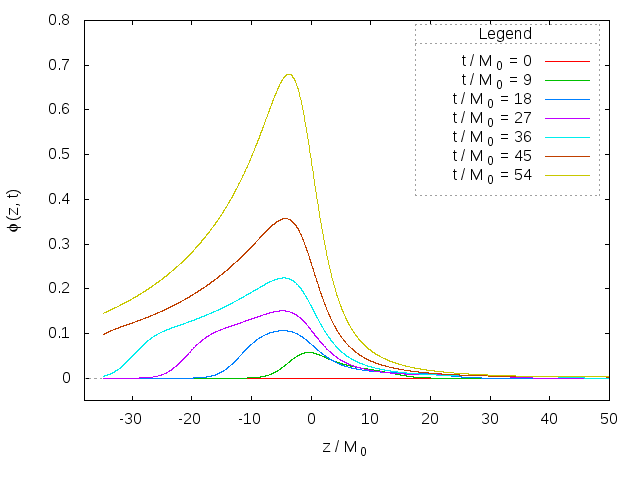}}
	\subfloat[][]
	{\includegraphics[width=0.50\textwidth]{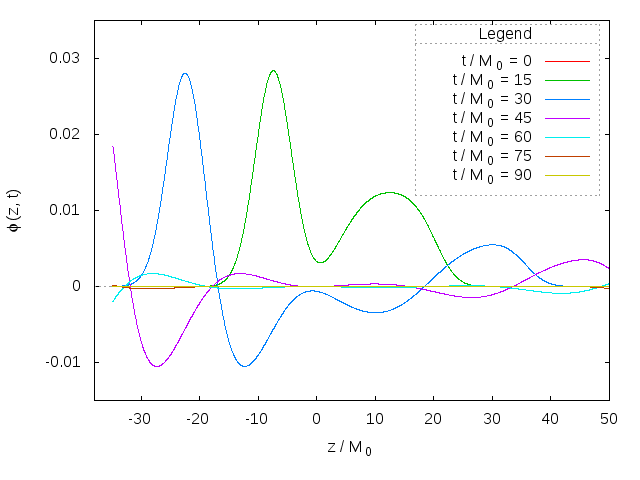}} \\
	\subfloat[][]
	{\includegraphics[width=0.50\textwidth]{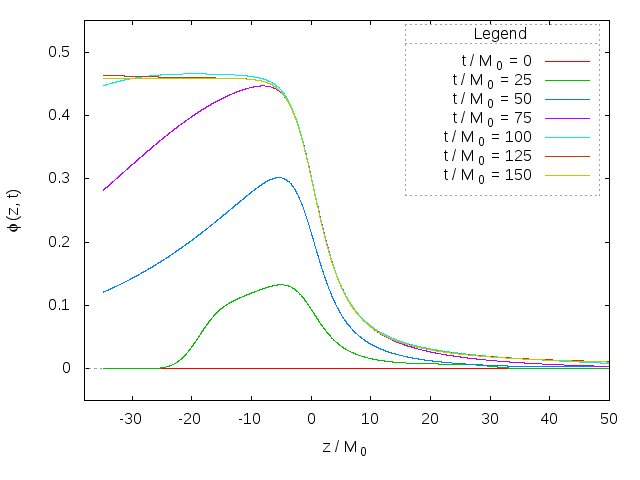}}
	\caption{Evolutions of the profile of the scalar field. In the upper left panel we show an unstable case with $\frac{\lambda}{M_0^2} = 0.5$ and $\frac{\gamma}{M_0^2} = 0.5$, in the upper right panel we show a stable case with $\frac{\lambda}{M_0^2} = -0.5$ and $\frac{\gamma}{M_0^2} = -0.5$, and in the lower panel we show the case with $\frac{\lambda}{M_0^2} = 0.5$ and $\frac{\gamma}{M_0^2} = -0.5$, in which the scalar field approaches a nontrivial configuration.}
	\label{fig:evolutions}
\end{figure}

We integrated numerically the field equations for values of $\frac{\lambda}{M_0^2}$ and $\frac{\gamma}{M_0^2}$ between $-1$ and $1$, and we analyzed the evolution of $\phi$ and $M_1$ at fixed $z$. First we checked that when $\frac{\gamma}{M_0^2} = 0$ the code produced the same results as in the case of quadratic EsGB that we studied in the previous chapter and we obtained behaviors identical to those represented in fig. \ref{fig:comparison}. \\
For $\frac{\gamma}{M_0^2} \neq 0$ we obtained that the Schwarzschild solution is unstable when $\frac{\lambda}{M_0^2} > \frac{\tilde \lambda}{M_0^2} = 0.363$ independently of the value of $\frac{\gamma}{M_0^2}$. This is in agreement with ref. \cite{MasatoTaishi}.\\
In the region with $\frac{\lambda}{M_0^2} > 0.363$ and $\frac{\gamma}{M_0^2} < 0$ we observed that the scalar field at $\frac{z}{M_0} = 0.01$ converges to a constant value. In fig. \ref{fig:equilibrium_comparison} we have shown the plots of the evolution of $\phi(z = 0.01 M_0, t)$ with three different choices of $\lambda$ and $\gamma$. The green curve represents a configuration where Schwarzschild solution is stable, and the others correspond to configurations where the Schwarzschild solution is unstable. In particular the red curve corresponds to a case with $\frac{\gamma}{M_0^2} > 0$ and the scalar field diverges, while the blue curve represents a case in which $\frac{\gamma}{M_0^2} < 0$ and the scalar field approaches a constant value.\\
In fig. \ref{fig:evolutions} we show the evolution of the complete profile of the scalar field corresponding to the same cases shown in fig. \ref{fig:equilibrium_comparison}. As we can see from the lower panel, in the case with $\frac{\lambda}{M_0^2} > 0.363$ and $\frac{\gamma}{M_0^2} < 0$ the scalar field approaches a nontrivial configuration, \ie a scalarized BH solution. We are going to discuss these asymptotic solutions in the next section.

\section{Scalarized Solutions}

\subsection{Characteristics and Scalar Charge}

\begin{figure}
	\centering
	\subfloat[][]
		{\includegraphics[width=0.5\textwidth]{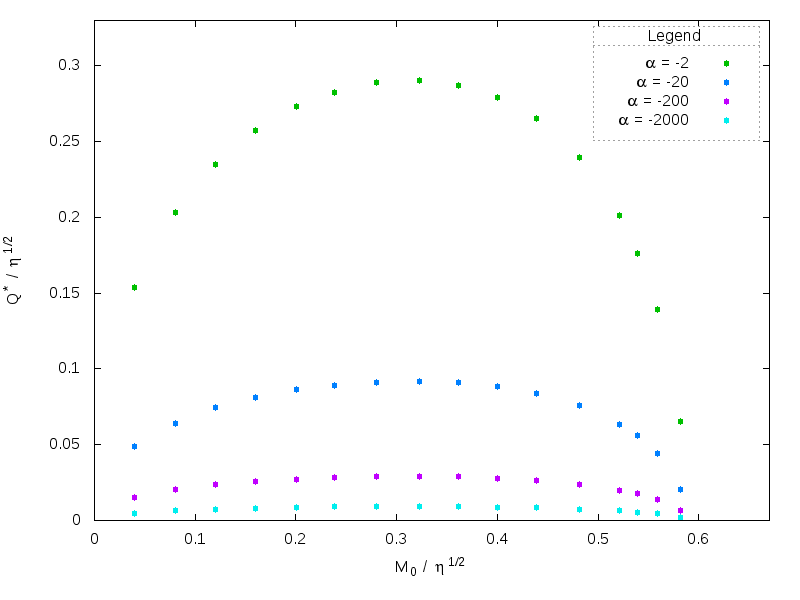}}
	\subfloat[][]
		{\includegraphics[width=0.5\textwidth]{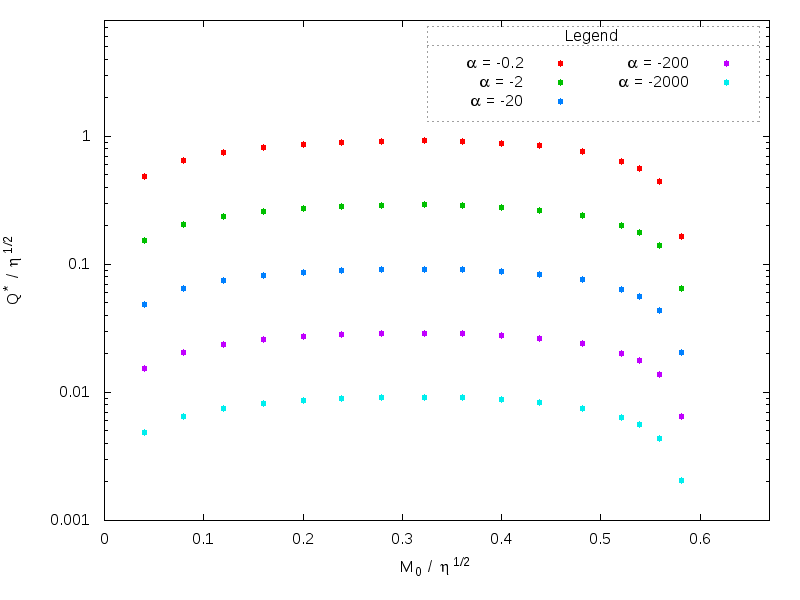}}
	\caption{Scalar charges obtained from the fit of the scalar field, classified by the value of $\alpha$, and shown in a plot of $\frac{Q^*}{\sqrt{\eta}}$ vs $\frac{M_0}{\sqrt{\eta}}$, where $\eta = 8\lambda$ and $Q^* = \sqrt{2} Q$ are respectively the coupling constant and the scalar charge used by Silva \ea in ref. \cite{Silva+Stability} and by Minamitsuji and Ikeda in ref. \cite{MasatoTaishi}. In the plot on the left the axes are on a linear scale, while in the right panel the data are shown in a semi-log plot. The set of points corresponding to $\alpha=-0.2$ (red) is only shown in the plot on the right.}
	\label{fig:scalar_charge}
\end{figure}

The scalarized BH solutions can be characterized with the number of nodes, which is zero for the solutions that we found, and with the scalar charge, which depends on the values of the parameters $\lambda$ and $\gamma$. \\
From a theoretical point of view the scalar charge can be computed as \cite{MasatoTaishi}
\begin{equation}
	Q = - \lim_{r \to +\infty} r^2 \frac{d \phi(r)}{dr}.
\end{equation}
However we cannot use this formula since at finite time the pulse of the scalar field propagates, and $\phi$ at infinity is null. Therefore we integrated the equations until $\frac{t}{M_0} = 575$ and with a spatial grid with $15000$ points that extends from $\frac{z_0}{M_0} = -34.84$ up to $\frac{z_\infty}{M_0} = 631.42$, in such a way that the pulse reaches position sufficiently far from the horizon. Then we fitted the profile of the scalar field with the function
\begin{equation}
	f(r) = a + \frac{b}{r} + \frac{c}{r^2}
	\label{eq:FitQ}
\end{equation}
in a region between the horizon and the pulse, and we obtained the scalar charge as $Q = b$. \\
We repeated this procedure for several values of $\lambda$ and $\alpha = \frac{\gamma}{\lambda}$. In fig. \ref{fig:scalar_charge} we show the scalar charges obtained, classified by the value of $\alpha$. On the x-axis there is $\frac{M_0}{\sqrt{\eta}}$ and on the y-axis there is $\frac{Q^*}{\sqrt{\eta}}$, where $\eta = 8\lambda$ and $Q^* = \sqrt{2} Q$ are respectively the coupling constant and the scalar charge used by Silva \ea in ref. \cite{Silva+Stability} and by Minamitsuji and Ikeda in ref. \cite{MasatoTaishi}. The results are in qualitative agreement with those shown in these references for the cases with $\alpha < -0.8$, in which, according to ref. \cite{Silva+Stability}, the nodeless scalarized modes are known to be stable. \\

As we can see from the right panel of fig. \ref{fig:scalar_charge}, when $\alpha = -0.2$ the behavior of the scalar charge is analogous to the other cases, with $\frac{Q^*}{\sqrt{\eta}}$ going to zero as $\frac{M_0}{\sqrt{\eta}}$ goes to the critical value $\frac{M_0}{\sqrt{8 \tilde \lambda}} = 0.587$ from the left. This is different from what is shown in ref. \cite{Silva+Stability} for $-0.8 < \alpha < 0$. \\
We identified two possible reasons for this discrepancy. Firstly, with our perturbative approach we can only find scalarized solutions whose scalar charge tends to zero, otherwise they cannot be considered as a perturbation of the Schwarzschild solution. On the other hand the expansion in the Newton's constant is not fully mathematically consistent, as we discussed earlier in this chapter, and this could be another reason for the appearance of this stable scalarized modes different from those found by Silva \ea.\\
In both cases the nodeless scalarized solutions found with our time-domain analysis could be an artifact of the perturbative approach, and it could be interesting to investigate in a future work whether these solutions can be found also with a nonperturbative analysis.

\subsection{Stability of the Scalarized Solutions}
\label{section:stability_scalarized}

To verify the stability of the scalarized asymptotic solutions we used the following procedure:
\begin{enumerate}
	\item we started from the initial profile of the scalar field written in eq. \eqref{eq:InitialProfileG}, and we integrated the field equations until $\frac{t}{M_0} = 200$ with the numerical procedure described for the analysis of the stability of the Schwarzschild solution;
	\item we considered the profile of the scalar field at $\frac{t}{M_0} = 100$ as an initial profile and we added to $\partial_t \phi$ a small perturbation which is the same as the initial profile \eqref{eq:InitialProfileG};
	\item we integrated eq. \eqref{constraint:Order1MassGTortoise} with the Simpson rule using the boundary conditions \eqref{eq:BoundaryG} to obtain the profile of $M_1$;
	\item we integrated eq. \eqref{eq:Order1FieldGTortoise} and \eqref{eq:Order1MassGTortoise} with the method of lines using the 4th order Runge-Kutta method for time evolution, until $\frac{t}{M_0} = 100$;
	\item we compared the profile of the field at $\frac{t}{M_0} = 100$ with the one obtained from the first integration until $\frac{t}{M_0} = 200$.
\end{enumerate}

\begin{figure}
	\centering
	\includegraphics[width=0.65\textwidth]{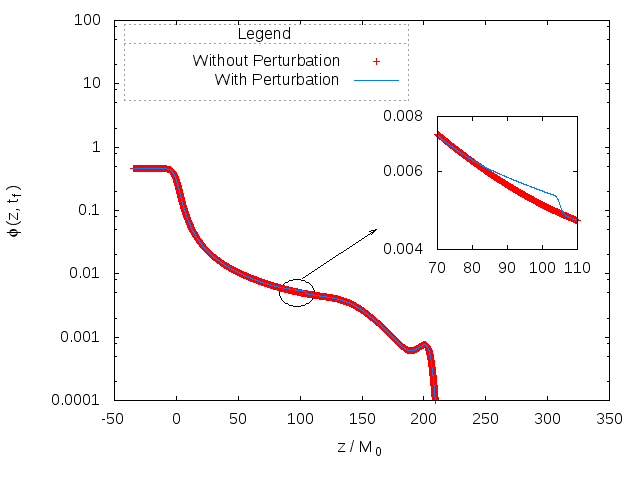}
	\caption{Comparison of the profiles of the scalar field obtained with the procedure described in section \ref{section:stability_scalarized}, in the case of $\frac{\lambda}{M_0^2} = 0.5$ and $\frac{\gamma}{M_0^2} = -0.5$. In red it is shown the profile of the scalar field at $\frac{t}{M_0} = \frac{t_f}{M_0} = 200$ obtained integrating the field equations using the initial profile in \eqref{eq:InitialProfileG}. In blue the profile of the scalar field obtained by adding a perturbation in the profile at $\frac{t}{M_0} = 100$ and integrating for $\frac{\Delta t}{M_0} = 100$.}
	\label{fig:stability_scalarized}
\end{figure}

The results of this computation are shown in fig. \ref{fig:stability_scalarized} for the case of $\frac{\lambda}{M_0^2} = 0.5$ and $\frac{\gamma}{M_0^2} = -0.5$, where the red profile is the one obtained from the first integration until $\frac{t}{M_0} = 200$, while the blue profile is the one obtained from the second integration, with a perturbation added in $\phi(z, t = 100 M_0)$.
As we can see the profile obtained with the second integration evolves in the same way as the scalarized profile obtained with the first integration except for a small perturbation, therefore we conclude that the nodeless scalarized solutions that we found in the region with $\frac{\lambda}{M_0^2} > 0.363$ and $\frac{\gamma}{M_0^2} < 0$ are stable. 

\chapter{Schwarzschild BHs in EsGB Gravity with Exponential Coupling}
\label{chapter:ExponentialCoupling}

In this chapter we are going to study the case of exponential coupling function $F[\phi] = \frac{\lambda}{3} \Bigl( 1 - e^{-3\phi^2} \Bigr)$, using the perturbative scheme with the expansion in the Newton's constant. As done for the other couplings, we will start by writing the field equations in tortoise coordinates, and then we will discuss the stability and the behavior of the scalar charge on the basis of the results of the numerical integration.

\section{Field Equations for Exponential Coupling}

The approach we will use is exactly the same that we used in the case of quartic coupling and therefore we are going to write the field equations at the first order, without repeating all the procedure and outlining only the main passages. \\

We consider a spherically symmetric system using a 1+1 decomposition:
\begin{equation}
	\phi = \phi(r, t), \qquad \qquad
g_{\mu\nu} = 
\begin{bmatrix}
   -A(r, t) & 0 & 0 & 0 \\
   0 & B(r, t) & 0 & 0 \\
   0 & 0 & r^2 & 0 \\
   0 & 0 & 0 & r^2 (\sin \theta)^2
\end{bmatrix}.
\end{equation}

Then we write the Newton's constant as $G = \epsilon + \OO(\epsilon^2)$, and we expand in the field equations.\\
At the order zero we obtain the Schwarzschild solution, provided that $GM \to M_0 \neq 0$ when $G \to 0$. \\

\subsection{Field Equations at the First Order in the Perturbative Expansion}

At the first order the equation for the scalar field is
\begin{equation}
	\epsilon \biggl[ \nabla_\mu \nabla^\mu \phi + 2\G \phi e^{-3\phi^2} \biggr] = 0.
	\label{eq:Order1FieldExponentialGeneral}
\end{equation}
The explicit form of eq. \eqref{eq:Order1FieldExponentialGeneral} can be computed using the Schwarzschild metric obtaining
\begin{multline}
	E^{(\phi)} = \epsilon  \Biggl(\frac{96 \lambda  M_0^2 e^{-3 \phi (r,t)^2} \phi (r,t)}{r^6}+\frac{r \phi ^{(0,2)}(r,t)}{2 M_0-r} + \\
	+ \frac{2 \left(r-M_0\right) \phi ^{(1,0)}(r,t)+r \left(r-2 M_0\right) \phi ^{(2,0)}(r,t)}{r^2}\Biggr)+\OO\left(\epsilon ^2\right) = 0.
\end{multline} \\

For the metric tensor instead we write $g_{tt}$ and $g_{rr}$ as
\begin{equation}
	\begin{cases}
		g_{tt} = -A(r, t) = -\Bigl(1 - \frac{2M_0}{r} + \epsilon A_1(r, t)\Bigr) \\
		g_{rr} = B(r, t) = \frac{1}{1 - \frac{2M(r, t)}{r}}
	\end{cases},
	\label{eq:gttgrrexpoenential}
\end{equation}
where $M(r, t) = M_0 + \epsilon M_1(r, t)$.\\
Substituting the expression \eqref{eq:gttgrrexpoenential} in 
\begin{multline}
	E^{(g)}_{\mu\nu} := R_{\mu\nu} - \frac{1}{2} g_{\mu\nu} R + \epsilon \frac{1}{2} \bigl( \nabla \phi \bigr)^2 g_{\mu\nu} - \epsilon\bigl( \nabla_\mu \phi \bigr) \bigl( \nabla_\nu \phi \bigr) + \\
	+ \frac{2\lambda}{3} \epsilon \biggl[ \nabla_\gamma \nabla^\alpha \Bigl(1 - e^{-3\phi^2} \Bigr) \biggr] \delta^{\gamma\delta\kappa\lambda}_{\alpha\beta\rho\sigma} \tensor{R}{^{\rho\sigma}_{\kappa\lambda}} \delta^\beta_\mu g_{\nu\delta} + \OO(\epsilon^2) = 0
\end{multline}
and expanding up to the first order we obtain the set of equations
\begin{multline}
	E^{(g)}_{tt} = \frac{1}{2} \epsilon  \Biggl[\frac{\left(r-2 M_0\right) e^{-3 \phi(r,t)^2}}{r^6} \Biggl(64 \lambda  M_0 \biggl(r \left(2 M_0-r\right) \phi(r,t) \phi^{(2,0)}(r,t)+ \\
	+ \phi^{(1,0)}(r,t) \Bigl(r \left(r-2 M_0\right) \left(6 \phi(r,t)^2-1\right) \phi^{(1,0)}(r,t)+ \\ 
	+ \left(r-3 M_0\right) \phi(r,t)\Bigr)\biggr) - r^3 e^{3 \phi (r,t)^2} \Bigl(r \left(r-2 M_0\right) \phi^{(1,0)}(r,t)^2+ \\
	- 4 M_1^{(1,0)}(r,t)\Bigr)\Biggr) - \phi^{(0,1)}(r,t)^2\Biggr]+\OO\left(\epsilon^2\right) = 0,
\end{multline}

\begin{multline}
	E^{(g)}_{tr} = \frac{\epsilon \,\, e^{-3 \phi(r,t)^2}}{r^4 \left(2 M_0-r\right)} \Biggl[-2 r^3 M_1^{(0,1)}(r,t) e^{3 \phi (r,t)^2}-32 \lambda  M_0^2 \phi (r,t) \phi ^{(0,1)}(r,t)+ \\
	+ r \left(r-2 M_0\right) \Biggl(\phi ^{(0,1)}(r,t) \phi ^{(1,0)}(r,t) \biggl(32 \lambda  M_0 \left(1-6 \phi (r,t)^2\right)+r^3 e^{3 \phi (r,t)^2}\biggr)+ \\
	+ 32 \lambda  M_0 \phi (r,t) \phi^{(1,1)}(r,t)\Biggr)\Biggr]+\OO\left(\epsilon ^2\right) = 0,
\end{multline}

\begin{equation}
	E^{(g)}_{t\theta} = 0 = 0,
\end{equation}

\begin{equation}
	E^{(g)}_{t\varphi} = 0 = 0,
\end{equation}

\begin{multline}
	E^{(g)}_{rr} = \frac{\epsilon}{2 r^4 \left(r-2 M_0\right)^2} \Biggl[64 \lambda  M_0 e^{-3 \phi (r,t)^2} \Biggl(r^3 \left(6 \phi(r,t)^2-1\right) \phi^{(0,1)}(r,t)^2+ \\
	+ \phi(r,t) \biggl(-\left(r-3 M_0\right) \left(r-2 M_0\right) \phi^{(1,0)}(r,t)-r^3 \phi^{(0,2)}(r,t)\biggr)\Biggr) + \\
	- r^3 \biggl(2 r \left(2 M_0-r\right) A_1^{(1,0)}(r,t)+r \left(r-2 M_0\right)^2 \phi^{(1,0)}(r,t)^2 + \\
	+ 4 M_0 A_1(r,t)+4 M_1(r,t)+r^3 \phi^{(0,1)}(r,t)^2\biggr)\Biggr]+\OO\left(\epsilon^2\right) = 0,
\end{multline}

\begin{equation}
	E^{(g)}_{r\theta} = 0 = 0,
\end{equation}

\begin{equation}
	E^{(g)}_{r\varphi} = 0 = 0,
\end{equation}

\begin{multline}
	E^{(g)}_{\theta\theta} = \frac{\epsilon}{2 r^3 \left(r-2 M_0\right)^2} \Biggl[-r^3 \left(2 M_0-r\right) \biggl(2 M_0 A_1(r,t)+ \\
	+ \left(2 M_0-r\right) \left(r A_1^{(1,0)}(r,t)-2 M_1^{(1,0)}(r,t)\right)+2 M_1(r,t)\biggr)+ \\
	+ r^3 \Biggl(\left(2 M_0-r\right) \biggl(r^2 \left(2 M_0-r\right) A_1^{(2,0)}(r,t)+r \left(5 M_0-2 r\right) A_1^{(1,0)}(r,t)+ \\
	+ \left(4 r-6 M_0\right) M_1^{(1,0)}(r,t)\biggr)+2 M_0^2 A_1(r,t)-2 r^3 M_1^{(0,2)}(r,t)+2 M_0 M_1(r,t)\Biggr)+ \\
	- 32 \lambda  M_0 \left(2 M_0-r\right) e^{-3 \phi (r,t)^2} \Biggl(\phi (r,t) \biggl(\left(2 M_0-r\right) \Bigl(\left(6 M_0-2 r\right) \phi^{(1,0)}(r,t)+ \\
	+ r \left(r-2 M_0\right) \phi^{(2,0)}(r,t)\Bigr)+r^3 \phi^{(0,2)}(r,t)\biggr)-r \left(r-2 M_0\right)^2 \phi^{(1,0)}(r,t)^2+ \\
	+ 6 r \left(r-2 M_0\right)^2 \phi(r,t)^2 \phi^{(1,0)}(r,t)^2+r^3 \left(1-6 \phi(r,t)^2\right) \phi^{(0,1)}(r,t)^2\Biggr)+ \\ 
	+ r^4 \left(2 M_0-r\right) \left(r^2 \phi^{(0,1)}(r,t)^2-\left(r-2 M_0\right)^2 \phi^{(1,0)}(r,t)^2\right)\Biggr]+\OO\left(\epsilon ^2\right) = 0,
\end{multline}

\begin{equation}
	E^{(g)}_{\theta\varphi} = 0 = 0,
\end{equation}

\begin{multline}
	E^{(g)}_{\varphi\varphi} = \frac{\epsilon \,\, \sin^2(\theta)}{2 r^3 \left(r-2 M_0\right)^2} \Biggl[-r^3 \left(2 M_0-r\right) \biggl(2 M_0 A_1(r,t) + \\
	+ \left(2 M_0-r\right) \left(r A_1^{(1,0)}(r,t)-2 M_1^{(1,0)}(r,t)\right)+2 M_1(r,t)\biggr)+ \\
	+ r^3 \Biggl(\left(2 M_0-r\right) \biggl(r^2 \left(2 M_0-r\right) A_1^{(2,0)}(r,t)+r \left(5 M_0-2 r\right) A_1^{(1,0)}(r,t)+ \\
	+ \left(4 r-6 M_0\right) M_1^{(1,0)}(r,t)\biggr)+2 M_0^2 A_1(r,t)-2 r^3 M_1^{(0,2)}(r,t)+2 M_0 M_1(r,t)\Biggr)+ \\
	- 32 \lambda  M_0 \left(2 M_0-r\right) e^{-3 \phi(r,t)^2} \Biggl(\phi(r,t) \biggl(\left(2 M_0-r\right) \Bigl(\left(6 M_0-2 r\right) \phi^{(1,0)}(r,t)+ \\
	+ r \left(r-2 M_0\right) \phi^{(2,0)}(r,t)\Bigr)+r^3 \phi^{(0,2)}(r,t)\biggr)-r \left(r-2 M_0\right)^2 \phi^{(1,0)}(r,t)^2+ \\
	+ 6 r \left(r-2 M_0\right)^2 \phi(r,t)^2 \phi^{(1,0)}(r,t)^2+r^3 \left(1-6 \phi(r,t)^2\right) \phi^{(0,1)}(r,t)^2\Biggr)+ \\
	+ r^4 \left(2 M_0-r\right) \left(r^2 \phi^{(0,1)}(r,t)^2-\left(r-2 M_0\right)^2 \phi^{(1,0)}(r,t)^2\right)\Biggr]+\OO\left(\epsilon^2\right) = 0.
\end{multline}

The equation $E^{(\phi)} = 0$ can be rewritten as
\begin{multline}
	\phi^{(0,2)}(r,t) = \frac{\left(r-2 M_0\right) e^{-3 \phi (r,t)^2}}{r^7} \Biggl[r^4 e^{3 \phi(r,t)^2} \biggl(2 \left(r-M_0\right) \phi^{(1,0)}(r,t)+ \\
	+ r \left(r-2 M_0\right) \phi^{(2,0)}(r,t)\biggr)+96 \lambda  M_0^2 \phi(r,t)\Biggr].
\end{multline}
After substituting the expression for $\phi^{(0,2)}(r,t)$ in the equations for the metric, we can isolate the term $M_1^{(0,1)}(r,t)$ in $E^{(g)}_{tr} = 0$, obtaining
\begin{multline}
	M_1^{(0,1)}(r,t) = \frac{16 \lambda  M_0 e^{-3 \phi(r,t)^2}}{r^3} \Biggl[\phi^{(0,1)}(r,t) \Biggl(-M_0 \phi(r,t) + \\
	- r \left(r-2 M_0\right) \left(6 \phi(r,t)^2-1\right) \phi^{(1,0)}(r,t)\Biggr)+r \left(r-2 M_0\right) \phi(r,t) \phi^{(1,1)}(r,t)\Biggr] + \\
	+ \frac{1}{2} r \left(r-2 M_0\right) \phi^{(0,1)}(r,t) \phi^{(1,0)}(r,t).
\end{multline}
Isolating the terms $M_1^{(1,0)}(r,t)$ and $A_1^{(1,0)}(r,t)$ in the equations $E^{(g)}_{tt} = 0$ and $E^{(g)}_{rr} = 0$ respectively we obtain the two constraints
\begin{multline}
	M_1^{(1,0)}(r,t) = -\frac{16 \lambda  M_0 e^{-3 \phi(r,t)^2}}{r^3} \Biggl[\phi^{(1,0)}(r,t) \Biggl(\left(r-3 M_0\right) \phi(r,t) + \\
	+ r \left(r-2 M_0\right) \left(6 \phi(r,t)^2-1\right) \phi^{(1,0)}(r,t)\Biggr)-r \left(r-2 M_0\right) \phi(r,t) \phi^{(2,0)}(r,t)\Biggr]+ \\
	+ \frac{r^3 \phi^{(0,1)}(r,t)^2}{4 r-8 M_0}+\frac{1}{4} r \left(r-2 M_0\right) \phi ^{(1,0)}(r,t)^2,
\end{multline}
\begin{multline}
	A_1^{(1,0)}(r,t) = \frac{1}{2 r^8 \left(r-2 M_0\right)} \Biggl[4 M_0 r^7 A_1(r,t)+r^8 \left(r-2 M_0\right)^2 \phi^{(1,0)}(r,t)^2 + \\
	+ 4 r^7 M_1(r,t)-64 \lambda  M_0 e^{-6 \phi(r,t)^2} \Biggl(96 \lambda  M_0^2 \left(2 M_0-r\right) \phi(r,t)^2 + \\
	- r^4 e^{3 \phi (r,t)^2} \biggl(\left(r-2 M_0\right) \phi(r,t) \Bigl(\left(3 r-5 M_0\right) \phi^{(1,0)}(r,t)+ \\
	+ r \left(r-2 M_0\right) \phi^{(2,0)}(r,t)\Bigr)+r^3 \left(1-6 \phi (r,t)^2\right) \phi^{(0,1)}(r,t)^2\biggr)\Biggr)+ \\
	+ r^{10} \phi^{(0,1)}(r,t)^2\Biggr].
\end{multline}

To summarize, we obtained two equations for the time-evolution of the scalar field $\phi$ and the correction to the mass $M_1$, and two constraints for $M_1$ and $A_1$.

\subsection{Equations in Tortoise Coordinates}

Before performing the transformation to tortoise coordinates, we compute the Kretschmann scalar $K = R^{\mu\nu\rho\sigma}R_{\mu\nu\rho\sigma}$ obtaining
\begin{multline}
	K = \frac{48 M_0^2}{r^6}+\frac{8 M_0 \, \epsilon \, e^{-6 \phi(r,t)^2}}{r^{12} \left(r-2 M_0\right)} \Biggl[r^6 e^{6 \phi(r,t)^2} \Biggl(r^2 \left(r-2 M_0\right)^2 \phi^{(1,0)}(r,t)^2+ \\
	+ 12 \left(r-2 M_0\right) M_1(r,t)-r^4 \phi^{(0,1)}(r,t)^2\Biggr)+9216 \lambda^2 M_0^3 \left(r-2 M_0\right) \phi(r,t)^2 + \\
	+ 96 \lambda  M_0 r^4 e^{3 \phi(r,t)^2} \Biggl(\left(r-2 M_0\right)^2 \phi^{(1,0)}(r,t) \biggl(r \left(6 \phi(r,t)^2-1\right) \phi^{(1,0)}(r,t) + \\
	+ 4 \phi(r,t)\biggr) +r^3 \left(1-6 \phi(r,t)^2\right) \phi^{(0,1)}(r,t)^2\Biggr)\Biggr]+\OO\left(\epsilon ^2\right).
\end{multline}
As in the previous cases $K$ is regular provided that the scalar field satisfies the condition
\begin{equation}
	\lim_{r \to 2M_0} \partial_t \phi(r, t) = \lim_{r \to 2M_0} \biggl( 1 - \frac{2M_0}{r} \biggr) \partial_r \phi(r, t).
\end{equation}\\

After the transformation in tortoise coordinates we obtain the equations
\begin{multline}
	\phi^{(0,2)}(z,t) = \frac{2 \phi ^{(1,0)}(z,t) \left(r(z,t)-2 M_0\right)}{r(z,t)^2}+ \\
	+ \frac{96 \lambda  M_0^2 e^{-3 \phi (z,t)^2} \phi (z,t) \left(r(z,t)-2 M_0\right)}{r(z,t)^7}+\phi ^{(2,0)}(z,t),
	\label{eq:Order1FieldExpTortoise}
\end{multline}
\begin{multline}
	M_1^{(0,1)}(z,t) = \frac{16 \lambda  M_0 e^{-3 \phi (z,t)^2}}{r(z,t)^3} \Biggl[r(z,t)^2 \phi (z,t) \phi ^{(1,1)}(z,t)+ \\
	+ \phi ^{(0,1)}(z,t) \left(r(z,t)^2 \left(1-6 \phi (z,t)^2\right) \phi ^{(1,0)}(z,t)-M_0 \phi (z,t)\right)\Biggr]+ \\
	+ \frac{1}{2} r(z,t)^2 \phi ^{(0,1)}(z,t) \phi ^{(1,0)}(z,t),
	\label{eq:Order1M1ExpTortoise}
\end{multline}
and the constraints
\begin{multline}
	M_1^{(1,0)}(z,t) = \frac{16 \lambda  M_0 e^{-3 \phi (z,t)^2}}{r(z,t)^3} \Biggl[\phi ^{(1,0)}(z,t) \biggl(\phi (z,t) \left(M_0-r(z,t)\right) + \\
	+ r(z,t)^2 \left(1-6 \phi (z,t)^2\right) \phi ^{(1,0)}(z,t)\biggr)+r(z,t)^2 \phi (z,t) \phi ^{(2,0)}(z,t)\Biggr] + \\
	+ \frac{1}{4} r(z,t)^2 \left(\phi ^{(0,1)}(z,t)^2+\phi ^{(1,0)}(z,t)^2\right),
	\label{constraint:Order1M1ExpTortoise}
\end{multline}
\begin{multline}
	A_1^{(1,0)}(z,t) = \frac{1}{2 r(z,t)^9}\Biggl[r(z,t)^7 \Biggl(4 M_0 A_1(z,t)+4 M_1(z,t)+ \\
	+ r(z,t)^3 \left(\phi ^{(0,1)}(z,t)^2+\phi ^{(1,0)}(z,t)^2\right)\Biggr) + \\
	+ 6144 \lambda ^2 M_0^3 e^{-6 \phi (z,t)^2} \phi (z,t)^2 \left(r(z,t)-2 M_0\right) + \\
	- 64 \lambda  M_0 r(z,t)^5 e^{-3 \phi (z,t)^2} \Biggl(7 M_0 \phi (z,t) \phi ^{(1,0)}(z,t)+ \\
	+ r(z,t)^2 \biggl(\left(6 \phi (z,t)^2-1\right) \phi ^{(0,1)}(z,t)^2-\phi (z,t) \phi ^{(2,0)}(z,t)\biggr)+ \\
	- 3 r(z,t) \phi (z,t) \phi ^{(1,0)}(z,t)\Biggr) \Biggr].
	\label{constraint:Order1A1ExpTortoise}
\end{multline}

\section{Stability of Schwarzchild Solution}

The trivial configuration of the scalar field that corresponds to the Schwarzschild solution is $\phi_0 = 0$, since it satisfies the equation
\begin{equation}
	0 = \frac{\delta F[\phi]}{\delta \phi}\Biggr|_{\phi = \phi_0} = 2\lambda \phi_0 e^{-3\phi^2}.
\end{equation}

In order to analyze the stability of the Schwarzschild solution we performed a numerical integration, imposing the boundary conditions
\begin{equation}
	\begin{cases}
		\partial_z \phi(z_0, t) = \partial_t \phi(z_0, t)\\
		M_1(z_0, t) = 0
	\end{cases},
	\label{eq:BoundaryExp}
\end{equation}
and taking a small perturbation around $\phi_0$ as initial profile of the scalar field
\begin{gather}
	\phi(z, t) = 0, \notag \\
	\partial_t \phi(z, t) = 
	\begin{cases}
		N e^{-\frac{(z-\mu)^2}{\sigma^2}} & \quad \text{if} \quad \mu - 5\sigma < z < \mu + 5\sigma\\
		0 				& \qquad \text{otherwise}
	\end{cases},
	\label{eq:InitialProfileExp}
\end{gather}
where $N = 0.01$, $\mu = 5 \, M_0$ and $\sigma = 4 \, M_0$. \\

For the numerical integration we used the same procedure as in the case of quartic coupling, using the Simpson rule on the constraint \eqref{constraint:Order1M1ExpTortoise} to obtain the initial profile of $M_1$, and integrating the equations \eqref{eq:Order1FieldExpTortoise} and \eqref{eq:Order1M1ExpTortoise} using the method of lines with the Runge-Kutta method for time evolution. The spatial grid was made of $10000$ points in the range $-34.84 \le \frac{z}{M_0} \le 330.11$, hence the grid step was $\frac{\Delta z}{M_0} = 0.036$. The timestep for the evolution was $\frac{\Delta t}{M_0} = 0.001$. \\

\begin{figure}
	\centering
	\subfloat[][$\phi$ \label{fig:comparison_field_exponential}]
		{\includegraphics[width=0.85\textwidth]{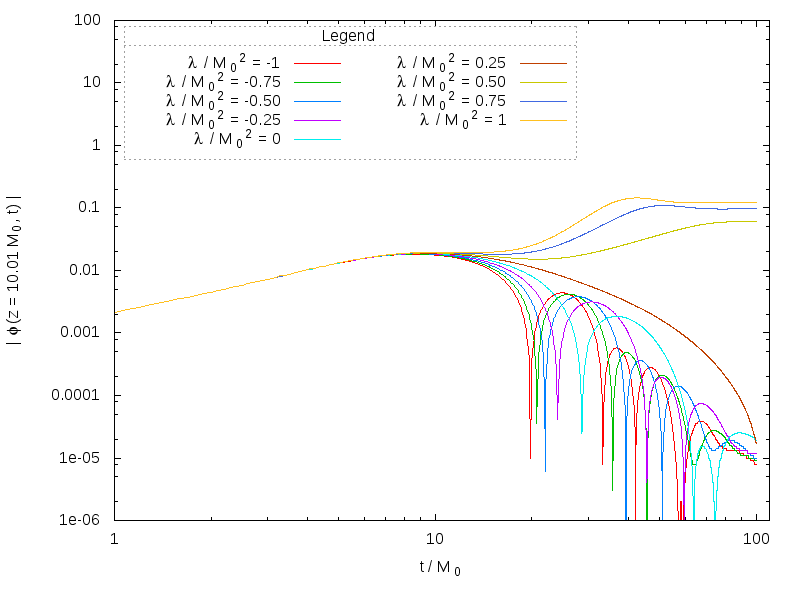}} \\
	\subfloat[][$M_1$ \label{fig:comparison_mass_exponential}]
		{\includegraphics[width=0.85\textwidth]{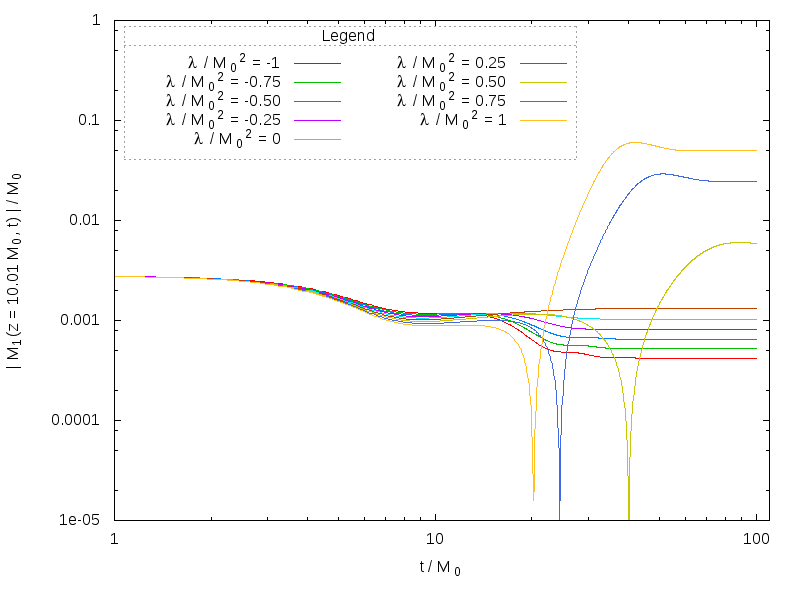}}
	\caption{Evolution of the scalar field and the mass correction at $\frac{z}{M_0} = 10.01$ for different values of $\lambda$. When $\frac{\lambda}{M_0^2} > \frac{\tilde \lambda}{M_0^2} = 0.363$ the Schwarzschild solution is unstable and a nontrivial stable configuration appears.}
	\label{fig:comparison_exponential}
\end{figure}

In fig. \ref{fig:comparison_exponential} we show the evolution of the profile of the scalar field and the mass correction at $\frac{z}{M_0} = 10.01$ for different values of $\lambda$. As we can see the Schwarzschild solution is stable below a threshold value of the coupling constant, that we estimated as $\frac{\tilde \lambda}{M_0^2} = 0.363$. This value is in agreement with the results of Bl\'azquez-Salcedo (ref. \cite{BlazquezSalcedo+}). \\

\section{Stability of Scalarized Solutions}

As we can see from fig. \ref{fig:comparison_exponential} when the Schwarzschild solution is unstable, a stable nontrivial solution appears. These scalarized solutions are nodeless and we computed their scalar charge fitting the profiles of the scalar field with the function
\begin{equation}
	f(r) = a + \frac{b}{r} + \frac{c}{r^2}
	\label{eq:fitExponential}
\end{equation}
in a region sufficiently far from the horizon and from the end of the pulse. The scalar charge was estimated as $Q = b$. \\
We performed numerical integrations until $\frac{t}{M_0} = 575$ on a grid with 15000 points that extends in the range $-34.84 \le \frac{z}{M_0} \le 631.42$, in such a way that we could consider an appropriate region in which execute the fit.\\

\begin{figure}
	\centering
	\includegraphics[width=0.6\textwidth]{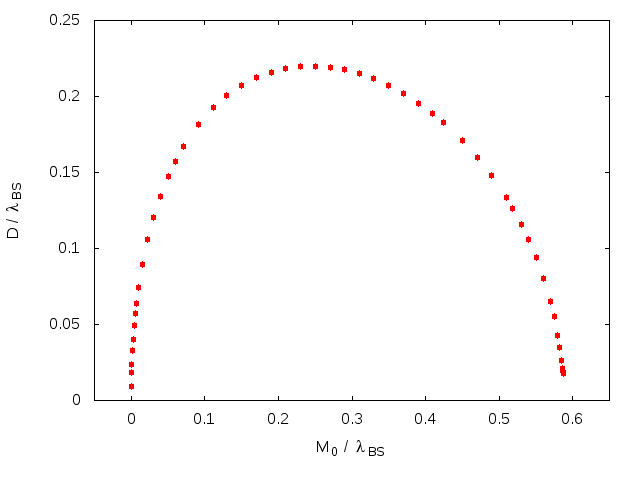}
	\caption{Scalar charge obtained from the fit of the profile of the scalar field at the end of the numerical integration. $\lambda_{BS} = \sqrt{8\lambda}$ and $D = \frac{Q}{\sqrt{2}}$ are respectively the coupling constant and the scalar charge used by Bl\'azquez-Salcedo \ea in ref. \cite{BlazquezSalcedo+}.}
	\label{fig:exponential_charge}
\end{figure}

The behavior of the scalar charge obtained with this procedure is shown in fig. \ref{fig:exponential_charge} in a plot of $\frac{D}{\lambda_{BS}}$ vs $\frac{M_0}{\lambda_{BS}}$, where $D = \frac{Q}{\sqrt{2}}$ and $\lambda_{BS} = \sqrt{8\lambda}$ are respectively the scalar charge and the coupling constant used in ref. \cite{BlazquezSalcedo+}, in such a way that we can compare the results. \\
As expected, $\frac{D}{\lambda_{BS}}$ tends to zero as $\frac{M_0}{\lambda_{BS}}$ goes to the critical value $\frac{M_0}{\sqrt{8 \tilde \lambda}} = 0.587$. The behavior of the scalar charge represented in fig. \ref{fig:exponential_charge} is not in agreement with the one found by Bl\'azquez-Salcedo \ea (ref. \cite{BlazquezSalcedo+}), since we found stable nodeless scalarized solution for every value of $\frac{M_0}{\lambda_{BS}}$ such that the Schwarzschild solution is unstable. \\
Similarly to the case of quartic coupling, the reason for this discrepancy could be related to the fact that the perturbative approach we used is not fully mathematically consistent, and it could be interesting to clarify this point in a further study.

\chapter{Conclusions}
\label{chapter:conclusions}

In this work we performed a time-domain analysis of Schwarzschild black holes in Einstein-scalar-Gauss-Bonnet gravity using a perturbative approach.\\

We started in chapter \ref{chapter:review} by reviewing some possible modifications of General Relativity and some results obtained in the context of EsGB gravity. In the first part we followed the presentation in ref. \cite{Berti+} and we outlined some possible reasons to introduce modifications in GR, that come mainly from theoretical and cosmological arguments. Among the possible extensions of General Relativity, which can be classified on the basis of the violations of the hypotheses of the Lovelock's theorem, we focused on scalar-tensor theories, in which a dynamical scalar field nonminimally coupled to the metric is considered.\\
In this context we considered two classes of theories: Horndeski theories and quadratic gravity. The first one contains theories whose field equations are of order two, and are free from the appearance of ghosts related to the Ostrogradsky instability. The second one contains theories that include in the action quadratic term in the curvature, and that can be considered as low-energy approximations of more general theories.\\
Then we introduced EsGB gravity, which is an Horndeski theory that contains quadratic terms in curvature, and therefore it combines the properties of these two classes of scalar-tensor theories. In this theories a no-hair theorem has been proven by Silva \ea in ref. \cite{Silva+Stability}, and the hypotheses are stationarity and asymptotic flatness of the solution, the stationarity of the scalar field, the existence of a constant $\phi_0$ such that $\frac{\delta F[\phi]}{\delta \phi}\Bigr|_{\phi_0} = 0$ and the requirement that $\frac{\delta^2 F[\phi]}{\delta \phi^2} \G < 0$. When the fourth condition is violated the GR black hole solution becomes unstable and a nontrivial stable configuration of the scalar field can appear, requiring the introduction of a scalar charge. This phenomenon is called \textit{spontaneous scalarization}.\\
We reviewed some results obtained in refs. \cite{EGBNoHair, DonevaYazadjiev, BlazquezSalcedo+, Silva+Stability, MasatoTaishi} performing a frequency-domain analysis with different choices of the coupling function: exponential $F[\phi] = \frac{\lambda}{3} \Bigl( 1 - e^{- 3 \phi^2} \Bigr)$, quadratic $F[\phi] = \lambda \phi^2$, and quartic $F[\phi] = \lambda \phi^2 + \gamma \phi^4$. In all the cases the Schwarzschild solution is stable for $\frac{\lambda}{M_0^2} < 0.363$, and the nontrivial configurations of the scalar field with $n \ge 1$ nodes are unstable. The stability properties of the scalarized solutions with $n = 0$ nodes are different between the three coupling functions. In the case of exponential coupling function the nodeless solution has been found to be stable only for $\frac{\lambda}{M^2}$ smaller than a critical value, with a quartic coupling it is stable for negative values of $\alpha = \frac{\gamma}{\lambda}$ that are below a critical value, while in the case of quadratic coupling there are not stable scalarized solutions.\\

In chapter \ref{chapter:QuadraticCoupling} we analyzed the stability of Schwarzschild black hole in EsGB gravity with quadratic coupling function using a perturbative approach. To perform this analysis we started by writing the scalar field and the metric in 1+1 variables, and we wrote $\phi$ as a small perturbation around the trivial solution of the equation for the scalar field, which is $\phi_0 = 0$, obtaining $\phi = \epsilon \varphi_1$. We then expanded the field equations obtaining at the order $0$ the Schwarzschild solution, at the first order a nontrivial evolution of the scalar field on the Schwarzschild background, and at the second order the backreaction of the scalar field on the metric.\\
From the field equations at the second order in $\epsilon$ we obtained a set of two equations and two constraints that we wrote in tortoise coordinates, and we integrated numerically with the method of lines using the Runge-Kutta method for time evolution.\\
Analyzing the solutions obtained for different values of $\frac{\lambda}{M_0^2}$ we observed that the Schwarzschild solution is unstable for $\frac{\lambda}{M_0^2}$ higher than a threshold value that we estimated to be in the interval $0.3267 < \frac{\tilde \lambda}{M_0^2} < 0.3268$, which is consistent with the results obtained in refs. \cite{BlazquezSalcedo+, Silva+Stability}. In the unstable case we estimated the time constant of the exponential growth for different values of $\frac{\lambda}{M_0^2}$ with a fit, and the results that we obtained are consistent with the frequencies of the Schwarzschild modes computed in ref. \cite{BlazquezSalcedo+}.\\
Then we computed the apparent horizon and the Null Energy Condition (NEC) to characterize its behavior. We observed that in the NEC there are $\OO(\epsilon^4)$ terms, that near the apparent horizon scale as $\OO(\epsilon^2)$ terms. This can be due to the fact that Schwarzschild coordinates are not regular and a different choice of coordinates may solve this critical issue.    \\ 

In chapter \ref{chapter:QuarticCoupling} we considered a quartic coupling function $F[\phi] = \lambda \phi^2 + \gamma \phi^4$ and we analyzed the stability of the Schwarzschild solution using a different perturbative approach and expanding in the Newton's constant. In this perturbative scheme the contributions that come from the quartic term are of the same order as the contributions that come from the quadratic term. The nonlinearities introduced in the equation for $\phi$ can quench the instability of the Schwarzschild solution and a stable nontrivial configuration of the scalar field can appear.\\
We started by writing the scalar field and the metric in 1+1 variables, we expressed the Newton's constant as $G = \epsilon + \OO(\epsilon^2)$ and we expanded in the field equations up to the first order in $\epsilon$. At the order $0$ we imposed that the term $GM = M_0$ is constant and different from zero, in such a way that the equations are solved by the Schwarzschild metric, instead of the flat one; however this assumption is not fully mathematically consistent since we are actually assuming that $M\to \infty$ as $G \to 0$. At the first order we obtained a nontrivial evolution of $\phi$ and the backreaction of the scalar field on the metric. \\
We integrated numerically the equations in tortoise coordinates using a small perturbation around $\phi = 0$ as initial condition for the scalar field, and we observed that the Schwarzschild solution is unstable for $\frac{\lambda}{M_0^2} > 0.363$, independently of the value of $\frac{\gamma}{M_0^2}$.\\
In the unstable case we observed that when $\alpha = \frac{\gamma}{\lambda} < 0$ the scalar field approaches a stable nodeless nontrivial configuration. We computed the scalar charges for different values of $\frac{\lambda}{M_0^2}$ and $\alpha$, obtaining a behavior that is consistent with the results contained in refs. \cite{Silva+Stability, MasatoTaishi} in the cases with $\alpha < -0.8$. Nevertheless when $\alpha = -0.2$ we obtained stable nodeless scalarized solutions different from those found by Silva \ea in ref. \cite{Silva+Stability}, and this can be due to the use of a perturbative approach which is not fully mathematically consistent.\\

In chapter \ref{chapter:ExponentialCoupling} we considered the exponential coupling $F[\phi] = \frac{\lambda}{3} \Bigl(1 - e^{-3\phi^2} \Bigr)$ using the same perturbative approach as in the case of quartic coupling.\\
We integrated the field equations at the first order using a small perturbation around $\phi_0 = 0$ as initial condition, and we obtained that the Schwarzschild solution is unstable for $\frac{\lambda}{M_0^2} > 0.363$, and in this case the scalar field reaches a nodeless nontrivial configuration. We computed the scalar charge of these scalarized solutions and we obtained a behavior different from the results of Bl\'azquez-Salcedo \ea shown in ref. \cite{BlazquezSalcedo+}. As in the previous case a possible reason for this discrepancy could be the inconsistency in the perturbative approach. \\

Concluding, with the time-domain analysis of EsGB gravity in a perturbative approach we found that the Schwarzschild solution is unstable when $\frac{\lambda}{M_0^2} > 0.363$ for all the three coupling functions analyzed. However there are some critical issues that have arisen during the analysis. Firstly, in the presence of the apparent horizon, the perturbative analysis in not applicable for the computation of $\tensor{T}{^t_t}$ and the Null Energy Condition using Schwarzschild coordinates; this issue may be overcome using different coordinates, which are regular at the horizon. Secondly, in the cases of quartic and exponential coupling, we found different stable nodeless scalarized solutions from what found in refs. \cite{BlazquezSalcedo+, Silva+Stability, MasatoTaishi}; it would be interesting to analyze in the future whether these solutions are an artifact of the perturbative approach or not. \\ 
We hope to come back on these issues in future works.

\appendix

\chapter{Field Equations of Einstein-scalar-Gauss-Bonnet Gravity}
\label{chapter:derivation}

In this section we will compute the field equations for EsGB gravity, using the stationary action principle. For this derivation we will follow the one contained in appendix A of ref. \cite{RipleyPretorius}. \\

The action we are considering is written in the form of eq. \eqref{eq:EsGBTheorem}:

\begin{equation}
S = \frac{1}{2} \int_\Omega d^4x\sqrt{-g}\Bigl(R - (\nabla \phi)^2 + 2 F[\phi] \mathcal{G} \Bigr).
\label{eq:action}
\end{equation}
We recall that the Gauss-Bonnet invariant is given by:
\begin{equation}
	\G = \frac{1}{4} \delta^{\alpha\beta\mu\nu}_{\rho\sigma\lambda\omega} \tensor{R}{^{\rho\sigma}_{\alpha\beta}} \tensor{R}{^{\lambda\omega}_{\mu\nu}} = R^2 - 4 R_{\mu\nu}R^{\mu\nu} + R_{\mu\nu\alpha\beta}R^{\mu\nu\alpha\beta},
	\label{eq:GaussBonnetInvariant}
\end{equation}
where $\delta^{\alpha\beta\mu\nu}_{\rho\sigma\lambda\omega} = \epsilon^{\alpha\beta\mu\nu} \epsilon_{\rho\sigma\lambda\omega}$ is the generalized Kronecker delta. \\

We start with the derivation of the equation for the scalar field and then we move to the equation for the metric.

\section{Equation for the Scalar Field}

In this case the field equation are given by the Euler-Lagrange equation in curved spacetime \cite{Ferrari-Gualtieri-Pani}:
\begin{equation}
	\nabla_\mu \frac{\delta \mathcal L}{\delta (\nabla_\mu \phi)} - \frac{\delta \mathcal L}{\delta \phi} = 0.
\end{equation}
Computing the derivatives we obtain
\begin{gather}
	\frac{\delta \mathcal L}{\delta (\nabla_\mu \phi)} = -\sqrt{-g} \nabla^\mu \phi \qquad \implies \qquad \nabla_\mu \frac{\delta \mathcal L}{\delta (\nabla_\mu \phi)} = -\sqrt{-g} \nabla_\mu\nabla^\mu \phi \\
	\frac{\delta \mathcal L}{\delta \phi} = \sqrt{-g} \frac{\delta F[\phi]}{\delta \phi} \G,
\end{gather}
hence the equation for the scalar field is:
\begin{equation}
	E^{(\phi)} := \nabla_\mu \nabla^\mu \phi + \frac{\delta F[\phi]}{\delta \phi} \mathcal{G} = 0.
\end{equation}

\section{Equation for the Metric}

Let us first consider Gauss-Bonnet term
\begin{equation}
	S_{GB} = \int_\Omega d^4x \sqrt{-g} F[\phi] \G = \frac{1}{4} \int_\Omega d^4x \sqrt{-g} F[\phi] \delta^{\alpha\beta\mu\nu}_{\rho\sigma\lambda\omega} \tensor{R}{^{\rho\sigma}_{\alpha\beta}} \tensor{R}{^{\lambda\omega}_{\mu\nu}}.
	\label{eq:actiongb}
\end{equation}
Varying the integrand we obtain
\begin{multline}
	\delta \biggl( \frac{1}{4} \sqrt{-g} F[\phi] \delta^{\alpha\beta\mu\nu}_{\rho\sigma\lambda\omega} \tensor{R}{^{\rho\sigma}_{\alpha\beta}} \tensor{R}{^{\lambda\omega}_{\mu\nu}} \biggr) = \\
	= \frac{1}{4} F[\phi] \delta^{\alpha\beta\mu\nu}_{\rho\sigma\lambda\omega} \Bigl( (\delta \sqrt{-g}) \tensor{R}{^{\rho\sigma}_{\alpha\beta}} \tensor{R}{^{\lambda\omega}_{\mu\nu}} + 2 \sqrt{-g} \tensor{R}{^{\rho\sigma}_{\alpha\beta}} \delta \tensor{R}{^{\lambda\omega}_{\mu\nu}} \Bigr) = \\
	= \frac{1}{4} F[\phi] \sqrt{-g} \delta^{\alpha\beta\gamma\delta}_{\rho\sigma\lambda\omega} \biggl( -\frac{1}{2} \tensor{R}{^{\rho\sigma}_{\alpha\beta}} \tensor{R}{^{\lambda\omega}_{\gamma\delta}}g_{\mu\nu} \delta g^{\mu\nu} + 2 \tensor{R}{^{\rho\sigma}_{\alpha\beta}} \delta \tensor{R}{^{\lambda\omega}_{\gamma\delta}} \biggr),
\end{multline}
where we used the relation \cite{Ferrari-Gualtieri-Pani}
\begin{equation}
	\delta \sqrt{-g} = - \frac{\sqrt{-g}}{2} g_{\mu\nu} \delta g^{\mu\nu}.
\end{equation}
Now
\begin{multline}
	\delta^{\rho\kappa\alpha\beta}_{\lambda\sigma\gamma\delta} \tensor{R}{^{\lambda\sigma}_{\rho\kappa}} \delta \tensor{R}{^{\gamma\delta}_{\alpha\beta}} = \delta^{\rho\kappa\alpha\beta}_{\lambda\sigma\gamma\delta} \tensor{R}{^{\lambda\sigma}_{\rho\kappa}} \delta \Bigl( g^{\omega\delta} \tensor{R}{^{\gamma}_{\omega\alpha\beta}} \Bigr) = \\
	= \delta^{\rho\kappa\alpha\beta}_{\lambda\sigma\gamma\delta} \biggl( \tensor{R}{^{\lambda\sigma}_{\rho\kappa}} \tensor{R}{^{\gamma}_{\omega\alpha\beta}} \delta g^{\omega\delta} + \tensor{R}{^{\lambda\sigma}_{\rho\kappa}} g^{\omega\delta} \delta \tensor{R}{^{\gamma}_{\omega\alpha\beta}} \biggr) = \\
	= \delta^{\rho\kappa\alpha\beta}_{\lambda\sigma\gamma\delta} \biggl( \tensor{R}{^{\lambda\sigma}_{\rho\kappa}} \tensor{R}{^{\gamma\epsilon}_{\alpha\beta}} g_{\epsilon\omega} \delta g^{\omega\delta} + \tensor{R}{^{\lambda\sigma}_{\rho\kappa}} g^{\omega\delta} \delta \tensor{R}{^{\gamma}_{\omega\alpha\beta}} \biggr).
\end{multline}

The tensor $\delta^{\rho\kappa\alpha\beta}_{\lambda\sigma\gamma\delta}g_{\epsilon\omega}$ is antisymmetric with respect to $\lambda\sigma\gamma\delta$, and from it we can construct a tensor which is antisymmetric with respect to $\lambda\sigma\gamma\delta\epsilon$ in the following way\footnote{For this technique, see also the appendices of ref. \cite{contodelta}.} 
\begin{equation}
	\delta^{\rho\kappa\alpha\beta}_{\lambda\sigma\gamma\delta}g_{\epsilon\omega} - \delta^{\rho\kappa\alpha\beta}_{\epsilon\sigma\gamma\delta}g_{\lambda\omega} - \delta^{\rho\kappa\alpha\beta}_{\lambda\epsilon\gamma\delta}g_{\sigma\omega} - \delta^{\rho\kappa\alpha\beta}_{\lambda\sigma\epsilon\delta}g_{\gamma\omega} - \delta^{\rho\kappa\alpha\beta}_{\lambda\sigma\gamma\epsilon}g_{\delta\omega}.
\end{equation}
Since this is an antisymmetric tensor with respect to 5 indices in a 4-dimensional space, it is equal to zero and so we have the identity
\begin{equation}
	\delta^{\rho\kappa\alpha\beta}_{\lambda\sigma\gamma\delta}g_{\epsilon\omega} = \delta^{\rho\kappa\alpha\beta}_{\epsilon\sigma\gamma\delta}g_{\lambda\omega} + \delta^{\rho\kappa\alpha\beta}_{\lambda\epsilon\gamma\delta}g_{\sigma\omega} + \delta^{\rho\kappa\alpha\beta}_{\lambda\sigma\epsilon\delta}g_{\gamma\omega} + \delta^{\rho\kappa\alpha\beta}_{\lambda\sigma\gamma\epsilon}g_{\delta\omega}.
\end{equation}
Now
\begin{multline}
	\delta^{\rho\kappa\alpha\beta}_{\epsilon\sigma\gamma\delta} g_{\lambda\omega} \tensor{R}{^{\lambda\sigma}_{\rho\kappa}} \tensor{R}{^{\gamma\epsilon}_{\alpha\beta}} \delta g^{\omega\delta} = \delta^{\rho\kappa\alpha\beta}_{\gamma\epsilon\sigma\delta} g_{\lambda\omega} \tensor{R}{^{\lambda\sigma}_{\rho\kappa}} \tensor{R}{^{\gamma\epsilon}_{\alpha\beta}} \delta g^{\omega\delta} = \\
	= \delta^{\alpha\beta\rho\kappa}_{\gamma\epsilon\sigma\delta} g_{\lambda\omega} \tensor{R}{^{\lambda\sigma}_{\rho\kappa}} \tensor{R}{^{\gamma\epsilon}_{\alpha\beta}} \delta g^{\omega\delta} = - \delta^{\alpha\beta\rho\kappa}_{\gamma\epsilon\sigma\delta} g_{\lambda\omega} \tensor{R}{^{\sigma\lambda}_{\rho\kappa}} \tensor{R}{^{\gamma\epsilon}_{\alpha\beta}} \delta g^{\omega\delta} = \\
	= - \delta^{\rho\kappa\alpha\beta}_{\lambda\sigma\gamma\delta} g_{\epsilon\omega} \tensor{R}{^{\gamma\epsilon}_{\alpha\beta}} \tensor{R}{^{\lambda\sigma}_{\rho\kappa}} \delta g^{\omega\delta}.
\end{multline}
Analogously we obtain
\begin{gather}
	\delta^{\rho\kappa\alpha\beta}_{\lambda\epsilon\gamma\delta}g_{\sigma\omega} \tensor{R}{^{\lambda\sigma}_{\rho\kappa}} \tensor{R}{^{\gamma\epsilon}_{\alpha\beta}} \delta g^{\omega\delta} = - \delta^{\rho\kappa\alpha\beta}_{\lambda\sigma\gamma\delta} g_{\epsilon\omega} \tensor{R}{^{\gamma\epsilon}_{\alpha\beta}} \tensor{R}{^{\lambda\sigma}_{\rho\kappa}} \delta g^{\omega\delta} \\
	\delta^{\rho\kappa\alpha\beta}_{\lambda\sigma\epsilon\delta}g_{\gamma\omega} \tensor{R}{^{\lambda\sigma}_{\rho\kappa}} \tensor{R}{^{\gamma\epsilon}_{\alpha\beta}} \delta g^{\omega\delta} = - \delta^{\rho\kappa\alpha\beta}_{\lambda\sigma\gamma\delta} g_{\epsilon\omega} \tensor{R}{^{\gamma\epsilon}_{\alpha\beta}} \tensor{R}{^{\lambda\sigma}_{\rho\kappa}} \delta g^{\omega\delta}.
\end{gather}
Hence
\begin{multline}
	\delta^{\rho\kappa\alpha\beta}_{\lambda\sigma\gamma\delta} \tensor{R}{^{\lambda\sigma}_{\rho\kappa}} \tensor{R}{^{\gamma\epsilon}_{\alpha\beta}} g_{\epsilon\omega} \delta g^{\omega\delta} = \\
	= \Bigl( \delta^{\rho\kappa\alpha\beta}_{\epsilon\sigma\gamma\delta}g_{\lambda\omega} + \delta^{\rho\kappa\alpha\beta}_{\lambda\epsilon\gamma\delta}g_{\sigma\omega} + \delta^{\rho\kappa\alpha\beta}_{\lambda\sigma\epsilon\delta}g_{\gamma\omega} + \delta^{\rho\kappa\alpha\beta}_{\lambda\sigma\gamma\epsilon}g_{\delta\omega} \Bigr) \tensor{R}{^{\lambda\sigma}_{\rho\kappa}} \tensor{R}{^{\gamma\epsilon}_{\alpha\beta}} \delta g^{\omega\delta} = \\
	= \Bigl( - \delta^{\rho\kappa\alpha\beta}_{\lambda\sigma\gamma\delta} g_{\epsilon\omega} - \delta^{\rho\kappa\alpha\beta}_{\lambda\sigma\gamma\delta} g_{\epsilon\omega} - \delta^{\rho\kappa\alpha\beta}_{\lambda\sigma\gamma\delta} g_{\epsilon\omega} + \delta^{\rho\kappa\alpha\beta}_{\lambda\sigma\gamma\epsilon}g_{\delta\omega} \Bigr) \tensor{R}{^{\lambda\sigma}_{\rho\kappa}} \tensor{R}{^{\gamma\epsilon}_{\alpha\beta}} \delta g^{\omega\delta},
\end{multline}
which gives (see ref. \cite{contodelta}) 
\begin{multline}
	\delta^{\rho\kappa\alpha\beta}_{\lambda\sigma\gamma\delta} \tensor{R}{^{\lambda\sigma}_{\rho\kappa}} \tensor{R}{^{\gamma\epsilon}_{\alpha\beta}} g_{\epsilon\omega} \delta g^{\omega\delta} = \frac{1}{4} \delta^{\rho\kappa\alpha\beta}_{\lambda\sigma\gamma\epsilon}g_{\delta\omega} \tensor{R}{^{\lambda\sigma}_{\rho\kappa}} \tensor{R}{^{\gamma\epsilon}_{\alpha\beta}} \delta g^{\omega\delta} = \\
	= \frac{1}{4} \delta^{\rho\kappa\alpha\beta}_{\lambda\sigma\gamma\delta}g_{\mu\nu} \tensor{R}{^{\lambda\sigma}_{\rho\kappa}} \tensor{R}{^{\gamma\delta}_{\alpha\beta}} \delta g^{\mu\nu}.
\end{multline}

Substituting this result in the variation of the Gauss-Bonnet term we obtain
\begin{multline}
	\delta S_{GB} = \int_\Omega d^4x \frac{1}{4} F[\phi] \sqrt{-g} \biggl( 2 \delta^{\rho\kappa\alpha\beta}_{\lambda\sigma\gamma\delta} \tensor{R}{^{\lambda\sigma}_{\rho\kappa}} \delta \tensor{R}{^{\gamma\delta}_{\alpha\beta}} - \frac{1}{2} \delta^{\rho\kappa\alpha\beta}_{\lambda\sigma\gamma\delta} \tensor{R}{^{\lambda\sigma}_{\rho\kappa}} \tensor{R}{^{\gamma\delta}_{\alpha\beta}} g_{\mu\nu} \delta g^{\mu\nu} \biggr) = \\
	= \int_\Omega d^4x \frac{1}{4} F[\phi] \sqrt{-g} \biggl( 2 \delta^{\rho\kappa\alpha\beta}_{\lambda\sigma\gamma\delta} \tensor{R}{^{\lambda\sigma}_{\rho\kappa}} g^{\omega\delta} \delta \tensor{R}{^{\gamma}_{\omega\alpha\beta}} + 2 \delta^{\rho\kappa\alpha\beta}_{\lambda\sigma\gamma\delta} \tensor{R}{^{\lambda\sigma}_{\rho\kappa}} \tensor{R}{^{\gamma}_{\omega\alpha\beta}} \delta g^{\omega\delta} + \\ - \frac{1}{2} \delta^{\rho\kappa\alpha\beta}_{\lambda\sigma\gamma\delta} \tensor{R}{^{\lambda\sigma}_{\rho\kappa}} \tensor{R}{^{\gamma\delta}_{\alpha\beta}} g_{\mu\nu} \delta g^{\mu\nu} \biggr) = \\
	= \int_\Omega d^4x \frac{1}{4} F[\phi] \sqrt{-g} \biggl( 2 \delta^{\rho\kappa\alpha\beta}_{\lambda\sigma\gamma\delta} \tensor{R}{^{\lambda\sigma}_{\rho\kappa}} g^{\omega\delta} \delta \tensor{R}{^{\gamma}_{\omega\alpha\beta}} + 2 \delta^{\rho\kappa\alpha\beta}_{\lambda\sigma\gamma\delta} \tensor{R}{^{\lambda\sigma}_{\rho\kappa}} g_{\omega\epsilon} \tensor{R}{^{\gamma\epsilon}_{\alpha\beta}} \delta g^{\omega\delta} + \\ - \frac{1}{2} \delta^{\rho\kappa\alpha\beta}_{\lambda\sigma\gamma\delta} \tensor{R}{^{\lambda\sigma}_{\rho\kappa}} \tensor{R}{^{\gamma\delta}_{\alpha\beta}} g_{\mu\nu} \delta g^{\mu\nu} \biggr) = \\
	= \int_\Omega d^4x \frac{1}{4} F[\phi] \sqrt{-g} \biggl( 2 \delta^{\rho\kappa\alpha\beta}_{\lambda\sigma\gamma\delta} \tensor{R}{^{\lambda\sigma}_{\rho\kappa}} g^{\omega\delta} \delta \tensor{R}{^{\gamma}_{\omega\alpha\beta}} + 2\frac{1}{4} \delta^{\rho\kappa\alpha\beta}_{\lambda\sigma\gamma\delta} \tensor{R}{^{\lambda\sigma}_{\rho\kappa}} \tensor{R}{^{\gamma\delta}_{\alpha\beta}} g_{\mu\nu} \delta g^{\mu\nu} + \\ - \frac{1}{2} \delta^{\rho\kappa\alpha\beta}_{\lambda\sigma\gamma\delta} \tensor{R}{^{\lambda\sigma}_{\rho\kappa}} \tensor{R}{^{\gamma\delta}_{\alpha\beta}} g_{\mu\nu} \delta g^{\mu\nu} \biggr) = \\
	= \int_\Omega d^4x \frac{1}{4} F[\phi] \sqrt{-g} \biggl( 2 \delta^{\rho\kappa\alpha\beta}_{\lambda\sigma\gamma\delta} \tensor{R}{^{\lambda\sigma}_{\rho\kappa}} g^{\omega\delta} \delta \tensor{R}{^{\gamma}_{\omega\alpha\beta}} \biggr).
	\label{eq:delta-S_GB}
\end{multline}

Now we need to evaluate $\delta \tensor{R}{^{\gamma}_{\omega\alpha\beta}}$, and to do this we use a procedure analogous to the one used in ref. \cite{Ferrari-Gualtieri-Pani} in the proof of the Palatini identity. \\
By definition the variation of the Riemann tensor is given by
\begin{multline}
	\delta \tensor{R}{^{\gamma}_{\omega\alpha\beta}} = \delta \Bigl( \Gamma^\gamma_{\omega\beta,\alpha} - \Gamma^\gamma_{\omega\alpha,\beta} + \Gamma^\sigma_{\omega\beta} \Gamma^\gamma_{\sigma\alpha} - \Gamma^\gamma_{\sigma\beta}\Gamma^\sigma_{\omega\alpha} \Bigr) = \\
	= \delta\Gamma^\gamma_{\omega\beta,\alpha} - \delta \Gamma^\gamma_{\omega\alpha,\beta} + \bigl( \delta \Gamma^\sigma_{\omega\beta} \bigr) \Gamma^\gamma_{\sigma\alpha} + \Gamma^\sigma_{\omega\beta} \bigl( \delta \Gamma^\gamma_{\sigma\alpha} \bigr) - \bigl( \delta \Gamma^\gamma_{\sigma\beta} \bigr) \Gamma^\sigma_{\omega\alpha} - \Gamma^\gamma_{\sigma\beta} \bigl(\delta \Gamma^\sigma_{\omega\alpha} \bigr).
\end{multline}
The variation of the Christoffel symbol 
\begin{equation}
	\delta \Gamma^\lambda_{\mu\nu} = \frac{1}{2} g^{\lambda\rho} \Bigl( \delta g_{\mu\rho;\nu} + \delta g_{\nu\rho;\mu} - \delta g_{\mu\nu;\rho} \Bigr)
	\label{eq:var_Christoffel}
\end{equation}
is a tensor (see ref. \cite{Ferrari-Gualtieri-Pani}) and therefore
\begin{multline}
	\bigl( \delta \Gamma^\gamma_{\omega\beta} \bigr)_{;\alpha} - \bigl( \delta \Gamma^\gamma_{\omega\alpha} \bigr)_{;\beta} = \bigl( \delta \Gamma^\gamma_{\omega\beta} \bigr)_{,\alpha} + \Gamma^\gamma_{\alpha\kappa} \delta \Gamma^\kappa_{\omega\beta} -\Gamma^\kappa_{\alpha\omega} \delta \Gamma^\gamma_{\kappa\beta} + \\
	- \Gamma^\kappa_{\alpha\beta} \delta \Gamma^\gamma_{\omega\kappa} - \bigl( \delta \Gamma^\gamma_{\omega\alpha} \bigr)_{,\beta} - \Gamma^\gamma_{\kappa\beta} \delta \Gamma^\kappa_{\omega\alpha} + \Gamma^\kappa_{\beta\omega} \delta \Gamma^\gamma_{\kappa\alpha} + \Gamma^\kappa_{\beta\alpha} \delta \Gamma^\gamma_{\omega\kappa} = \\
	= \delta \tensor{R}{^{\gamma}_{\omega\alpha\beta}}.
\end{multline}
Contracting with the generalized Kronecker delta we obtain
\begin{equation}
	\delta^{\rho\kappa\alpha\beta}_{\lambda\sigma\gamma\delta} \delta \tensor{R}{^{\gamma}_{\omega\alpha\beta}} = \delta^{\rho\kappa\alpha\beta}_{\lambda\sigma\gamma\delta} \Bigl( \bigl( \delta \Gamma^\gamma_{\omega\beta} \bigr)_{;\alpha} - \bigl( \delta \Gamma^\gamma_{\omega\alpha} \bigr)_{;\beta} \Bigr) = 2 \delta^{\rho\kappa\alpha\beta}_{\lambda\sigma\gamma\delta} \bigl( \delta \Gamma^\gamma_{\omega\beta} \bigr)_{;\alpha}.
	\label{eq:deltadR}
\end{equation}
Substituting eq. \eqref{eq:var_Christoffel}
\begin{equation}
	\delta^{\rho\kappa\alpha\beta}_{\lambda\sigma\gamma\delta} \delta \tensor{R}{^{\gamma}_{\omega\alpha\beta}} = \delta^{\rho\kappa\alpha\beta}_{\lambda\sigma\gamma\delta} g^{\gamma\epsilon} \bigl( \delta g_{\omega\epsilon;\beta\alpha} + \delta g_{\beta\epsilon;\omega\alpha} - \delta g_{\omega\beta;\epsilon\alpha} \bigr).
	\label{eq:delta_varR}
\end{equation}
The variation of the Gauss-Bonnet term contains also a $g^{\omega\delta}$, so $\delta^{\rho\kappa\alpha\beta}_{\lambda\sigma\gamma\delta} g^{\gamma\epsilon}g^{\omega\delta}$ is antisymmetric with respect to $\omega$ and $\epsilon$, and when contracted with $\delta g_{\omega\epsilon;\beta\alpha}$ gives zero. The remaining terms give
\begin{multline}
	\delta^{\rho\kappa\alpha\beta}_{\lambda\sigma\gamma\delta} g^{\omega\delta} g^{\gamma\epsilon} \bigl( \delta g_{\beta\epsilon;\omega\alpha} - \delta g_{\omega\beta;\epsilon\alpha} \bigr) = 2 \delta^{\rho\kappa\alpha\beta}_{\lambda\sigma\gamma\delta} g^{\omega\delta} g^{\gamma\epsilon} \delta g_{\beta\epsilon;\omega\alpha} = \\
	= 2 \delta^{\rho\kappa\alpha\beta}_{\lambda\sigma\gamma\delta} g^{\omega\delta} g^{\gamma\epsilon} \nabla_\alpha \nabla_\omega \delta g_{\beta\epsilon}.
\end{multline}
Substituting in eq. \eqref{eq:delta-S_GB} we obtain
\begin{equation}
	\delta S_{GB} = \int_\Omega d^4x \frac{1}{2} F[\phi] \sqrt{-g} 2 \delta^{\rho\kappa\alpha\beta}_{\lambda\sigma\gamma\delta} \tensor{R}{^{\lambda\sigma}_{\rho\kappa}} g^{\omega\delta} g^{\gamma\epsilon} \nabla_\alpha \nabla_\omega \delta g_{\beta\epsilon}.
	\label{eq:var_S_GB}
\end{equation}
Integrating by parts and using the fact that $\delta g(\partial \Omega) = 0$,
\begin{multline}
	\delta S_{GB} = - \int_\Omega d^4x F[\phi] \sqrt{-g} \delta^{\rho\kappa\alpha\beta}_{\lambda\sigma\gamma\delta} \Bigl( \nabla_\alpha \tensor{R}{^{\lambda\sigma}_{\rho\kappa}} \Bigr) g^{\omega\delta} g^{\gamma\epsilon} \nabla_\omega \delta g_{\beta\epsilon} + \\
	- \int_\Omega d^4x \Bigl( \nabla_\alpha F[\phi] \Bigr) \sqrt{-g} \delta^{\rho\kappa\alpha\beta}_{\lambda\sigma\gamma\delta} \tensor{R}{^{\lambda\sigma}_{\rho\kappa}} g^{\omega\delta} g^{\gamma\epsilon} \nabla_\omega \delta g_{\beta\epsilon}.
	\label{eq:var_parts}
\end{multline}
For the second Bianchi identity \cite{Ferrari-Gualtieri-Pani}
\begin{equation}
	R_{\alpha\beta\mu\nu;\lambda} + R_{\alpha\beta\nu\lambda;\mu} + R_{\alpha\beta\lambda\mu;\nu} = 0.
	\label{eq:Bianchi_id}
\end{equation}
In the term $\delta^{\rho\kappa\alpha\beta}_{\lambda\sigma\gamma\delta} \nabla_\alpha \tensor{R}{^{\lambda\sigma}_{\rho\kappa}} = \delta^{\rho\kappa\alpha\beta}_{\lambda\sigma\gamma\delta} \tensor{R}{^{\lambda\sigma}_{\rho\kappa;\alpha}}$ the cyclic permutation of $\rho$, $\kappa$ and $\alpha$ are summed and hence this term goes to zero for the second Bianchi identity, and in eq. \eqref{eq:var_parts} only the term which contains $\nabla_\alpha F[\phi]$ survives.\\
Integrating by parts once again
\begin{multline}
	\delta S_{GB} = \int_\Omega d^4x \Bigl( \nabla_\omega \nabla_\alpha F[\phi] \Bigr) \sqrt{-g} \delta^{\rho\kappa\alpha\beta}_{\lambda\sigma\gamma\delta} \tensor{R}{^{\lambda\sigma}_{\rho\kappa}} g^{\omega\delta} g^{\gamma\epsilon} \delta g_{\beta\epsilon} = \\
	= \int_\Omega d^4x \Bigl( \nabla_\alpha \nabla_\omega F[\phi] \Bigr) \sqrt{-g} \delta^{\rho\kappa\alpha\beta}_{\lambda\sigma\gamma\delta} \tensor{R}{^{\lambda\sigma}_{\rho\kappa}} g^{\omega\delta} g^{\gamma\epsilon} \delta g_{\beta\epsilon} = \\
	= \int_\Omega d^4x \Bigl( \nabla_\alpha \nabla^\delta F[\phi] \Bigr) \sqrt{-g} \delta^{\rho\kappa\alpha\beta}_{\lambda\sigma\gamma\delta} \tensor{R}{^{\lambda\sigma}_{\rho\kappa}} g^{\gamma\epsilon} \delta g_{\beta\epsilon}.
	\label{eq:var_second-parts}
\end{multline}
In the second line of eq. \eqref{eq:var_second-parts} we exchanged the two covariant derivatives applied to $F[\phi]$. This can be done since $F[\phi]$ is scalar and therefore 
\begin{multline}
	\nabla_\omega \nabla_\alpha F[\phi] = \nabla_\omega \partial_\alpha F[\phi] = \partial_\omega \partial_\alpha F[\phi] - \Gamma^\gamma_{\omega\alpha} \partial_\gamma F[\phi] = \\
	= \partial_\alpha \partial_\omega F[\phi] - \Gamma^\gamma_{\alpha\omega} \partial_\gamma F[\phi] = \nabla_\alpha \partial_\omega F[\phi] = \nabla_\alpha \nabla_\omega F[\phi].
\end{multline}

Relabeling some indices in eq. \eqref{eq:var_second-parts} we obtain
\begin{multline}
	\delta S_{GB} = \int_\Omega d^4x \Bigl( \nabla_\gamma \nabla^\alpha F[\phi] \Bigr) \sqrt{-g} \delta^{\gamma\delta\kappa\lambda}_{\beta\alpha\rho\sigma} \tensor{R}{^{\rho\sigma}_{\kappa\lambda}} g^{\beta\epsilon} \delta g_{\delta\epsilon} = \\
	= \int_\Omega d^4x \Bigl( \nabla_\gamma \nabla^\alpha F[\phi] \Bigr) \sqrt{-g} \delta^{\gamma\delta\kappa\lambda}_{\beta\alpha\rho\sigma} \tensor{R}{^{\rho\sigma}_{\kappa\lambda}} \delta^\beta_\mu g^{\mu\nu} \delta g_{\delta\nu}.
\end{multline}
Since \cite{Ferrari-Gualtieri-Pani}
\begin{equation}
	g^{\mu\nu} \delta g_{\nu\delta} = -g_{\nu\delta} \delta g^{\mu\nu},
\end{equation}
we finally obtain, for the Gauss-Bonnet term,
\begin{equation}
	\delta S_{GB} = - \int_\Omega d^4x \Bigl( \nabla_\gamma \nabla^\alpha F[\phi] \Bigr) \sqrt{-g} \delta^{\gamma\delta\kappa\lambda}_{\beta\alpha\rho\sigma} \tensor{R}{^{\rho\sigma}_{\kappa\lambda}} \delta^\beta_\mu g_{\delta\nu} \delta g^{\mu\nu}.
	\label{eq:var_S_GB_fin}
\end{equation}
\\

Instead for the field term 
\begin{multline}
	\delta S_\phi = \delta \biggl( - \frac{1}{2} \int_\Omega d^4x \sqrt{-g} \bigl( \nabla \phi \bigr)^2 \biggr) = \delta \biggl( - \frac{1}{2} \int_\Omega d^4x \sqrt{-g} \bigl( \nabla_\mu \phi \bigr) \bigl( \nabla_\nu \phi \bigr) g^{\mu\nu} \biggr) = \\
	= - \frac{1}{2} \int_\Omega d^4x \sqrt{-g} \biggl( - \frac{1}{2} g_{\mu\nu} \biggr) \bigl( \nabla_\alpha \phi \bigr) \bigl(\nabla^\alpha \phi \bigr) \, \delta g^{\mu\nu} - \frac{1}{2} \int_\Omega d^4x \sqrt{-g} \bigl( \nabla_\mu \phi \bigr) \bigl( \nabla_\nu \phi \bigr) \, \delta g^{\mu\nu}.
\end{multline}
\\

Finally, for the Einstein-Hilbert term we use the derivation that can be found in ref. \cite{Ferrari-Gualtieri-Pani}, which consists in what follows.
Varying the Einstein-Hilbert action we obtain
\begin{multline}
	\delta S_{EH} = \delta \biggl( \frac{1}{2} \int_\Omega d^4x \sqrt{-g} R \biggr) = \delta \biggl( \frac{1}{2} \int_\Omega d^4x \sqrt{-g} R_{\mu\nu} g^{\mu\nu} \biggr) = \\
	= \frac{1}{2} \int_\Omega d^4x \biggl(- \frac{\sqrt{-g}}{2} R g_{\mu\nu} \delta g^{\mu\nu} + \sqrt{-g} R_{\mu\nu} \delta g^{\mu\nu} + \sqrt{-g} \delta R_{\mu\nu} g^{\mu\nu} \biggr).
	\label{eq:var_EH}
\end{multline}

For the Palatini identity
\begin{equation}
	\delta R_{\mu\nu} = \bigl( \delta \Gamma^\lambda_{\mu\nu} \bigr)_{;\lambda} - \bigl( \delta \Gamma^\lambda_{\mu\lambda} \bigr)_{;\nu},
	\label{eq:Palatini_id}
\end{equation}
hence
\begin{multline}
	\int_\Omega d^4x \sqrt{-g} \delta R_{\mu\nu} g^{\mu\nu} = \int_\Omega d^4x \sqrt{-g} \Bigl[ \bigl( \delta \Gamma^\lambda_{\mu\nu} \bigr)_{;\lambda} - \bigl( \delta \Gamma^\lambda_{\mu\lambda} \bigr)_{;\nu} \Bigr] g^{\mu\nu} = \\
	= \int_\Omega d^4x \sqrt{-g} \Bigl[ \bigl( \delta \Gamma^\lambda_{\mu\nu} g^{\mu\nu} \bigr)_{;\lambda} - \bigl( \delta \Gamma^\lambda_{\mu\lambda} g^{\mu\nu} \bigr)_{;\nu} \Bigr] = \\
	= \int_{\partial \Omega} dS_\lambda \delta \Gamma^\lambda_{\mu\nu} g^{\mu\nu} - \int_{\partial \Omega} dS_\nu \delta \Gamma^\lambda_{\mu\lambda} g^{\mu\nu} = 0,
\end{multline}
where in the first line we used the relation $g_{\mu\nu;\lambda} = 0$, in the second we used the Gauss' theorem for curved spaces and in the final step the condition $\delta \Gamma (\partial \Omega) = 0$.\\
Therefore for the Einstein-Hilbert term
\begin{equation}
	\delta S_{EH} = \frac{1}{2} \int_\Omega d^4x \sqrt{-g} \biggl(R_{\mu\nu} - \frac{1}{2} g_{\mu\nu} R \biggr) \delta g^{\mu\nu}.
\end{equation}
\\

Summing the three contributions we obtain
\begin{multline}
	\delta S = \delta S_{EH} + \delta S_\phi + \delta S_{GB} = \\
	= \frac{1}{2} \int_\Omega d^4x \sqrt{-g} \biggl[ R_{\mu\nu} - \frac{1}{2} g_{\mu\nu} R + \frac{1}{2} \bigl( \nabla \phi \bigr)^2 g_{\mu\nu} - \bigl( \nabla_\mu \phi \bigr) \bigl( \nabla_\nu \phi \bigr) + \\
	- 2 \Bigl( \nabla_\gamma \nabla^\alpha F[\phi] \Bigr) \delta^{\gamma\delta\kappa\lambda}_{\beta\alpha\rho\sigma} \tensor{R}{^{\rho\sigma}_{\kappa\lambda}} \delta^\beta_\mu g_{\delta\nu} \biggr] \delta g^{\mu\nu}, 
\end{multline}
from which, once exchanged $\alpha$ and $\beta$ in the generalized Kronecker delta, we obtain the field equation for the metric tensor
\begin{multline}
	E^{(g)}_{\mu\nu} := R_{\mu\nu} - \frac{1}{2} g_{\mu\nu} R + \frac{1}{2} \bigl( \nabla \phi \bigr)^2 g_{\mu\nu} - \bigl( \nabla_\mu \phi \bigr) \bigl( \nabla_\nu \phi \bigr) + \\
	+ 2 \Bigl( \nabla_\gamma \nabla^\alpha F[\phi] \Bigr) \delta^{\gamma\delta\kappa\lambda}_{\alpha\beta\rho\sigma} \tensor{R}{^{\rho\sigma}_{\kappa\lambda}} \delta^\beta_\mu g_{\nu\delta} = 0.
\end{multline}

\backmatter


\begin{thebibliography}{99} 

\bibitem{Eddington}
Dyson Frank Watson , Eddington Arthur Stanley and Davidson C. 1920IX. A determination of the deflection of light by the sun's gravitational field, from observations made at the total eclipse of May 29, 1919. Philosophical Transactions of the Royal Society of London. Series A, Containing Papers of a Mathematical or Physical Character 220:291-333. http://doi.org/10.1098/rsta.1920.0009

\bibitem{LIGO1}
B.~P.~Abbott \textit{et al.} [LIGO Scientific and Virgo], ``Observation of Gravitational Waves from a Binary Black Hole Merger,'' Phys. Rev. Lett. \textbf{116} (2016) no.6, 061102, doi:10.1103/PhysRevLett.116.061102, [arXiv:1602.03837 [gr-qc]].

\bibitem{LIGO2}
B.~P.~Abbott \textit{et al.} [LIGO Scientific and Virgo], ``GW151226: Observation of Gravitational Waves from a 22-Solar-Mass Binary Black Hole Coalescence,'' Phys. Rev. Lett. \textbf{116} (2016) no.24, 241103, doi:10.1103/PhysRevLett.116.241103, [arXiv:1606.04855 [gr-qc]].

\bibitem{LIGO3}
B.~P.~Abbott \textit{et al.} [LIGO Scientific and VIRGO], ``GW170104: Observation of a 50-Solar-Mass Binary Black Hole Coalescence at Redshift 0.2,'' Phys. Rev. Lett. \textbf{118} (2017) no.22, 221101, [erratum: Phys. Rev. Lett. \textbf{121} (2018) no.12, 129901], doi:10.1103/PhysRevLett.118.221101, [arXiv:1706.01812 [gr-qc]].

\bibitem{LIGO4}
B.~P.~Abbott \textit{et al.} [LIGO Scientific and Virgo], ``GW170608: Observation of a 19-solar-mass Binary Black Hole Coalescence,'' Astrophys. J. \textbf{851} (2017) no.2, L35, doi:10.3847/2041-8213/aa9f0c, [arXiv:1711.05578 [astro-ph.HE]].

\bibitem{LIGO5}
B.~P.~Abbott \textit{et al.} [LIGO Scientific and Virgo], ``GW170814: A Three-Detector Observation of Gravitational Waves from a Binary Black Hole Coalescence,'' Phys. Rev. Lett. \textbf{119} (2017) no.14, 141101, doi:10.1103/PhysRevLett.119.141101, [arXiv:1709.09660 [gr-qc]].

\bibitem{LIGO6}
B.~P.~Abbott \textit{et al.} [LIGO Scientific and Virgo], ``GW170817: Observation of Gravitational Waves from a Binary Neutron Star Inspiral,'' Phys. Rev. Lett. \textbf{119} (2017) no.16, 161101, doi:10.1103/PhysRevLett.119.161101, [arXiv:1710.05832 [gr-qc]].

\bibitem{LIGO7}
B.~P.~Abbott \textit{et al.} [LIGO Scientific and Virgo], ``GW190425: Observation of a Compact Binary Coalescence with Total Mass $\sim 3.4 M_{\odot}$,'' Astrophys. J. Lett. \textbf{892} (2020) no.1, L3, doi:10.3847/2041-8213/ab75f5, [arXiv:2001.01761 [astro-ph.HE]].

\bibitem{LIGO8}
R.~Abbott \textit{et al.} [LIGO Scientific and Virgo], ``GW190412: Observation of a Binary-Black-Hole Coalescence with Asymmetric Masses,'' Phys. Rev. D \textbf{102} (2020) no.4, 043015, doi:10.1103/PhysRevD.102.043015, [arXiv:2004.08342 [astro-ph.HE]].

\bibitem{LIGO9}
R.~Abbott \textit{et al.} [LIGO Scientific and Virgo], ``GW190814: Gravitational Waves from the Coalescence of a 23 Solar Mass Black Hole with a 2.6 Solar Mass Compact Object,'' Astrophys. J. Lett. \textbf{896} (2020) no.2, L44, doi:10.3847/2041-8213/ab960f, [arXiv:2006.12611 [astro-ph.HE]].

\bibitem{LIGO10}
R.~Abbott \textit{et al.} [LIGO Scientific and Virgo], ``GW190521: A Binary Black Hole Merger with a Total Mass of $150 ~ M_{\odot}$,'' Phys. Rev. Lett. \textbf{125} (2020) no.10, 101102, doi:10.1103/PhysRevLett.125.101102, [arXiv:2009.01075 [gr-qc]].

\bibitem{EHTC1}
K.~Akiyama \textit{et al.} [Event Horizon Telescope], ``First M87 Event Horizon Telescope Results. I. The Shadow of the Supermassive Black Hole,'' Astrophys. J. \textbf{875} (2019) no.1, L1, doi:10.3847/2041-8213/ab0ec7, [arXiv:1906.11238 [astro-ph.GA]].

\bibitem{EHTC2}
K.~Akiyama \textit{et al.} [Event Horizon Telescope], ``First M87 Event Horizon Telescope Results. II. Array and Instrumentation,'' Astrophys. J. Lett. \textbf{875} (2019) no.1, L2, doi:10.3847/2041-8213/ab0c96, [arXiv:1906.11239 [astro-ph.IM]].

\bibitem{EHTC3}
K.~Akiyama \textit{et al.} [Event Horizon Telescope],  ``First M87 Event Horizon Telescope Results. III. Data Processing and Calibration,'' Astrophys. J. Lett. \textbf{875} (2019) no.1, L3, doi:10.3847/2041-8213/ab0c57, [arXiv:1906.11240 [astro-ph.GA]].

\bibitem{EHTC4}
K.~Akiyama \textit{et al.} [Event Horizon Telescope], ``First M87 Event Horizon Telescope Results. IV. Imaging the Central Supermassive Black Hole,'' Astrophys. J. Lett. \textbf{875} (2019) no.1, L4, doi:10.3847/2041-8213/ab0e85, [arXiv:1906.11241 [astro-ph.GA]].

\bibitem{EHTC5}
K.~Akiyama \textit{et al.} [Event Horizon Telescope], ``First M87 Event Horizon Telescope Results. V. Physical Origin of the Asymmetric Ring,'' Astrophys. J. Lett. \textbf{875} (2019) no.1, L5, doi:10.3847/2041-8213/ab0f43, [arXiv:1906.11242 [astro-ph.GA]].

\bibitem{EHTC6}
K.~Akiyama \textit{et al.} [Event Horizon Telescope], ``First M87 Event Horizon Telescope Results. VI. The Shadow and Mass of the Central Black Hole,'' Astrophys. J. Lett. \textbf{875} (2019) no.1, L6, doi:10.3847/2041-8213/ab1141, [arXiv:1906.11243 [astro-ph.GA]].

\bibitem{Weinberg1}
Weinberg, S. (1995). The Quantum Theory of Fields. Cambridge: Cambridge University Press. doi:10.1017/CBO9781139644167 .

\bibitem{DarkEnergy}
R.~Bean, S.~M.~Carroll and M.~Trodden, ``Insights into dark energy: interplay between theory and observation,'' [arXiv:astro-ph/0510059 [astro-ph]].

\bibitem{CosmologicalConstantMeasure}
A.~G.~Riess \textit{et al.} [Supernova Search Team], ``Observational evidence from supernovae for an accelerating universe and a cosmological constant,'' Astron. J. \textbf{116} (1998), 1009-1038, doi:10.1086/300499, [arXiv:astro-ph/9805201 [astro-ph]].

\bibitem{CosmologicalConstantCarroll}
S.~M.~Carroll, ``The Cosmological constant,'' Living Rev. Rel. \textbf{4} (2001), 1, doi:10.12942/lrr-2001-1, [arXiv:astro-ph/0004075 [astro-ph]].

\bibitem{WeinbergCosmologicalConstant}
S.~Weinberg, ``The Cosmological Constant Problem,'' Rev. Mod. Phys. \textbf{61} (1989), 1-23, doi:10.1103/RevModPhys.61.1 .

\bibitem{Berti+}
E.~Berti, E.~Barausse, V.~Cardoso, L.~Gualtieri, P.~Pani, U.~Sperhake, L.~C.~Stein, N.~Wex, K.~Yagi and T.~Baker, \textit{et al.}, ``Testing General Relativity with Present and Future Astrophysical Observations,'' Class. Quant. Grav. \textbf{32} (2015), 243001, doi:10.1088/0264-9381/32/24/243001, [arXiv:1501.07274 [gr-qc]].

\bibitem{NS-NSmerger}
J.~Sakstein and B.~Jain, ``Implications of the Neutron Star Merger GW170817 for Cosmological Scalar-Tensor Theories,'' Phys. Rev. Lett. \textbf{119} (2017) no.25, 251303, doi:10.1103/PhysRevLett.119.251303, [arXiv:1710.05893 [astro-ph.CO]].

\bibitem{GWcorrections}
H.~Witek, L.~Gualtieri, P.~Pani and T.~P.~Sotiriou, ``Black holes and binary mergers in scalar Gauss-Bonnet gravity: scalar field dynamics,'' Phys. Rev. D \textbf{99} (2019) no.6, 064035, doi:10.1103/PhysRevD.99.064035, [arXiv:1810.05177 [gr-qc]].

\bibitem{Lovelock}
D.~Lovelock, ``The four-dimensionality of space and the einstein tensor,'' J. Math. Phys. \textbf{13} (1972), 874-876, doi:10.1063/1.1666069 .

\bibitem{Ferrari-Gualtieri-Pani}
Ferrari, Gualtieri, Pani; ``General Relativity: the physical theory of
gravity - with applications to compact objects and gravitational-wave
sources'' (Taylor \& Francis 2020).

\bibitem{HorndeskiOriginal}
G.~W.~Horndeski, ``Second-order scalar-tensor field equations in a four-dimensional space,'' Int. J. Theor. Phys. \textbf{10} (1974), 363-384, doi:10.1007/BF01807638 .

\bibitem{QuirosScalarTensor}
I.~Quiros, ``Selected topics in scalar\textendash{}tensor theories and beyond,'' Int. J. Mod. Phys. D \textbf{28} (2019) no.07, 1930012, doi:10.1142/S021827181930012X, [arXiv:1901.08690 [gr-qc]].

\bibitem{Representations}
T.~P.~Sotiriou, V.~Faraoni and S.~Liberati, ``Theory of gravitation theories: A No-progress report,'' Int. J. Mod. Phys. D \textbf{17} (2008), 399-423, doi:10.1142/S0218271808012097, [arXiv:0707.2748 [gr-qc]].

\bibitem{WoodardOstrogradski}
R.~P.~Woodard, Scholarpedia \textbf{10} (2015) no.8, 32243, doi:10.4249/scholarpedia.32243, [arXiv:1506.02210 [hep-th]].

\bibitem{KobayashiHorndeski}
T.~Kobayashi, ``Horndeski theory and beyond: a review'', Rept. Prog. Phys. \textbf{82} (2019) no.8, 086901, doi:10.1088/1361-6633/ab2429, [arXiv:1901.07183 [gr-qc]].

\bibitem{StelleRenormalization}
K.~S.~Stelle, ``Renormalization of Higher Derivative Quantum Gravity,'' Phys. Rev. D \textbf{16} (1977), 953-969, doi:10.1103/PhysRevD.16.953 .

\bibitem{RipleyPretorius}
J.~L.~Ripley and F.~Pretorius, ``Gravitational collapse in Einstein dilaton-Gauss\textendash{}Bonnet gravity,'' Class. Quant. Grav. \textbf{36} (2019) no.13, 134001, doi:10.1088/1361-6382/ab2416, [arXiv:1903.07543 [gr-qc]].

\bibitem{CardosoGualtieri}
V.~Cardoso and L.~Gualtieri, ``Testing the black hole `no-hair' hypothesis,'' Class. Quant. Grav. \textbf{33} (2016) no.17, 174001, doi:10.1088/0264-9381/33/17/174001,	[arXiv:1607.03133 [gr-qc]].

\bibitem{SotiriouNoHair}
T.~P.~Sotiriou and V.~Faraoni, ``Black holes in scalar-tensor gravity,'' Phys. Rev. Lett. \textbf{108} (2012), 081103, doi:10.1103/PhysRevLett.108.081103, [arXiv:1109.6324 [gr-qc]].

\bibitem{GrahamNoHair}
A.~A.~H.~Graham and R.~Jha, ``Stationary Black Holes with Time-Dependent Scalar Fields,'' Phys. Rev. D \textbf{90} (2014) no.4, 041501, doi:10.1103/PhysRevD.90.041501, [arXiv:1407.6573 [gr-qc]].

\bibitem{Torii}
T.~Torii, K.~I.~Maeda and T.~Tachizawa, ``NonAbelian black holes and catastrophe theory. 1. Neutral type,'' Phys. Rev. D \textbf{51} (1995), 1510-1524, doi:10.1103/PhysRevD.51.1510, [arXiv:gr-qc/9406013 [gr-qc]].

\bibitem{ShiftSymmetricNoHair}
L.~Hui and A.~Nicolis, ``No-Hair Theorem for the Galileon,'' Phys. Rev. Lett. \textbf{110} (2013), 241104, doi:10.1103/PhysRevLett.110.241104, [arXiv:1202.1296 [hep-th]].

\bibitem{ShiftControesempio}
T.~P.~Sotiriou and S.~Y.~Zhou, ``Black hole hair in generalized scalar-tensor gravity,'' Phys. Rev. Lett. \textbf{112} (2014), 251102, doi:10.1103/PhysRevLett.112.251102, [arXiv:1312.3622 [gr-qc]].

\bibitem{Kanti}
P.~Kanti, N.~E.~Mavromatos, J.~Rizos, K.~Tamvakis and E.~Winstanley, ``Dilatonic black holes in higher curvature string gravity,'' Phys. Rev. D \textbf{54} (1996), 5049-5058, doi:10.1103/PhysRevD.54.5049, [arXiv:hep-th/9511071 [hep-th]].

\bibitem{EGBNoHair}
H.~O.~Silva, J.~Sakstein, L.~Gualtieri, T.~P.~Sotiriou and E.~Berti, ``Spontaneous scalarization of black holes and compact stars from a Gauss-Bonnet coupling,'' Phys. Rev. Lett. \textbf{120} (2018) no.13, 131104, doi:10.1103/PhysRevLett.120.131104, [arXiv:1711.02080 [gr-qc]].

\bibitem{tachyon}
G.~Feinberg, ``Possibility of Faster-Than-Light Particles,'' Phys. Rev. \textbf{15} (1967), 1089-1105, doi:10.1103/PhysRev.159.1089 .

\bibitem{DonevaYazadjiev}
D.~D.~Doneva and S.~S.~Yazadjiev, ``New Gauss-Bonnet Black Holes with Curvature-Induced Scalarization in Extended Scalar-Tensor Theories,'' Phys. Rev. Lett. \textbf{120} (2018) no.13, 131103, doi:10.1103/PhysRevLett.120.131103, [arXiv:1711.01187 [gr-qc]].

\bibitem{DEF1}
T.~Damour and G.~Esposito-Farese,``Nonperturbative strong field effects in tensor - scalar theories of gravitation,'' Phys. Rev. Lett. \textbf{70} (1993), 2220-2223, doi:10.1103/PhysRevLett.70.2220 .

\bibitem{DEF2}
T.~Damour and G.~Esposito-Farese, ``Tensor - scalar gravity and binary pulsar experiments,'' Phys. Rev. D \textbf{54} (1996), 1474-1491, doi:10.1103/PhysRevD.54.1474, [arXiv:gr-qc/9602056 [gr-qc]].

\bibitem{DonevaMassive}
D.~D.~Doneva, K.~V.~Staykov and S.~S.~Yazadjiev, ``Gauss-Bonnet black holes with a massive scalar field,'' Phys. Rev. D \textbf{99} (2019) no.10, 104045, doi:10.1103/PhysRevD.99.104045, [arXiv:1903.08119 [gr-qc]].

\bibitem{BlazquezSalcedo+}
J.~L.~Bl\'azquez-Salcedo, D.~D.~Doneva, J.~Kunz and S.~S.~Yazadjiev, ``Radial perturbations of the scalarized Einstein-Gauss-Bonnet black holes,'' Phys. Rev. D \textbf{98} (2018) no.8, 084011, doi:10.1103/PhysRevD.98.084011, [arXiv:1805.05755 [gr-qc]].

\bibitem{Silva+Stability}
H.~O.~Silva, C.~F.~B.~Macedo, T.~P.~Sotiriou, L.~Gualtieri, J.~Sakstein and E.~Berti, ``Stability of scalarized black hole solutions in scalar-Gauss-Bonnet gravity,'' Phys. Rev. D \textbf{99} (2019) no.6, 064011, doi:10.1103/PhysRevD.99.064011, [arXiv:1812.05590 [gr-qc]].

\bibitem{MasatoTaishi}
M.~Minamitsuji and T.~Ikeda, ``Scalarized black holes in the presence of the coupling to Gauss-Bonnet gravity,'' Phys. Rev. D \textbf{99} (2019) no.4, 044017, doi:10.1103/PhysRevD.99.044017, [arXiv:1812.03551 [gr-qc]].

\bibitem{Macedo+}
C.~F.~B.~Macedo, J.~Sakstein, E.~Berti, L.~Gualtieri, H.~O.~Silva and T.~P.~Sotiriou, ``Self-interactions and Spontaneous Black Hole Scalarization,'' Phys. Rev. D \textbf{99} (2019) no.10, 104041, doi:10.1103/PhysRevD.99.104041, [arXiv:1903.06784 [gr-qc]].

\bibitem{DonevaCharged}
D.~D.~Doneva, S.~Kiorpelidi, P.~G.~Nedkova, E.~Papantonopoulos and S.~S.~Yazadjiev, ``Charged Gauss-Bonnet black holes with curvature induced scalarization in the extended scalar-tensor theories,'' Phys. Rev. D \textbf{98} (2018) no.10, 104056, doi:10.1103/PhysRevD.98.104056, [arXiv:1809.00844 [gr-qc]].

\bibitem{Charged2}
Y.~Brihaye and B.~Hartmann, ``Spontaneous scalarization of charged black holes at the approach to extremality,'' Phys. Lett. B \textbf{792} (2019), 244-250, doi:10.1016/j.physletb.2019.03.043, [arXiv:1902.05760 [gr-qc]].

\bibitem{Dima+}
A.~Dima, E.~Barausse, N.~Franchini and T.~P.~Sotiriou, ``Spin-induced black hole spontaneous scalarization,'' Phys. Rev. Lett. \textbf{125} (2020) no.23, 231101, doi:10.1103/PhysRevLett.125.231101, [arXiv:2006.03095 [gr-qc]].

\bibitem{Hod}
S.~Hod, ``Onset of spontaneous scalarization in spinning Gauss-Bonnet black holes,'' Phys. Rev. D \textbf{102} (2020) no.8, 084060, doi:10.1103/PhysRevD.102.084060, [arXiv:2006.09399 [gr-qc]].

\bibitem{contodelta}
G.~'t Hooft and M.~J.~G.~Veltman, ``One loop divergencies in the theory of gravitation,'' Ann. Inst. H. Poincare Phys. Theor. A \textbf{20} (1974), 69-94.

\bibitem{Mathematica}
Wolfram Research, Inc., Mathematica, Version 12.0, Champaign, IL (2019).

\bibitem{LaplaceDeRahm}
Fecko, M. (2006). \textit{Differential Geometry and Lie Groups for Physicists}. Cambridge: Cambridge University Press. doi:10.1017/CBO9780511755590 .

\bibitem{ScholarpediaMOL}
Samir Hamdi et al. (2007) Method of lines. Scholarpedia, 2(7):2859, doi:10.4249/scholarpedia.2859 .

\bibitem{NumericalRecipes}
William H. Press, Saul A. Teukolsky, William T. Vetterling, and Brian P. Flannery. 1992. Numerical Recipes in C: The Art of Scientific Computing (2nd. ed.). Cambridge University Press, USA.

\bibitem{NumericDerivatives}
Fornberg, Bengt, ``Generation of Finite Difference Formulas on Arbitrarily Spaced Grids,'' \textit{Mathematics of Computation} \textbf{51} (1988), no. 184, 699-706, doi:10.2307/2008770 .

\bibitem{ApparentHorizonFoliation}
V.~Faraoni, G.~F.~R.~Ellis, J.~T.~Firouzjaee, A.~Helou and I.~Musco, ``Foliation dependence of black hole apparent horizons in spherical symmetry,'' Phys. Rev. D \textbf{95} (2017) no.2, 024008, doi:10.1103/PhysRevD.95.024008, [arXiv:1610.05822 [gr-qc]].

\bibitem{Hayward}
S.~A.~Hayward, ``General laws of black hole dynamics,'' Phys. Rev. D \textbf{49} (1994), 6467-6474, doi:10.1103/PhysRevD.49.6467 [arXiv:gr-qc/9303006].

\end{thebibliography}
\end{document}